\documentclass[aps,prx,superscriptaddress,amsmath,amssymb,floatfix,twocolumn,showpacs,amsfonts,longbibliography]{revtex4-1}
\usepackage{times}
\usepackage[varg]{txfonts}
\usepackage[utf8x]{inputenc}
\usepackage{textcomp}
\usepackage{graphicx}
\usepackage{subfigure}
\usepackage{tabu}
\usepackage{color}
\usepackage[colorlinks=true,citecolor=blue,urlcolor=blue,linkcolor=blue,hyperindex]{hyperref}
\usepackage{braket}
\usepackage{overpic}
\usepackage{amssymb}
\makeatletter

\input epsf

\def\be{\begin{equation}}
\def\ee{\end{equation}}
\def\ba{\begin{eqnarray}}
\def\ea{\end{eqnarray}}
\makeatother

\allowdisplaybreaks

\begin{document}
\title{Deconfined quantum criticality in spin-1/2 chains with long-range interactions}
\author{Sibin Yang}
\affiliation{State Key Laboratory of Optoelectronic Materials and Technologies, School of Physics, Sun Yat-Sen University, Guangzhou 510275, China}
\affiliation{Department of Physics, Boston University, 590 Commonwealth Avenue, Boston, Massachusetts 02215, USA}

\author{Dao-Xin Yao}
\email{yaodaox@mail.sysu.edu.cn}
\affiliation{State Key Laboratory of Optoelectronic Materials and Technologies, School of Physics, Sun Yat-Sen University, Guangzhou 510275, China}

\author{Anders W. Sandvik}
\email{sandvik@bu.edu}
\affiliation{Department of Physics, Boston University, 590 Commonwealth Avenue, Boston, Massachusetts 02215, USA}
\affiliation{Beijing National Laboratory of Condensed Matter Physics and Institute of Physics, \\ Chinese Academy of Sciences, Beijing 100190, China}

\date{\today}

\begin{abstract}
We study spin-$1/2$ chains with long-range power-law decaying unfrustrated (bipartite) Heisenberg exchange
$J_r \propto r^{-\alpha}$ and a competing multi-spin interaction $Q$ favoring a dimerized (valence-bond solid, VBS) ground state. Employing
quantum Monte Carlo techniques and Lanczos diagonalization, we analyze order parameters and excited-state level crossings
to characterize quantum phase transitions between the different ground states in the $(Q,\alpha)$ plane. For weak multi-spin coupling
$Q$ and sufficiently slowly decaying Heisenberg interactions (small $\alpha$), the system has a long-range-ordered antiferromagnetic (AFM) ground state, and upon
increasing $\alpha$ there is a direct, continuous transition into a quasi long-range ordered (QLRO) critical state of the type in the standard Heisenberg
chain. This transition has been studied previously in other models and we further characterize it here. For rapidly decaying long-range interactions the
system undergoes a transition between QLRO and VBS ground states of the same kind as in the frustrated $J_1$-$J_2$ Heisenberg chain.
Our most important finding is a direct continuous quantum phase transition between the AFM and VBS states---a close analogy to the two-dimensional deconfined
quantum-critical point. In previous one-dimensional analogies of deconfined quantum criticality the two ordered phases both have gapped fractional
excitations, and the gapless critical point can be described by conventional Luttinger-Liquid theory. In contrast, in our model the excitations
fractionalize upon transitioning from the gapless AFM state, changing from spin waves to deconfined spinons.
We extract critical exponents at the AFM--VBS transition and use order-parameter distributions to study emergent symmetries. We find that
the O($3$) AFM and scalar VBS order parameters combine into an O($4$) vector at the critical point, but this symmetry
is only apparent after a scale transformation is applied to one of the order parameters. Thus, the order parameter fluctuations exhibit covariance
of a distribution in an uniaxially deformed O($4$) sphere (a so-called elliptical symmetry), with the anisotropy increasing with the length scale on which 
it is observed. This unusual quantum phase transition does not yet have any known field theory description, and our detailed results can
serve to guide its construction. We also discuss possibilities to detect the quantum phases and quantum phase transitions of the model experimentally,
e.g., in trapped-ion or Rydberg-atom systems, where long-range spin interactions in linear chains can be engineered.
\end{abstract}

\maketitle

\section{Introduction}

The possibility of direct, continuous quantum phase transitions between antiferromagnetic (AFM) and spontaneously dimerized valence-bond solid (VBS) ground
states in two-dimensional (2D) quantum spin systems has been under intense scrutiny during the past several years. Following numerical results pointing to the
existence of such unusual order--order transitions \cite{Sandvik02,Motrunich04} and prior field-theory descriptions of both AFM and VBS states in 2D quantum
magnets \cite{Haldane83,Chakravarty89,Read90,Murthy90}, the deconfined quantum critical point (DQCP) \cite{Senthil04a,Senthil04b,Sachdev08} was proposed as
a scenario for a generically continuous AFM--VBS quantum phase transition. In contrast, within the standard Landau-Ginzburg-Wilson (LGW) paradigm, such
a phase transition with simultaneous breaking of two unrelated symmetries should require fine-tuning of parameters in order to avoid a first-order
transition or a coexistence phase. In this paper we explore an analogy to the 2D DQCP in a one-dimensional (1D) quantum spin chain with competing
long-range AFM interactions and short-range couplings favoring VBS formation.

\subsection{Deconfined quantum criticality}

The 2D DQCP is described field-theoretically by spin $S=1/2$ carrying spinon degrees of freedom coupled to a non-compact U($1$) gauge field \cite{Senthil04a}.
The AFM and VBS order parameters should be understood as composites of these objects and not as independent order parameters. While this fundamental difference
from the LGW formulation with two separate order parameters is indicative of a direct AFM--VBS transition, the unusual continuous nature of this transition
has been proven rigorously within the field theory only in a limit where the SU($2$) symmetry of the spinons is enhanced to SU($N$) with large $N$. Numerous
numerical studies of 2D quantum spin Hamiltonians designed to host
AFM--VBS transitions \cite{Sandvik07,Melko08,Jiang08,Lou09,Sandvik10a,Kaul11,Harada13,Chen13,Block13,Pujari15,Shao16,Qin17,Ma18}, and of related 3D classical
lattice models \cite{Kragset06,Sreejith14,Nahum15a,Nahum15b,Sreejith19}, have been carried out in order to test the theory for $N=2$ and other small values of $N$.
Signatures of deconfined spinon excitations are observable on long length scales in both isotropic and anisotropic $S=1/2$ systems \cite{Shao16,Ma18}, and
for SU($N$) models there is a striking agreement between $1/N$ expansions for the critical exponents and simulation results for lattice models with moderate
values of $N$ \cite{Kaul12,Dyer15}. However, a consensus on the ultimate nature of the transition for small $N$---continuous or very weakly first-order---is still
lacking \cite{Wang17,Ma19,Nahum19}.

In the case of 1D systems, quantum phase transitions beyond the LGW description have been well understood for a long time and are generally described within
the framework of the Luttinger Liquid (LL) \cite{Voit95}. This description also allows for continuous order--order transitions
\cite{Nakamura99,Nakamura00,Sengupta02,Sandvik04,Tsuchiizu04}. Such 1D transitions were initially not discussed explicitly in terms of deconfinement, because also the
ordered phases have deconfined excitations---domain-wall-like topological defects whose intrinsic size diverges as the LL critical point is approached
\cite{Tang11}. Recently it was again pointed out that the field-theory descriptions of these quantum phase transitions share some similarities with their putative 2D
DQCP counterparts, and alternative, dual field-theories can be constructed which make the analogies more explicit than the standard LL description \cite{Jiang19}.
This development has stimulated additional numerical studies of a specific 1D model exhibiting a transition between a ferromagnet and a VBS \cite{Roberts19,Huang19},
confirming a continuous transition and finding LL behavior at the gapless point separating the ordered phases. Previous studies of an extended 1D Hubbard model
had also found LL criticality separating two ordered phases (charge-density-wave and dimerized) \cite{Sandvik04}, and this type of transition as well should
be described by the alternative dual field theories.

Here we take the studies of 1D DQCP analogies in another direction by considering an $S=1/2$ spin chain with Heisenberg exchange interactions
decaying with distance as a power law, thus enabling true long-range AFM order to form (which is ruled out by the Mermin-Wagner theorem \cite{Mermin66} when
the interactions are short-ranged). We add a local multi-spin coupling favoring dimer order and study the ground state phases and quantum
phase transitions of the system as the parameters controlling the two types of couplings are varied. For sufficiently slowly decaying Heisenberg interactions
we find a direct, continuous AFM--VBS ground-state transition at which the elementary low-energy excitations change from $S=1$ spin waves with anomalous nonlinear
dispersion to deconfined $S=1/2$ spinons.

Though here we will focus on models, we note that long-range interacting spin chains and the phenomena we investigate are not merely of theoretical 
interest. Long-range Heisenberg interactions, with or without frustrated signs of the exchange couplings, can in principle be realized in linear arrays 
of metallic atoms \cite{Tung11}. Greater tunability and design of specific interactions is possible with trapped ion systems and Rydberg atoms in optical lattices, 
which are currently among the most promising platforms for quantum simulators \cite{Bohnet16,Zeiher17,Nguyen18}. Our results should provide useful guides
to possible exotic 1D states and transitions in these experimental settings.

In the reminder of this section we provide further background information and a brief expos\'e of the main results. In Sec.~\ref{sub:spinons} we further elaborate
on the place our work in the context of deconfined quantum criticality scenarios and emergent symmetries in one and two dimensions. In Sec.~\ref{sub:longrange} we 
summarize previous works on Heisenberg spin chains with long-range interactions. We define the new model in Sec.~\ref{sub:jqlong}, where we also preview the ground 
state phase diagram and our main results for the quantum phase transitions. In Sec.~\ref{sub:outline} we outline the orgaznization of the later sections.

\subsection{Spinons and emergent symmetries}
\label{sub:spinons}

Spinons were first discussed in the context of a 1D frustrated $S=1/2$ quantum spin models with long-range VBS order \cite{Shastry81} and the standard
Heisenberg chain with exact Bethe Ansatz solution \cite{Faddeev81}. In the conventional Heisenberg spin chain with nearest-neighbor interactions $J_1$, there is no
long-range AFM or VBS order; the ground state is critical, with both spin and dimer correlations decaying with distance $r$ as $r^{-1}$, up to multiplicative 
logarithmic (log) corrections \cite{Giamarchi89}. This quasi-long-range ordered (QLRO) state undergoes a transition into a two-fold degenerate ordered VBS state once
sufficiently strong frustrated next-nearest-neighbor interactions $J_2$ are introduced \cite{Majumdar69a,Majumdar69b}. The same QLRO--VBS transition can also be
realized in spin chains with phonons (the spin-Peierls mechanism) \cite{Uhrig97,Sandvik99,Suwa15}, or in $J$-$Q$ models with certain multi-spin interactions
$Q$ (projectors of locally correlated singlets) instead of the $J_2$ couplings \cite{Tang11,Sanyal11,Patil18}.

In a field-theory description, the dimerization transition is driven by a perturbation which is marginally irrelevant in the QLRO phase (causing 
the log corrections) and becomes marginally relevant in the VBS phase (leading to the opening of an initially exponentially small gap) \cite{Affleck85,Affleck87}.
The marginal perturbation vanishes at the transition point, i.e., it changes sign. Unlike the case of the analogous 2D systems, the spinons are deconfined 
in both phases, because of the lack of confining potential (the presence of which in 2D is directly related to the higher dimensionality \cite{Sulejman17}). 
In the VBS phase, the spinons are finite domain walls between the two degenerate dimer patterns, while in the QLRO phase they can be regarded as critical 
AFM or VBS domain walls (i.e., these domain walls do not have finite extent but are characterized by a power-law shape \cite{Tang11}).
  
If long-range unfrustrated Heisenberg interactions are included, true long-range AFM order can also be stabilized in a 1D system, since the Mermin-Wagner
theorem only rules out breaking of the spin-rotation symmetry when the interactions are short-ranged \cite{Mermin66}. In a Heisenberg chain with power-law
decaying long-range unfrustrated interactions of strength $J_r \propto (-1)^{r-1}r^{-\alpha}$, AFM order appears if the decay power is sufficiently small---the
critical value of $\alpha$ is non-universal, depending on details of the short-range interactions \cite{Laflorencie05,Sandvik10b}. In the AFM phase, spinons
no longer exist as elementary excitations and the ground state can be regarded as a spinon condensate with gapless $S=1$ excitations---spin waves with anomalous,
sublinear dispersion relation \cite{Yusuf04}. It has been confirmed numerically that spinons are not well-defined quasiparticles
in the AFM state induced by long-range interactions \cite{Tang11}.

The QLRO--VBS transition in the $J_1$-$J_2$ and $J$-$Q$ chains shares some similarities with the 2D DQCP, even though only the VBS phase has true
long-range order. The analogy is manifested in the way the AFM and VBS order parameters are not independent but arise out of common spinon degrees of
freedom that interact in different ways on the two sides of the phase transition. In both cases the critical point is associated with an emergent
symmetry: In 1D, it has been known for a long time that the O(3) AFM order parameter and the scalar VBS order parameter decay according to the same power law
and form a critical O($4$) symmetric order parameter. The higher symmetry is explicit in the Wess-Zumino-Witten conformal field theory (CFT) of $S=1/2$
spin chains \cite{Affleck85,Affleck87}, where three components of the vector field correspond to the AFM order parameter and the fourth component
represents the VBS order parameter. A recent numerical study of a $J$-$Q$ spin chain at the dimerization transition has demonstrated how violations
of the symmetry vanish with increasing distance or system size \cite{Patil18}. 

In a four-fold degenerate columnar VBS state on the 2D square-lattice, the $Z_4$ symmetric order parameter is akin to the $Z_q$ order parameter of a $q$-state
classical 3D clock model, which for $q\ge 4$ exhibits emergent U($1$) symmetry and an XY-universal phase transition \cite{Levin04}. The prediction of emergent
U($1$) symmetry in the neighborhood of the DQCP been confirmed in numerical studies of the $J$-$Q$ model \cite{Sandvik07,Jiang08}. In one variant of the DQCP
scenario for SU($2$) spins \cite{Senthil06,Wang17}, the O($3$) AFM order parameter and the emergent U($1$) VBS order parameter further combine into an SO($5$)
symmetric pseudovector, in direct analogy with the 1D CFT discussed above. While evidence for SO($5$) symmetry has also been observed numerically
\cite{Nahum15a,Suwa16}, it is not yet clear whether the symmetry is asymptotically exact or broken on some large length scale, bringing the DQCP down
to the lower O($3$)$\times$U($1$) symmetry of the original proposal \cite{Senthil04a}.

2D systems harboring two-fold degenerate singlet patterns in their ground states have also been studied. In the Shastry-Sutherland model, frustrated 
interactions cause a plaquette singlet solid (PSS) state for a narrow range of ratios $J_2/J_1$ of the second and first neighbor interactions \cite{Koga00,Corboz13},
and a similar state has been realized with a ``checker-board'' $J$-$Q$ model \cite{Zhao19}. It has been argued that the DQCP phenomenon is also realized at
the AFM--PSS transition in the Shastry-Sutherland model \cite{Lee19}, though in the checker-board $J$-$Q$ model, which can be studied with reliable quantum
Monte Carlo (QMC) simulations, a first-order transition was found between the AFM and PSS phases \cite{Zhao19}. An emergent O($4$) symmetry was also found, which
would not be expected at a first-order transition driven by conventional mechanisms. A similar phenomenon was observed in a related 3D loop model \cite{Serna19},
and in a different context it was also recently argued that supersymmetry between bosonic and fermionic degrees of freedom may emerge at certain
first-order transition \cite{Yu19}. A very interesting aspect of the two-fold degenerate PSS state is that it can be realized experimentally in SrCu$_2$(BO$_3$)$_2$
under high pressure \cite{Zayed17}, and the expected AFM order expected (within the Shastry-Sutherland scenario) adjacent to the PSS phase
has also been identified recently at still higher pressures \cite{Guo19}.

Here our primary aim is to investigate a direct AFM--VBS transition in a 1D system with long-range interactions, in order to explore potential close
1D analogies to the 2D DQCP. Such analogies can also further our understanding of the broader phenomenon of non-LGW quantum phase transitions. In this regard,
the AFM--VBS transition that we identify here is fundamentally different from the LL transitions between two gapped states recently promoted as DQCP analogies
\cite{Jiang19,Roberts19,Huang19}. The gapless--gapped nature of the transition in our model is closer to the 2D DQCP scenario, at least on a phenomenological
level, as the gapless spin-wave excitations fractionalize at the critical point when the VBS phase is entered. Beyond the line of DQCPs identified here,
the phase diagram of the long-range interacting $J$-$Q$ model also contains other interesting quantum phase transitions. We will study all the transitions
and pay particular attention to a potential emergent O($4$) symmetry of the combined O($3$) AFM and scalar VBS order parameters.

\subsection{Long-range interacting Heisenberg chains}
\label{sub:longrange}

In Ref.~\cite{Laflorencie05} Laflorencie et al.~used QMC simulations and field-theory techniques to study a Heisenberg chain with unfrustrated long-range
interactions, defined by the Hamiltonian
\begin{equation}
H=\sum_{i=1}^L \mathbf{S}_i \cdot\mathbf{S}_{i+1} +
  \lambda \sum_{r=2}^{L/2} \frac{(-1)^{r-1}}{r^\alpha} \sum_{i=1}^L \mathbf{S}_i \cdot\mathbf{S}_{i+r}.
\label{hlong}
\end{equation}
They identified an interesting quantum phase transition where the QLRO critical state undergoes a direct, continuous transformation into a long-range ordered AFM
state when $\alpha$ is taken below a critical value. This critical $\alpha$ value depends on the relative strength $\lambda$ of the long-range part
of the interaction. From previous works within spin-wave theory \cite{Yusuf04}, it had already been predicted that the AFM phase has gapless spin wave
excitations with nonlinear low-energy dispersion. Laflorencie et al.~\cite{Laflorencie05} further studied the quantum phase transition using large-$N$
SU($N$) calculations within the nonlinear $\sigma$-model and QMC calculations for $N=2$, and found the dispersion relation $\omega \propto |k|^z$ with
continuously varying dynamic exponent in the range $3/4 \le z \le 1$, with $z \to 3/4$ for large $\alpha$. Interestingly, in later work using Lanczos
exact diagonalization (ED), a level crossing between $S=0$ and $S=2$ excitations was identified at the QLRO--AFM transition \cite{Sandvik10b,Sandvik10d,S2comment},
and the scaling of the finite-size gaps gave $z$ in very good agreement with the previous results even though only system sizes up to $L=32$ were used.
The results were later confirmed also by DMRG calculations on larger systems \cite{Wang18}. These established results for the QLRO--AFM transition
form one of the corner stones of our work presented in this paper, where we will add multi-spin interactions to a long-range interaction similar
to Eq.~(\ref{hfrustlong}), with the main aim of studying a possible direct quantum phase transition from the AFM state to a spontaneously
dimerized VBS state.

In the previous Lanczos ED study mentioned above \cite{Sandvik10b}, a Heisenberg chain with both long-range interactions and frustrated short-range
($J_2$) interactions was studied. The Hamiltonian of this model was defined as
\begin{equation}
H=\sum_{r=1}^{N/2} J_r \sum_{i=1}^L \mathbf{S}_i \cdot\mathbf{S}_{i+r},
\label{hfrustlong}
\end{equation}
where the distance ($r$) dependent coupling strengths are
\begin{equation}
 J_2=g, \quad J_{r\neq2}=G\frac{(-1)^{r-1}}{r^\alpha},~~~G=\left (1+\sum_{r=3}^{N/2}\frac{1}{r^\alpha}\right )^{-1}.
\label{j2long}
\end{equation}
Here the factor $G$ provides a suitable normalization, ensuring that the sum over of all the magnitudes $|J_r|$ of the non-frustrated couplings (i.e., all $r \not=2$)
equals unity. Using system sizes up to $L=32$, three phases were identified in the $(g,\alpha^{-1})$ plane;
AFM, VBS and QLRO. However, where the AFM--VBS transition is direct, without an intervening
QLRO phase, it is strongly discontinuous. Moreover, the VBS phase was difficult to characterize completely on the accessible small systems, as the
long-range dimerization appeared to coexist with either long-range or slowly decaying period-four spin correlation when the long-range interactions decayed
slowly with $r$. Slightly larger system sizes were reached in subsequent DMRG calculations \cite{Wang18}, but there the focus was on the QLRO--AFM transition.
Another DMRG study argued for a ``sublattice decoupled'' phase at large frustration and slowly decaying interactions \cite{Kumar13}. The frustration 
induced by the $J_2$ term in Eq.~(\ref{j2long}) prohibits large-scale explorations using QMC methods.

\subsection{Long-range interacting J-Q chain}
\label{sub:jqlong}

Here, in order to possibly realize an analogue of the DQCP in a 1D model amenable to QMC simulations, we introduce a $J$-$Q$ Hamiltonian where the frustrated
$J_2$ interaction in Eq.~(\ref{hfrustlong}) is replaced by a six-spin interaction $Q$, which, when sufficiently strong, drives the system into a VBS phase. We
find similarities with the frustrated model defined in Eq.~(\ref{hfrustlong}), but also important differences, especially as regards the nature of the AFM--VBS
transition. The model and physics we aim to investigate are inspired by the original 2D $J$-$Q$ model, which we briefly review before turning to the 1D model.

The 2D $S=1/2$ $J$-$Q$ model hosts a direct quantum phase transition between an AFM and a four-fold degenerate columnar VBS ground state, thus possibly realizing
the DQCP scenario \cite{Sandvik07}. The Hamiltonian of the simplest variant of the model is
\begin{equation}
 H=-J\sum_{\langle ij \rangle}P_{i,j}-Q\sum_{\langle ijkl\rangle}P_{i,j}P_{k,l},
\label{hjq2d}
\end{equation}
where $P_{ij}$ is the singlet projector on the spins on sites $i,j$;
\begin{equation}
P_{i,j}=1/4-\mathbf{S}_i \cdot\mathbf{S}_j.
\end{equation}
The summations in Eq.~(\ref{hjq2d}) are over nearest neighbors $\langle ij\rangle$ in the first term, and in the second term the four-tuples
$\langle ijkl \rangle$ correspond to sites on $2\times 2$ plaquettes such that $ij$ and $kl$ form two horizontal or vertical nearest-neighbor links. The
$J$-$Q$ Hamiltonian thus retains all the  symmetries of the square lattice, and its VBS ordering for large $Q/J$ in the thermodynamic limit is associated with spontaneous
symmetry breaking into one out of four equivalent columnar patterns. The VBS ordering is clearly driven by the locally correlated singlets induced by
the $Q$ terms.

In a 1D system, the two-fold degenerate VBS ordering can similarly be driven by $Q$ terms projecting two or more correlated singlets along the chain 
\cite{Tang11,Sanyal11}. For large $Q/J$, the strength of the VBS order increases with the number of singlet projectors. Here, to obtain a robust VBS state we 
use three singlet projectors ($Q_3$ interaction) and combine this six-spin interaction with long-range antiferromagnetic interactions in the Hamiltonian
\begin{equation}
\begin{split}
 H =-\sum_{{\rm odd}~r}J_r \sum_{i=1}^{L} P_{i,{i+r}}  -Q \sum_{i=1}^L P_{i,i+1}P_{i+2,i+3}P_{i+4,i+5},
\label{hjrjq}
\end{split}
\end{equation}
for $L$ spins on a chain with periodic boundary conditions. Here, the first sum is only over odd distances, $r=1,3,\ldots$, up to $L/2$ or  $L/2-1$ for even
system sizes $L$, unlike Eq.~(\ref{hfrustlong}) where also even distances are included and the coupling is antiferromagnetic for odd $r$ and ferromagnetic
for even $r$. This difference only represents a convenience in the QMC simulations and does not qualitatively affect the physics. We use the same power-law
form of the interaction strength as in Eq.~(\ref{j2long}), including the same normalization constant $G$ (i.e., for legacy reasons the couplings were summed
over both even and $r$).

We study the model using both a ground state projector QMC (PQMC) method operating in the valence bond basis \cite{Sandvik10c,Beach06} and the stochastic series
expansion (SSE) finite-temperature QMC method \cite{Sandvik10d}. Both these QMC simulation techniques incorporate sampling of the long-range interactions in such a
way that the scaling of the number of operations for each complete Monte Carlo update scales with the system size as $L \ln(L)$ \cite{Sandvik03}, instead of the
$L^2$ scaling obtaining with conventional summations over the interactions. With the SSE method the temperature is chosen low enough to obtain ground state results.
We also use standard Lanczos ED for small systems. We analyze the relevant order parameters in the ground state and also study excitation
energies, which exhibit characteristic level crossings at the quantum phase transitions identified here. We use the PQMC method to generate joint
AFM and VBS order parameter distributions, which can give information on emergent higher symmetries.

When $\alpha \to \infty$, the interaction strengths $J_r \propto r^{-\alpha}$ vanish for $r>1$ and the model Eq.~(\ref{hjrjq}) reduces to the conventional
$J$-$Q_3$ chain, which is known to host a dimerization transition of the same kind as in the frustrated $J_1$-$J_2$ chain \cite{Tang11,Sanyal11,Patil18}.
We therefore study the phase diagram in the plane $(Q,\alpha^{-1})$. The conventional dimerization transition then extends for $\alpha^{-1}>0$ up from
the point $(Q_c,0)$, with $Q_c \approx 0.165$, and an interesting question is how this transition evolves when $\alpha^{-1}$ increases further toward $1$
(which we do not exceed because the energy becomes super-extensive at this point). From the previous works on the Hamiltonians in Eqs.~(\ref{hlong}) and
(\ref{hfrustlong}) \cite{Laflorencie05,Sandvik10b,Wang18}, we also know that the QLRO state at the Heisenberg point $(0,0)$ transforms into a long-range AFM 
state upon increasing $\alpha^{-1}$ beyond a critical value. Thus, we expect, and confirm, a QLRO--AFM boundary $\alpha^{-1}_c(Q)$ for some range of $Q>0$. Our
most interesting finding is that this QLRO--AFM boundary merges with the QLRO--VBS phase boundary at a point $(Q,\alpha^{-1}) \approx (0.55,0.7)$, so
that for $\alpha^{-1} \agt 0.7$ there is a direct continuous AFM--VBS transition. 

\begin{figure}[t]
  \includegraphics[width=7cm]{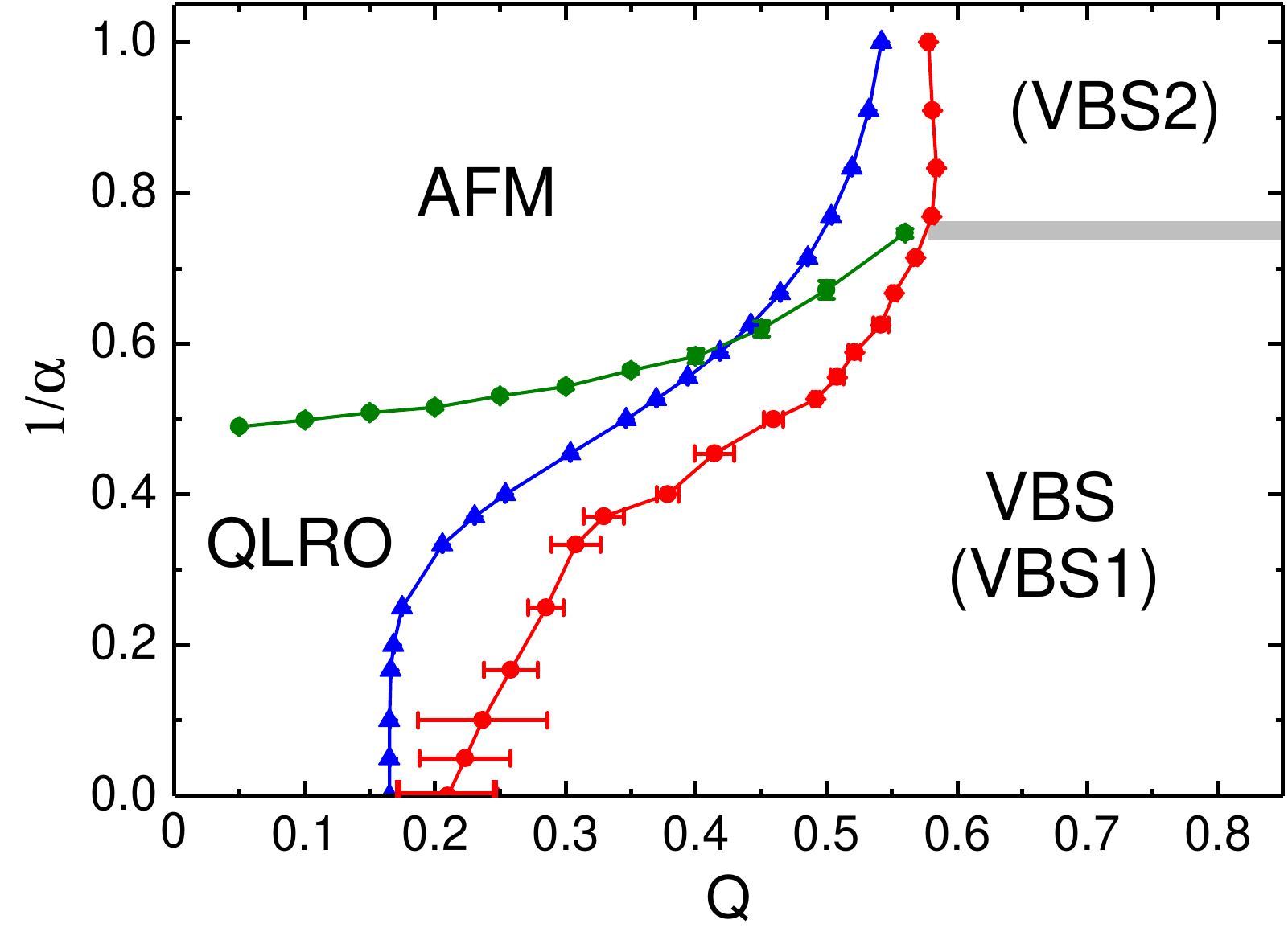}
  \caption{Phase diagram obtained by PQMC and ED calculations of the 1D long-range $J$-$Q$ model, Eq.~(\ref{hjrjq}).
   Extrapolated Binder-cumulant crossing points (illustrated in Fig.~\ref{Fig.Bindercross} and \ref{Fig.Binderdb}) for chains with $L$ up
   to $256$ were used to determine the phase boundaries shown with red and green circles. The blue triangles show the QLRO--VBS phase boundary obtained from
   singlet-triplet excited level crossing points from ED data up to $L=32$. Both estimates for the QLRO--VBS boundary are affected by remaining finite-size
   corrections, and further analysis with larger systems at some points show that the true phase boundary falls between the two curves here.
   The QMC results are less affected by corrections for $\alpha^{-1} \agt 0.7$ and the ED results are better for $\alpha^{-1} \alt 0.6$. The true phase boundaries
   for $0.6 \alt \alpha^{-1} \alt 0.7$ are more uncertain. In the phase marked VBS2, long-range dimerization coexists with algebraically decaying spin
   correlations. The boundary between the conventional VBS phase with exponentially decaying spin correlations and the VBS2 phase is difficult to determine,
   and the thick grey line is only a rough estimate. We some times refer to the conventional VBS phase as the VBS1 phase.}
  \label{Fig.QMC}
\end{figure}

For a concrete overview of our findings, Fig.~\ref{Fig.QMC} shows the phase diagram with QLRO, AFM, and VBS phases, the approximate phase boundaries of which
were obtained from PQMC results for the order parameters of chains with up to $256$ spins. The figure also indicates the QLRO--VBS phase boundary obtained from
level spectroscopy on smaller systems, $L\le 32$, where excitation energies can be computed by Lanczos ED and the crossing between the lowest singlet and triplet
excitations can be analyzed. This level crossing is known to mark the QLRO--VBS transition \cite{Sandvik10b,Wang18}. It is apparent that the results of the two methods
show some disagreement, though qualitatively the two  phase boundaries look similar. We will explain the quantitative disagreements by remaining finite-size
corrections which make it difficult to extrapolate some parts of the phase boundaries to infinite size, as explained further in the caption of Fig.~\ref{Fig.QMC}.
In further tests we study larger systems with the PQMC method, and also extract the singlet and triplet gaps for systems with up to $L=96$ from imaginary-time
correlations computed with the SSE method. We will demonstrate increasingly good agreement between the cumulant and level crossing method as the system
size increases, in particular at the important AFM--VBS transition.

By studying spin correlations, we also find that the VBS phase further divides into two different regions; for large $\alpha$ it is a conventional VBS phase
with exponentially decaying spin correlations, while for smaller values of $\alpha$ these correlations appear to be algebraic. In the latter case we also expect
gapless spin excitations. This coexistence between long-range VBS order and algebraic spin correlations in the upper-right part of the phase diagram in
Fig.~\ref{Fig.QMC} is similar to what was previously observed in the frustrated system, Eq.~(\ref{hfrustlong}), but in that case the spin correlations
are peaked at wave-number $q=\pi/2$ \cite{Sandvik10b,Sandvik10d}, instead of the dominantly staggered ($q=\pi$) correlations in the $J$-$Q$ chain. We will
here use the terms VBS1 and VBS2 when we need to distinguish between the conventional VBS phase and that (VBS2) coexisting with algebraic spin correlations, and
use the term VBS to collectively refer to both of them or any of them.

All the phase transitions in Fig.~\ref{Fig.QMC} appear to be continuous. On the AFM--VBS2 boundary, we find that the critical AFM and VBS order parameters
become locked to each other and form an $O(4)$ vector, but this higher symmetry is apparent in the order-parameter distribution generated with the PQMC method
only if the length scale $r$ (or system size $L$) for one of the order parameters is rescaled according to a power law. In other words, the decays of the two
order parameters are governed by different anomalous dimensions, $\eta_{\rm A}$ (AFM) and $\eta_{\rm V}$ (VBS), but they still exhibit a highly non-trivial
covariance reflecting an emergent higher symmetry. The emergent symmetry likely arises from a more fundamental spinon degree of freedom underlying the two
order parameters. At the QLRO--VBS1 transition the well known O($4$) symmetry \cite{Affleck85,Affleck87,Patil18} is apparent without such a rescaling
(beyond a trivial, size-independent factor) because $\eta_{\rm A}=\eta_{\rm V}$.

\subsection{Outline of the paper}
\label{sub:outline}

The remaining sections of the paper are organized as follows:
In Sec.~\ref{Sec:Binder} we describe how the phase boundaries indicated with red and green circles in Fig.~\ref{Fig.QMC}
were determined using the Binder cumulant method and also discuss the correlation functions that further positively identify the phases. In Sec.~\ref{Sec:level}
we identify characteristic finite-size gap crossings associated with the phase transitions, using Lanczos ED for small systems. These calculations resulted in
the QLRO--AFM phase boundary shown with blue triangles in Fig.~\ref{Fig.QMC}. We study the singlet-triplet level crossing for larger systems with gaps extracted
from SSE-computed imaginary-time correlations, and explain the discrepancies between the QMC and ED phase boundaries in Fig.~\ref{Fig.QMC}. We also
discuss level crossings associated with the other phase transitions. In Sec.~\ref{Sec:AFMVBS} we determine the critical exponents $z$ (the dynamic exponent),
$\eta_{\rm A,V}$ (the anomalous dimension for both the AFM and VBS order parameters), and $\nu_{{\rm A,V}}$ (the correlation length exponents) on some parts of
the phase boundaries. We also study emergent symmetries using order-parameter distributions generated with the PQMC method. We
summarize and further discuss the results and their implications in Sec.~\ref{Sec:summary}. Some further auxiliary calculations are reported in two
appendicies.

\section{Phase boundaries and correlation functions}
\label{Sec:Binder}

We have used the valence-bond PQMC method \cite{Sandvik10c} to efficiently simulate the ground state of our model and determine the phase boundaries from the
finite-size scaling behaviors of the Binder cumulants of the AFM and VBS order parameters. In the PQMC method a singlet-sector amplitude-product state~\cite{Liang88}
is used as a ``trial state'', and $(-H)^{m}$ is applied to this state to project out the ground state. To sample the configurations, the paths of evolving
valence bonds are first expressed in the basis of the $S^z_i$ spin components, resulting in diagonal path-integral-like configurations. The sampling procedures
in this configuration space include loop updates, for which it is very easy to incorporate sampling of the long-range interactions in such a way that the scaling
of the computational effort involved in a complete Monte Carlo update of all the degrees of freedom is $\propto L\ln(L)$ \cite{Sandvik03}, instead of $\propto L^2$ 
in systems where the interactions have to be summed over exactly. For ``measuring'' observables, the valence bonds of the trial state are evolved with the operator 
strings with the restriction to the singlet space maintained, so that spin-rotational invariant quantities are obtained \cite{Beach06}. Convergence to the ground 
state is ensured by carrying out calculations for increasing values of $m$ until no changes in computed expectation values are apparent. The results reported here 
were obtained with $m$ of order $L^2$ and should not be affected by any remaining errors beyond statistical errors. We have also confirmed that the results are 
independent of details of the trial state.

\subsection{Phase boundaries}
\label{sub:cumulant}

We analyze the AFM Binder cumulant to determine the phase boundaries between the AFM phase and the other phases, employing curve-crossing techniques that have been
extensively used in the past; see, e.g., the supplementary materials in Ref.~\cite{Sandvik10d} for a detailed discussion. The cumulant is defined as ~\cite{Binder81}
\begin{equation}
  U_{\bf A}=\frac{5}{2}\left (1-\frac{3}{5}\frac{\langle {m}^{4}_s\rangle}{\langle {m}^{2}_s\rangle^{2}} \right ),
\label{uadef}
\end{equation}
where the coefficients are chosen for the O($3$) symmetric order parameter, which is the staggered magnetization,
\begin{equation}
\label{msdef}
  \mathbf{m}_s=\frac{1}{L}\sum_{i=1}^L(-1)^i\mathbf{S}_{i}.
\end{equation}
If there is long-range AFM order in the thermodynamic limit,
then $U_{\bf A}\rightarrow 1$ when $L \rightarrow \infty$, while $U_A\rightarrow 0$ for a magnetically disordered VBS state.
For a finite system the step function is rounded, and curves for two different sizes, e.g., $L_1=L$ and $L_2=2L$, exhibit a crossing point when drawn versus
the relevant control parameter. The crossing point flows toward the transition point as $L \to \infty$.

\begin{figure}[t]
  \centering
  \includegraphics[width=65mm]{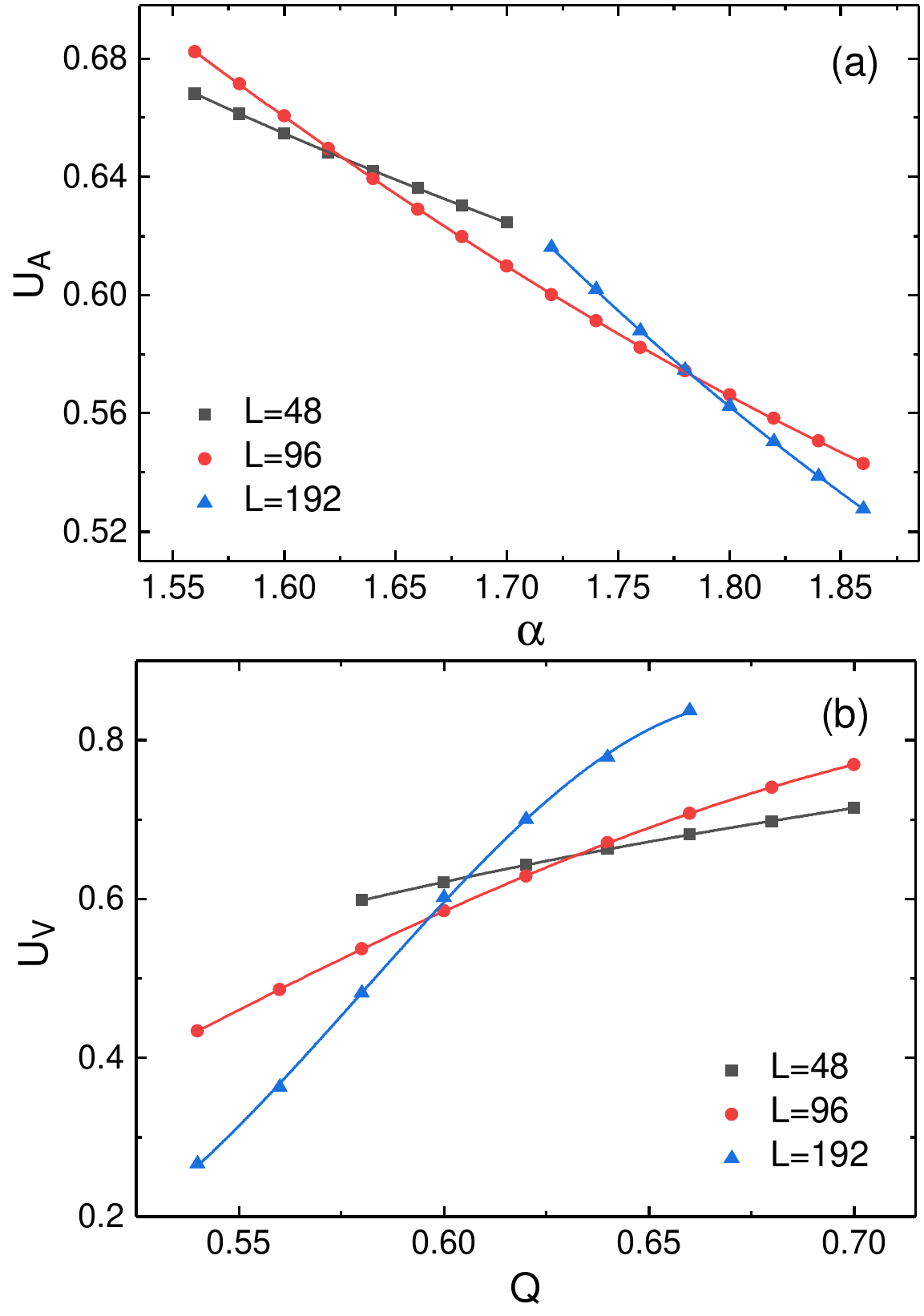}
  \caption{Illustration of the cumulant-crossing method for determining phase boundaries. (a) AFM Binder cumulants versus the long-range interaction
    exponent for three different system sizes, $L=48,96$, and $192$, at fixed $Q=0.15$. (b) VBS Binder cumulants for the same system sizes graphed versus
    the parameter $Q$ at fixed $\alpha=1$. In both (a) and (b), the error bars are smaller than the graph symbols. Cubic polynomial fits to the data sets
    (shown as the solid curves) deliver the crossing points between the computed quantity for two different system sizes (here using sizes $L_1=L$ and
    $L_2=2L$). Error bars on the crossing points are estimated by repeated fits to data with added Gaussian noise (with the standard deviation for
    a given point equal to the error bar on the original data).}
  \label{Fig.Bindercross}
\end{figure}

\begin{figure*}
  \includegraphics[width=17.5cm]{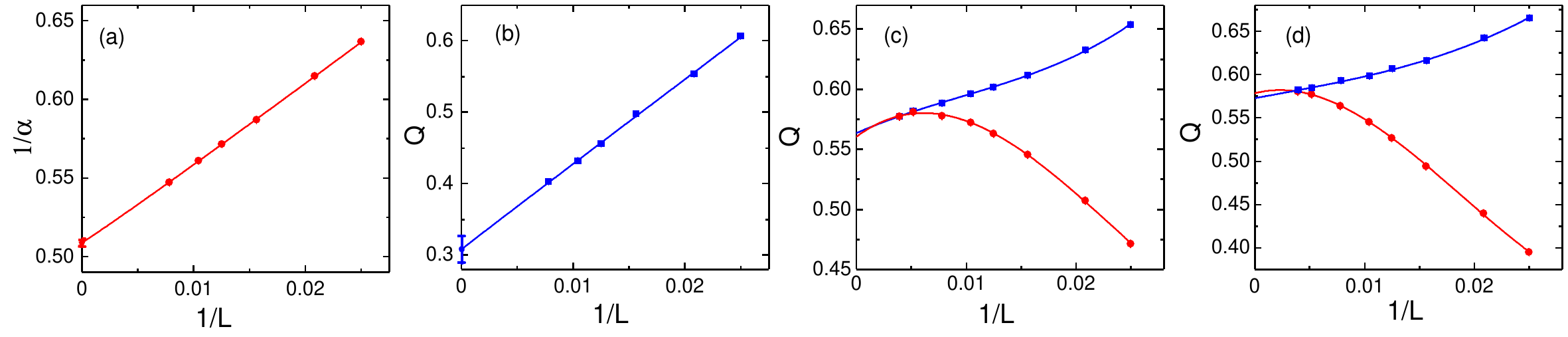}
  \caption{Examples of AFM and VBS cumulant crossing points extracted from system size pairs $(L,2L)$ using the fitting procedures illustrated
    in Fig.~\ref{Fig.Bindercross} and graphed versus $1/L$. (a) $U_{\rm A}$ crossing
    points extrapolating to the critical value of $\alpha$ on the QLRO--AFM phase boundary at fixed $Q=0.15$. (b) $U_V$ crossing points giving the critical
    value of $Q$ on the QLRO--VBS phase boundary at $\alpha=3$. In (c) and (d) results of both cumulants are shown at $\alpha=1$ and $\alpha=1.2$, respectively,
    with the blue squares and red circles for $U_{\rm V}$ and $U_{\rm A}$ crossings, respectively. The curves in (a) and (b) are power-law fits of the form
    $p=a+bL^{-c}$ ($p=\alpha$ or $p=Q$) with parameters $a,b,c$, and these fits result in the critical point estimates $\alpha_c^{-1}=0.509(3)$ and $Q_c=0.31(2)$,
    respectively. In (c) and (d), the two sets of crossing points approach each other as $L$ increases, indicating direct AFM--VBS transitions. The size
    dependences can not be well fit with a single power-law correction, and instead cubic polynomials were used. Error bars on the extrapolated $Q$ values
    are not shown in (c) and (d) but are approximately $0.01$.}
    \label{Fig.Binderdb}
\end{figure*}

We define the VBS Binder cumulant as
\begin{equation}
  U_V=\frac{3}{2}\left (1-\frac{1}{3}\frac{\langle D^{4}\rangle}{\langle D^{2}\rangle^{2}} \right ),
\label{uvdef}
\end{equation}
where $D$ is the VBS order parameter \cite{dnote},
\begin{equation}
\label{ddef}
  D=\frac{1}{L}\sum_{i=1}^L (-1)^{i} \mathbf{S}_i\cdot\mathbf{S}_{i+1},  
\end{equation}
and the coefficients are those appropriate for a scalar order parameter, so that $U_V\rightarrow 1$ in the VBS phase and $U_V\rightarrow 0$ in a phase
with no such order.

In a disordered state with power-law decaying spin and dimer correlations it is not immediately clear whether the Binder cumulants decay to zero. In the QLRO
state, the squared order parameters should scale as $1/L$ with a multiplicative log correction with known exponents (which are different for the two order
parameters) \cite{Giamarchi89}, but the log corrections of the fourth powers in Eqs.~(\ref{uadef}) and \ref{uvdef}) are not known, as far as we are aware.
In Appendix \ref{app:heisenberguauv} we show that $U_{\rm A}$ and $U_{\rm V}$ for the standard Heisenberg chain decay logarithmically to zero as $L \to \infty$.
The existence of $(L,2L)$ crossing points corresponding to the AFM--QLRO and VBS--QLRO phase transitions will be demonstrated below. We refer to
Ref.~\cite{Liu18} for another recent example where the cumulant method was used to identify a phase transition between AFM and critical states.

We here summarize the procedures we have used to determine the quantum critical points forming the red and green phase boundaries in Fig.~\ref{Fig.QMC}. First, we 
calculate the cumulants $U_A$ and $U_V$ for different system sizes $L$ up to $L=256$ (in some cases up to $L=512$), scanning versus one of the control parameters
$p$, $p=\alpha$ or $p=Q$, with the other one held fixed. Second, we extract the value of $p$ at which $U_{\rm X}(p)$, X=A or X=V, evaluated for two
different system sizes, $L_1=L$ and $L_2=2L$, cross each other; $U_{\rm X}(p,L_1)=U_{\rm X}(p,L_2)$. We use polynomial fits to points in the relevant parameter
regime to extract the crossing $\alpha$ or $Q$ values and their statistical errors. In Fig.~\ref{Fig.Bindercross} we show examples of such data sets and fits.
Finally, we extrapolate the crossing points to the thermodynamic limit, and the so obtained values represent points on the phase boundaries (the red and green
points connected by lines in Fig.~\ref{Fig.QMC}).

Examples of extrapolations of crossing points versus $1/L$ are shown in Fig.~\ref{Fig.Binderdb} (where $L$ is the smaller of the two system sizes used to extract 
each point). Here the (a) and (b) panels correspond to points on the QLRO--AFM and QLRO--VBS boundaries, respectively, obtained from the Binder cumulant of the 
relevant long-range order parameter. We have used power-law fits to the data points (in both cases the correction to the infinite-size value is close to $\propto 1/L$)
and estimated the statistical errors on the final extrapolated parameter values by repeating the fits many times with Gaussian noise added to the data. The (d) and 
(c) panels in Fig.~\ref{Fig.Binderdb} each show crossing $Q$ points extracted from both the AFM and VBS cumulants at fixed $\alpha$. In Fig.~\ref{Fig.Binderdb}(c), 
for $\alpha=1$ the size dependence of the AFM points is non-monotonic, and also the VBS points exhibit a non-trivial finite-size behavior that
can not be fitted to a single power law correction. In Fig.~\ref{Fig.Binderdb}(d), for $\alpha=1.2$ we do not observe any non-monotonic behavior, but the
flattening-out of the $U_{\rm A}$ cross-points for the larger sizes suggests that the behavior is qualitatively the same as at $\alpha=1$, with a likely maximum
close to the largest $L$ available here. Such complicated finite-size behaviors that necessitate the use of two corrections have previously been observed in a
2D spin model \cite{Ma18b}. Here we do not have data for sufficiently large systems to carry out reliable fits with two arbitrary powers of $1/L$, and instead we
use polynomial fits to obtain approximate critical $Q$ values. It is visually clear that, for both $\alpha=1$ in Fig.~\ref{Fig.Binderdb}(c) and $\alpha=1.2$ in
Fig.~\ref{Fig.Binderdb}(d), the $U_{\rm A}$ and $U_{\rm V}$ crossing points approach each other with increasing $L$, and this behavior represents the first
indication of a direct AFM--VBS transition. In later sections we will investigate this transition for $\alpha^{-1} \in [0.7,1]$ in more detail and present
additional evidence for a single continuous transition and no intervening QLRO state or coexistence phase.

In Fig.~\ref{Fig.QMC} the QLRO--VBS1 and AFM--VBS2 phase boundaries (both shown as red circles) were obtained from extrapolations of the VBS cumulant
crossing points for system sizes up to $L=256$, i.e., the points for larger system sizes in Figs.~\ref{Fig.Binderdb}(c) and (d) were not included in order to
use consistent procedures for all cases (given that we have data up to $L=512$ only for a small number of points). As is clear from Figs.~\ref{Fig.Binderdb}(c)
and \ref{Fig.Binderdb}(d), in the case of the AFM--VBS2 boundary, the crossing points from AFM cumulants are harder to extrapolate reliably. The available
VBS cumulant results for systems larger than $L=256$ show that the remaining finite-size corrections shift the AFM--VBS2 part of the boundary
in Fig.~\ref{Fig.QMC} only marginally. We will show that QLRO--VBS1 phase boundary for $\alpha^{-1} \alt 0.7$ is more significantly affected by
finite size corrections.

\begin{figure}[t]
\includegraphics[width=65mm]{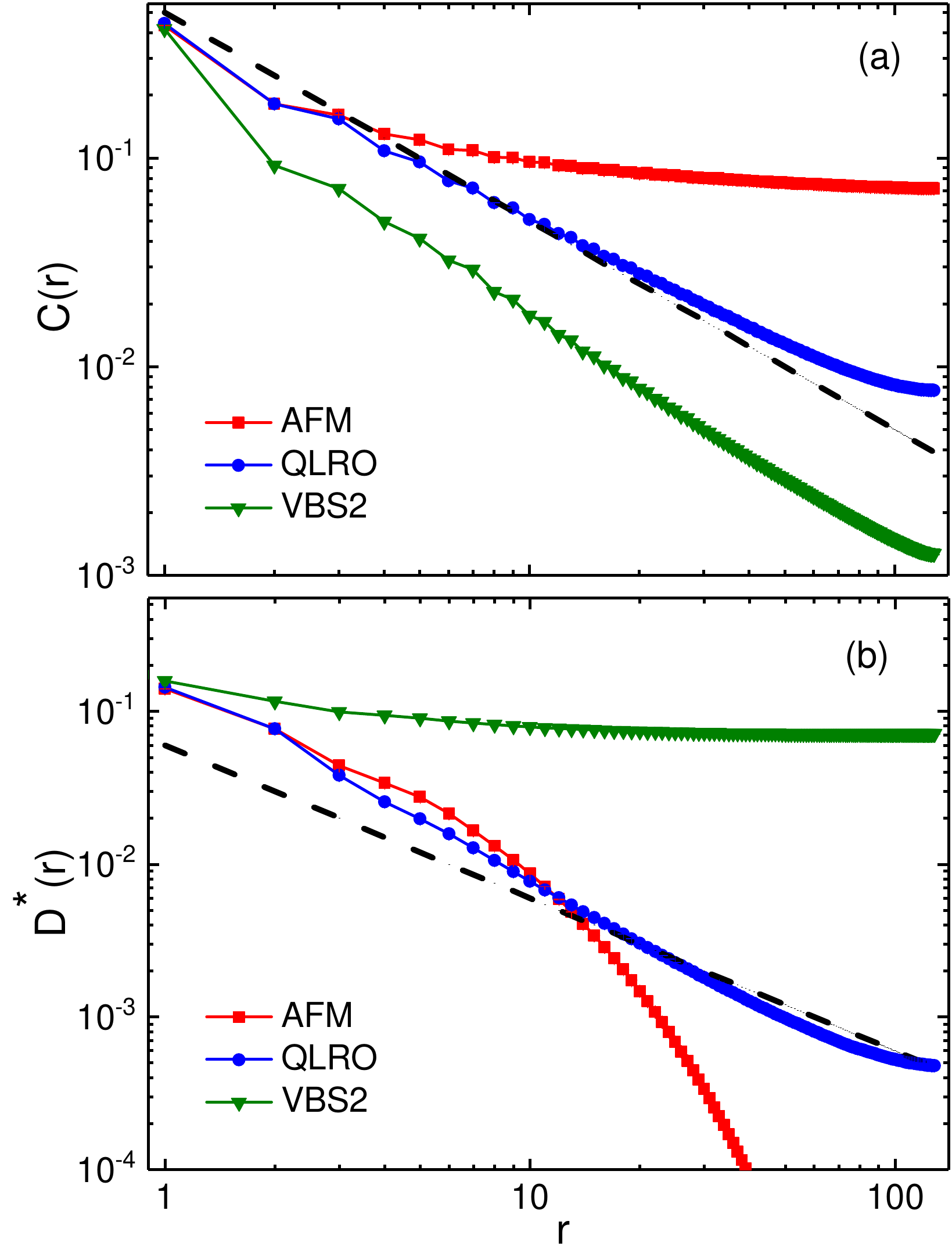}
\caption{Log-log plots of spin (a) and dimer (b) correlation functions calculated on $L=256$ chains at $\alpha=1.2, Q=0.3$ (AFM phase), $\alpha = 2.5, Q=0.1$ 
(QLRO phase), and $\alpha=1.2, Q=0.8$ (VBS2 phase). The dashed lines show the form $\propto r^{-1}$.}
\label{Fig.spindimer}
\end{figure}

We point out that the statistical error bars on the QMC-computed QLRO--VBS1 phase boundary in Fig.~\ref{Fig.QMC}
for $\alpha^{-1} \alt 0.2$ are large, which likely reflects the fact that the
VBS order parameter should be exponentially small in the VBS phase close to the phase boundary for this type phase transition (thus leading to large error bars
in the Binder cumulant). In Sec.~\ref{Sec:level} we will describe the excited-state level crossing method, which is known to work well (converge rapidly with
increasing system size) at the QLRO--VBS transition in the $J_1$-$J_2$ chain \cite{Eggert96,Sandvik10b} as well as in the $J$-$Q_2$ and $J$-$Q_3$ chains with only
short-range interactions \cite{Tang11}. Based on such calculations with the Lanczos ED method for $L$ up to $32$, the phase boundary shown with the blue
triangles in Fig.~\ref{Fig.QMC} is obtained; it is shifted to smaller $Q$ values relative to the QMC estimated points, but the overall shape of the boundary
is similar. We will later demonstrate that the phase boundaries from the two different methods approach each other as the system sizes are further increased,
with the level crossing results in Fig.~\ref{Fig.QMC} being better for $\alpha^{-1} \alt 0.6$ and the cumulant results being better for $\alpha^{-1} \agt 0.7$. In
the remaining intermediate region, extrapolations of results obtained with both methods are challenging with the currently available system sizes.

\subsection{Correlation functions}
\label{sub:correlations}

To confirm that the identification of the three phases in Fig.~\ref{Fig.QMC} is correct, we here study spin and dimer correlation functions at
selected points inside the phases. Results for system size $L=256$ are shown in Fig.~\ref{Fig.spindimer}. Here panel (a) shows the distance dependence
of the staggered spin-spin correlation function, defined as
\begin{equation}
C(r)=(-1)^r\langle \mathbf{S}_i\cdot\mathbf{S}_j\rangle,
\end{equation}
while panel (b) shows the staggered dimer-dimer correlation function defined as
\begin{equation}
  D^*(r)=\left [D(r)-\frac{1}{2}D(r-1)-\frac{1}{2}D(r+1) \right ](-1)^r,
\end{equation}
were, $D(r)$ is the PQMC computed full dimer-dimer correlation function
\begin{equation}
  D(r)=\langle B_iB_{i+r}\rangle,
\end{equation}
with the dimer operator $B_i=\mathbf{S}_i\cdot\mathbf{S}_{i+1}$.
Though the results in Fig.~\ref{Fig.spindimer} represent three points clearly inside the phases according to the phase diagram in Fig.~\ref{Fig.QMC}
the points are still relatively close to the phase boundaries and not extreme cases deep inside the phases.

\begin{figure}[t]
  \centering
  \includegraphics[width=63mm]{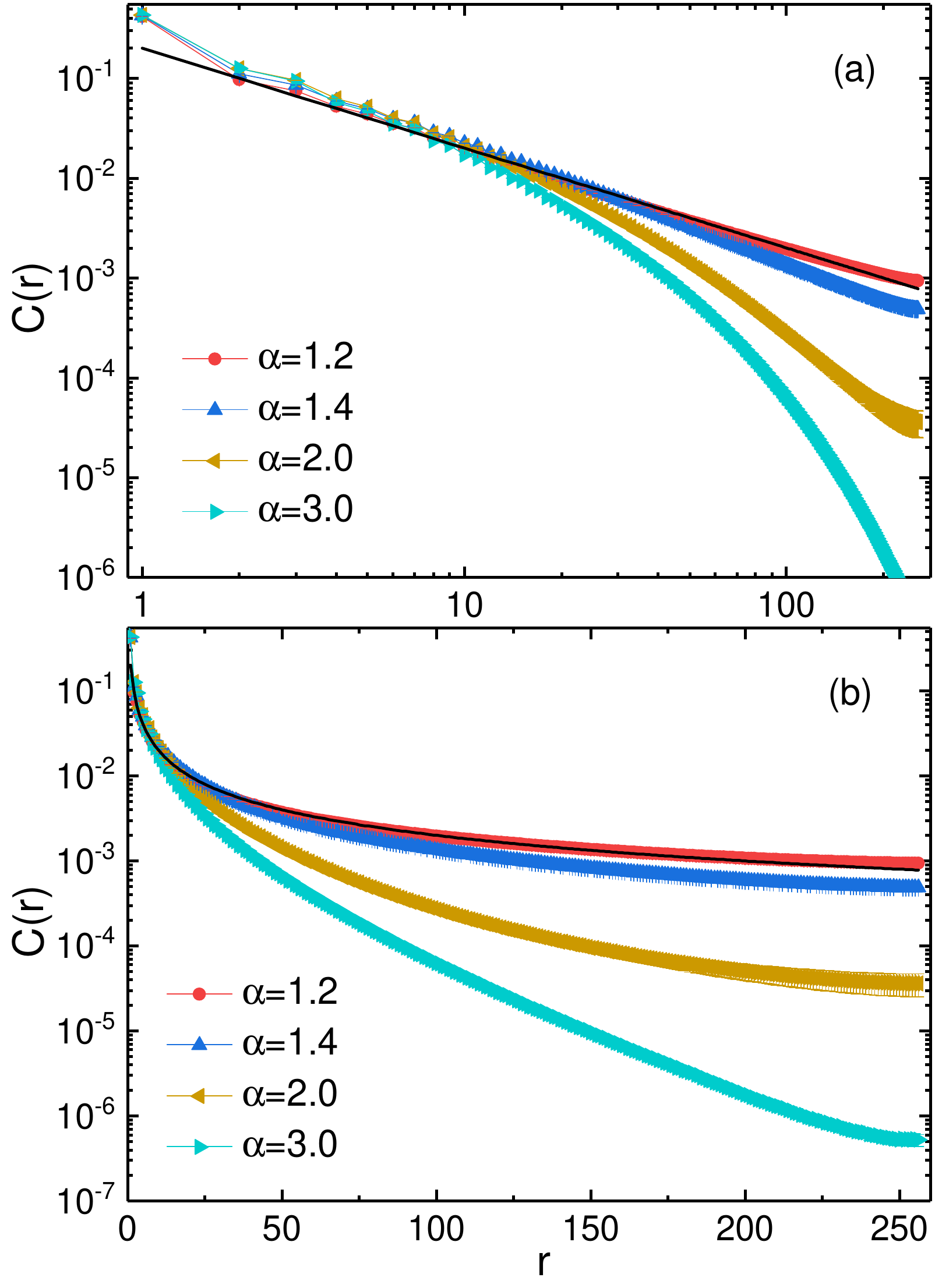}
  \caption{Log-log (a) and lin-log (b) plot of the distance dependent spin correlation function calculated on $L=512$ chains at $Q=0.7$ and
   different values of $\alpha$. The black curve almost coinciding with the $\alpha=1.2$ data for a range of
   distances is of the form $\propto r^{-1}$.}
\label{Fig.corl512}
\end{figure}

Long-range order in the AFM and VBS phases is reflected in Fig.~\ref{Fig.spindimer}
in the correlation functions $C(r)$ and $D^*(r)$, respectively, which flatten out and approach non-zero constants at long distances. In
the QLRO phase both correlation functions decay approximately as $r^{-1}$, as expected. Note again that these correlation functions also have different
multiplicative log corrections, and a pure $r^{-1}$ form should only be expected exactly on the QLRO--VBS boundary (where the marginal operator responsible for
the logs vanishes). The slower than $r^{-1}$ decay of $C(r)$ and faster than $r^{-1}$ of $D^*(r)$ (before the enhancement of the correlations due to
the periodic boundaries set in close to $r=L/2$) are consistent with the log correction $\ln^{1/2}(r)$ in the former and $\ln^{-3/2}(r)$ in the latter
\cite{Giamarchi89}.

In the AFM phase the dimer correlations decay very rapidly, most likely exponentially---the form was difficult to ascertain in a previous study at
$Q=0$ \cite{Tang11} and also in the present case. In the VBS phase, the spin correlations do not decay exponentially but instead appear to follow
a power-law form, close to $r^{-1}$ at the chosen point but decaying faster as $Q$ is further increased or $\alpha$ increased. In the standard VBS phase, e.g., in
the $J_1$-$J_2$ chain or the $J$-$Q_3$ model without the long-range interaction, the spin correlations decay exponentially. In the present system we
also find exponentially decaying spin correlations for smaller values of $\alpha^{-1}$.

In Ref.~\cite{Sandvik10b} the VBS phase adjacent to the long-range ordered AFM phase of the frustrated Hamiltonian Eq.~(\ref{hfrustlong}) was
found to have dimer order coexisting with strong spin correlations peaked at wave-number $q=\pi/2$, and it was conjectured that this phase is different from the
standard gapped VBS phase with exponentially decaying spin correlations. In the long-range $J$-$Q$ model the spin correlations are always peaked at $q=\pi$,
but also here there appears to be a transition from exponentially decaying to algebraic spin correlations in the dimerized systems. The different decay
forms are illustrated in Fig.~\ref{Fig.corl512}, where we have fixed $Q=0.7$ in chains with $L=512$ and graph the distance dependence for several
values of $\alpha$. To more clearly distinguish between exponential and power-law decays, we use log-log and lin-log scales in Fig.~\ref{Fig.corl512}(a) and
Fig.~\ref{Fig.corl512}(b), respectively. At $\alpha=1.2$ the form of $C(r)$ is very close to $r^{-1}$ over a substantial range of distances, while at
$\alpha=1.4$ the decay is somewhat faster but still appears to be algebraic. For even larger $\alpha$ the decay is much faster and not well described by
a power law. There is also no clear-cut linear regime on the lin-log plot (pure exponential decay) in Fig.~\ref{Fig.corl512}(b), but the form could be a
stretched exponential. Thus, we posit that there is both a standard gapped VBS phase (VBS1), for large $\alpha$, and a phase with long-range VBS order coexisting
with algebraic spin correlations with varying exponent (VBS2). The latter phase likely has gapless spin excitation.

Because of intricate finite-size effects and cross-overs,
we have not been able to accurately determine the phase boundaries between the two VBS phases, neither from the change in the form of the spin correlations nor
from the opening up of a gap, but the behaviors observed are consistent with the boundary between the two phases initially extending out almost horizontally
from the point $(Q\approx 0.55,\alpha\approx 1.4)$ where the QLRO phase ends, and we have indicated schematically such a boundary between the two VBS phases
in Fig.~\ref{Fig.QMC}. 

\section{Level spectroscopy}
\label{Sec:level}

We here employ the level crossing method to locate quantum phase transitions. The idea underlying this spectroscopic approach is that different
kinds of ground states typically have disparate lowest excitations, characterized by different quantum numbers. Thus, when traversing a ground state transition,
some of the lowest excitation gaps may cross each other, and the crossing point of two crossing levels for a given system size constitutes a finite-size estimate
of the critical point, which can be extrapolated to infinite system size. The method was first proposed in the context of the $J_1$-$J_2$ Heisenberg chain
\cite{Nomura92}, where very precise results for the dimerization transition can be obtained by extrapolating the crossing point between the lowest singlet and triplet
gaps \cite{Eggert96,Sandvik10d}. The method has also been used for other transitions, e.g., for Hubbard chains in Refs.~\cite{Nakamura99,Nakamura00},
for long-range interacting Heisenberg chains in Refs.~\cite{Sandvik10b,Sandvik10d,Wang18}, and for spin-phonon chains in Ref.~\cite{Suwa15}. We here compute
the excitation gaps for levels with relevant quantum numbers corresponding to the phase transitions of the long-range $J$-$Q$ chain, using ED
of the Hamiltonian with the Lanczos algorithm and also by analyzing the decay rate of imaginary-time correlation functions computed with the SSE QMC method.
With the latter approach we can reach larger system sizes.

\subsection{Lanczos diagonalization}
\label{Subsec:LCED}

\begin{figure}[t]
\includegraphics[width=7cm]{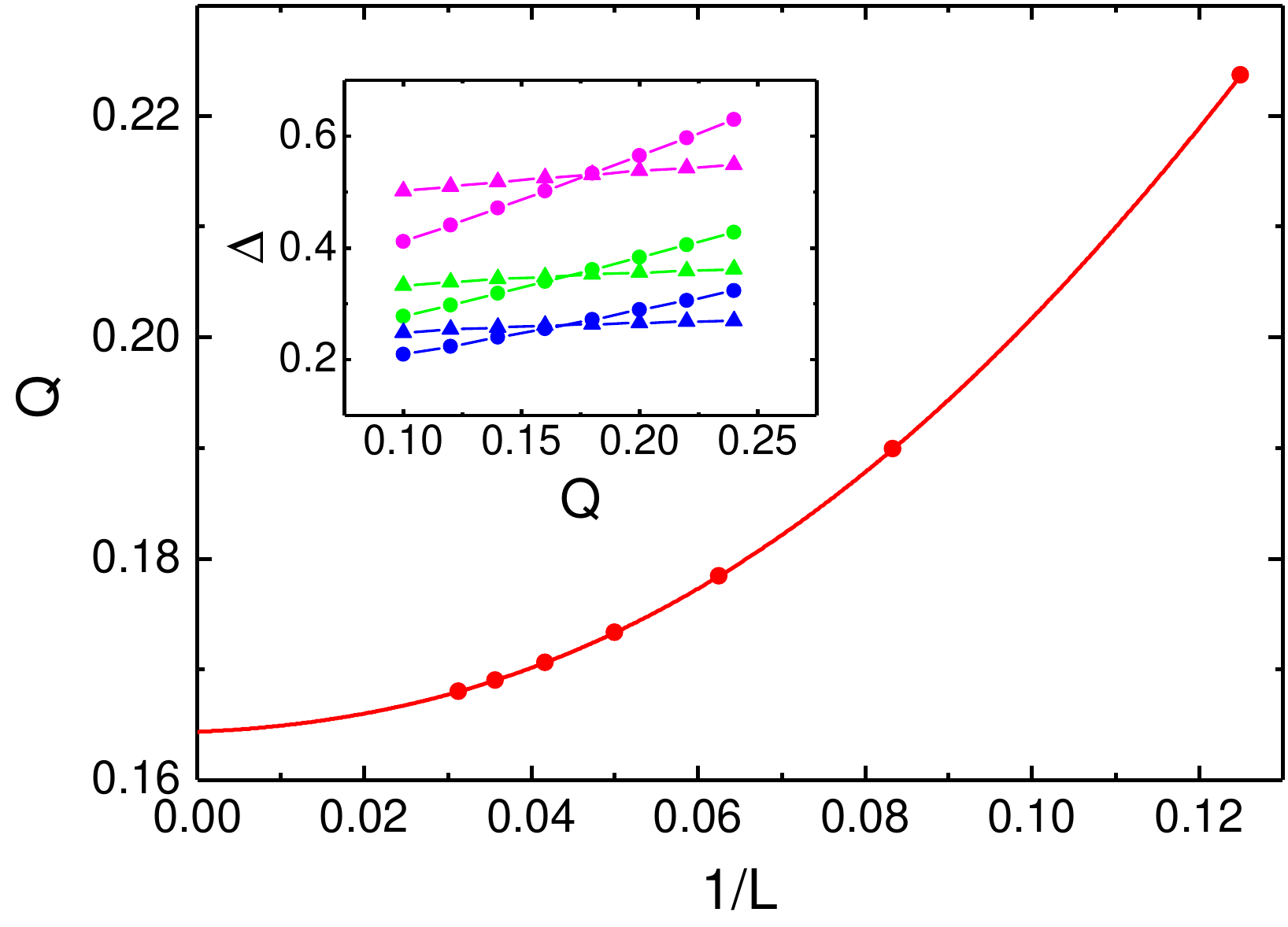}
\caption{Level crossing points for the $J$-$Q$ model without long-range interactions ($\alpha^{-1}=0)$ along with a fit to a power-law correction,
  $Q_c(L)=Q_c(\infty)+aL^{-b}$ (with adjustable parameters $a$ and $b$). The extrapolated critical point is $Q_c=0.16478(5)$ and the correction exponent
  $b \approx 2.1$. The inset shows examples of the crossing singlet (triangles) and  triplet (circles) gaps for system sizes $L=16,24,32$ from top to bottom.}
\label{JQ3new}
\end{figure}

When performing Lanczos ED of the Hamiltonian we exploit all possible lattice symmetries on periodic rings as well as spin-inversion symmetry in the
standard way (see, e.g., Ref.~\cite{Sandvik10d}). We do not implement total-spin conservation but compute ${\bf S}^2$ of the low-lying states generated in the
process. The ground state has $S=0$ and momentum $k=0$ when $L$ is a multiple of $4$, which we choose here for sizes up to $L=32$ (for $L$ of the form $4n+2$,
with $n$ an integer, the ground state has $k=\pi$ \cite{Sandvik10d}). The relevant low-energy levels have momentum $k=0$ or $k=\pi$, and the spin is $S=0,1$,
or $2$.

\begin{figure}[t]
\includegraphics[width=6.25cm]{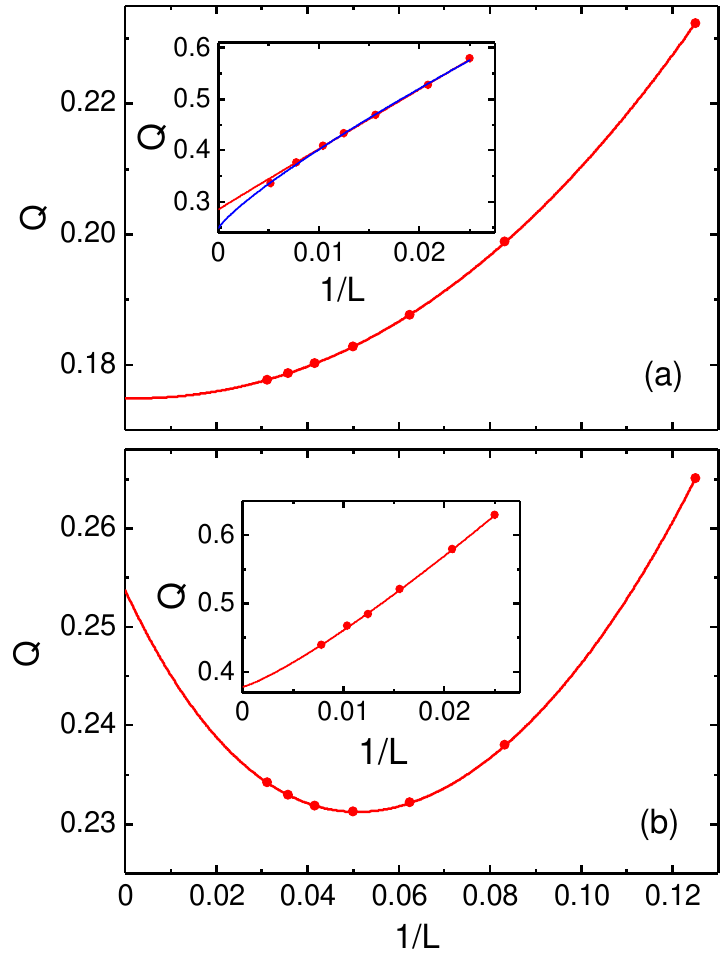}
\caption{Singlet-triplet crossing points graphed vs inverse system size for two cases of the QLRO--VBS transition. The long range parameter is fixed at
$\alpha=4$ in (a) and $\alpha=2.5$ in (b). In (a) a fit with a single power-law correction $\propto L^{-b}$ gives $Q_c \approx 0.175$ and the exponent
$b \approx 2.2$, while a fourth-order polynomial fit in (b) gives $Q_c \approx 0.254$. The insets show the results from the QMC Binder (VBS) cumulant method
for the same $\alpha$ values, along with power law fits. In (a) two fits are shown, with the red curve based on $(L,2L)$ cumulant crossing points with $L$ up
to $128$ and the blue curve including also $L=192$. The disagreement between the cumulant and level-crossing results can be explained by large remaining
finite-size corrections, as discussed in the text.}
\label{a4a25}
\end{figure}

\subsubsection{QLRO--VBS transition}

We first consider the QLRO--VBS transition, which has been studied with the level crossing method in other systems in the past
\cite{Nakamura99,Nakamura00,Eggert96,Sandvik10d}. We will first keep $\alpha$ fixed and scan the relevant gaps to the ground state versus $Q$.
Based on the previous studies, we expect the lowest and second-lowest
excitation for $Q < Q_c$ to have quantum numbers $(k=\pi,S=1)$ and $(k=\pi,S=0)$, respectively, while for $Q>Q_c$ the order should be switched. In
Fig.~\ref{JQ3new} we confirm this behavior with extracted singlet-triplet gap crossing points for the model with $\alpha^{-1}=0$, graphing the crossing $Q$
values versus $1/L$ along with a power-law fit. The inset shows examples of the $Q$ dependent singlet and triplet gaps. It is known that the correction
to the infinite-size $Q_c$ in the $J_1$-$J_2$ Heisenberg chain is $\propto L^{-2}$ \cite{Eggert96}, and also in the present case the fit indeed delivers
an exponent very close to the expected value $-2$. The extrapolated critical point is $Q_c=0.16478(5)$, which is consistent with the value quoted in
Ref.~\cite{Tang11} and also agrees well with a result from finite-size scaling of the order parameter obtained with QMC simulations of much larger systems
\cite{Sanyal11}. As seen in Fig.~\ref{Fig.QMC}, our cumulant crossing point $Q_c(\alpha^{-1}=0)$ also agrees roughly with the level crossing result,
thought the error bars in the former are large.

We now follow the dimerization transition as the long-range interactions are turned on. In Fig.~\ref{a4a25} crossing $Q$ values extracted at $\alpha=4$
and $\alpha=2.5$ are analyzed. At $\alpha=4$ in Fig.~\ref{a4a25}(a), the behavior is seemingly qualitatively similar to the case $\alpha=\infty$ in
Fig.~\ref{JQ3new}, while at $\alpha=2.5$ in Fig.~\ref{a4a25}(b) the size dependence is non-monotonic. Here it can be noted that we can not fit
the non-monotonic form in Fig.~\ref{a4a25}(b) to a single power law---there are not enough data points beyond the minimum to fit to only this part.
Instead, in this case, and all cases henceforth where the behavior is non-monotonic, we carry out a polynomial fit.

\begin{figure}[t]
\includegraphics[width=6.5cm]{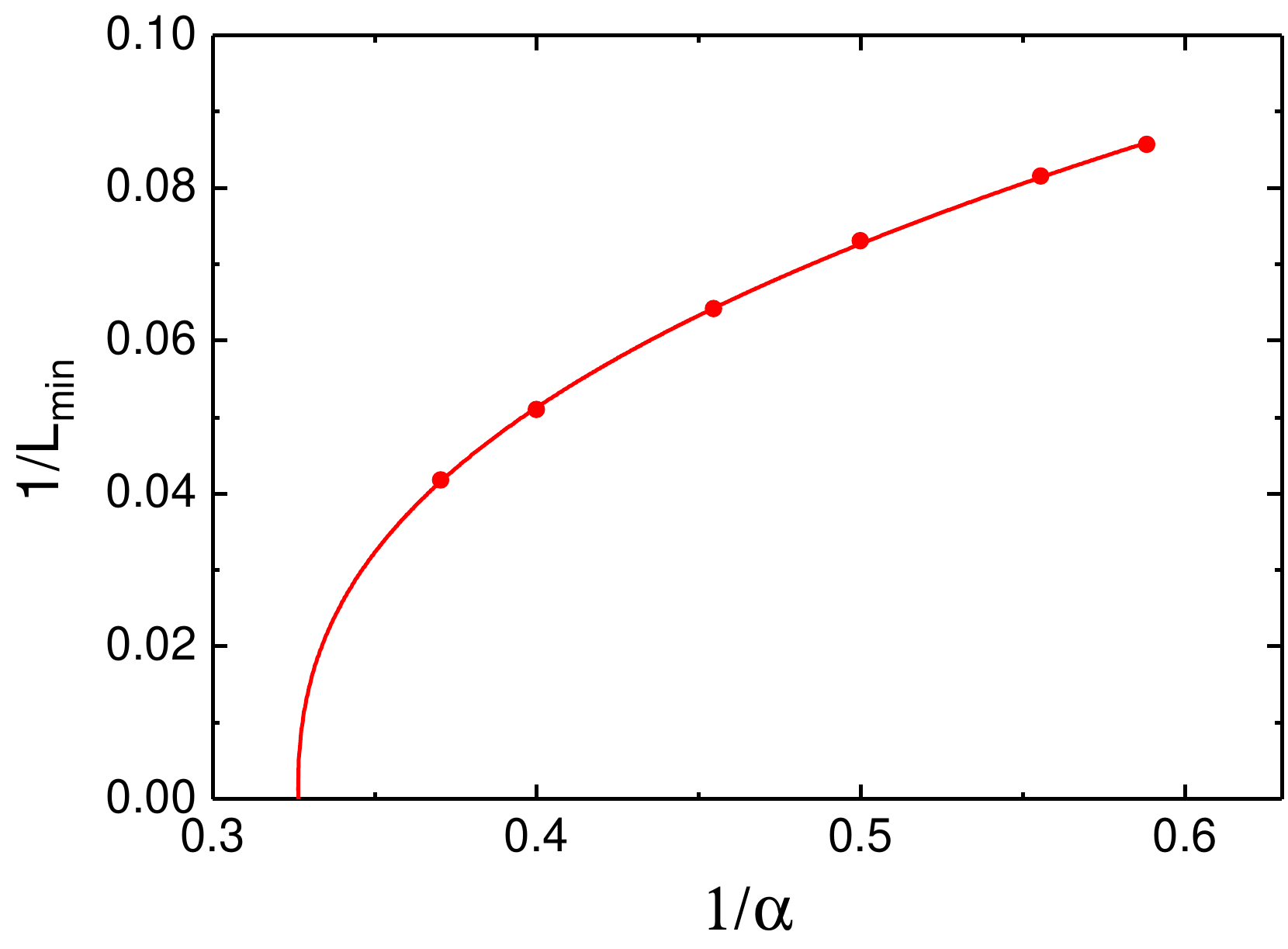}
\caption{Dependence on $\alpha^{-1}$ of the inverse system size at which the singlet--triplet gap crossing is at its minimum $Q$ value, obtained
  by interpolating data such as those in Fig.~\ref{a4a25}(b). The curve is a fit to the form $L^{-1}_{\rm min} = c(\alpha^{-1}-\alpha^{-1}_0)^b$,
  with $\alpha^{-1}_0 \approx 0.33$ and $b \approx 0.4$.}
  \label{Lminfit}
\end{figure}

One might expect that, asymptotically the approach to the infinite-size $Q_c$ should be of the same form on the entire
QLRO--VBS boundary, though it is possible that the long-range interaction could change the behavior, perhaps inducing a correction of the form
$\propto L^{-1}$ (thus motivating the polynomial fit). One may then also wonder whether the non-monotonic behavior actually is present for any finite
value of $\alpha$, with the minimum in Fig.~\ref{a4a25}(b) moving toward smaller system sizes as $\alpha$ is decreased. If so, the fitted form
in Fig.~\ref{a4a25}(a), where no minimum is yet seen with the available system sizes, would somewhat underestimate the critical $Q$ value.

In order to investigate the change from leading monotonic size corrections $\propto L^{-2}$ to (likely) $\propto L^{-1}$ (plus higher order powers of
$L^{-1}$ in both cases), we study the drift of the minimum in Fig.~\ref{a4a25}(b) as a function of $\alpha$. We use the fitted polynomial to
extract the size $L_{\rm min}$ of the minimum $Q$ value. Results are shown in Fig.~\ref{Lminfit}, along with a power-law fit that
extrapolates to $\alpha_0^{-1} \approx 0.325$ for the special value at which the non-monotonic behavior first sets in. It is tempting
to speculate that $\alpha_0=3$ exactly, though we have no insights into why that would be the case. In any case, it seems likely that the non-monotonic
form only sets in for $\alpha \alt 3$ and the use of different fitting forms above and below this value is justified.

\begin{figure}[t]
\includegraphics[width=6.5cm]{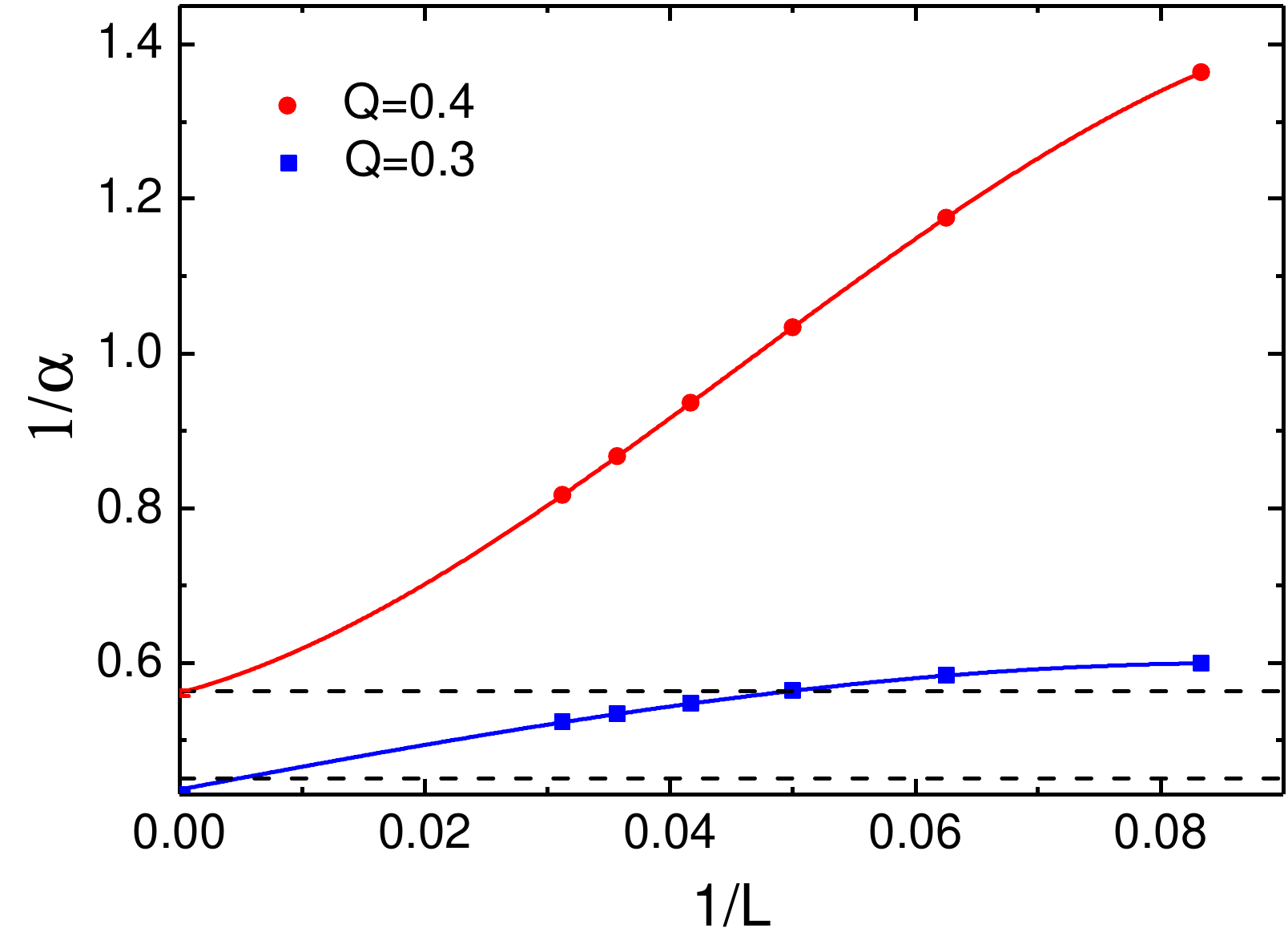}
\caption{ED results for singlet-triplet gap crossing points when $Q$ is held fixed at $0.3$ (blue squares) and $0.4$ (red circles). The curves
are fourth-order polynomial fits. The dashed lines show the corresponding values obtained by interpolating the ED data set shown in Fig.~\ref{Fig.QMC},
where the gaps were scanned versus $Q$ at fixed $\alpha$ values.}
\label{Q0304test}
\end{figure}

To test the reliability of the extrapolation method in the region where a minimum is present, we show in Fig.~\ref{Q0304test} results of scanning the gaps
versus $\alpha$ at fixed $Q=0.3$ and $0.4$. Apart from holding a different variable fixed, the procedures for extracting the gap crossing points are the same
as in the above analysis with $\alpha$ fixed. Here
we have again used polynomial fits, and the results agree very well with those of the previous scans versus $Q$. However, going to larger $Q$ (e.g., $Q=0.45$),
scanning versus $\alpha$ no longer works as no level crossing is found, or the crossing points appear at very large $\alpha$ and sensible extrapolations can
not be carried out. This behavior is related to the fact that we are moving close to the almost vertical phase boundary in Fig.~\ref{Fig.QMC}, which is
also where the extrapolations versus $Q$ with $\alpha$ fixed become difficult because of the strongly non-monotonic behavior. In Sec.~\ref{sub:LCQMC} we
will show further results in this regime based on SSE studies of larger system sizes.

\begin{figure}
\includegraphics[width=6.75cm]{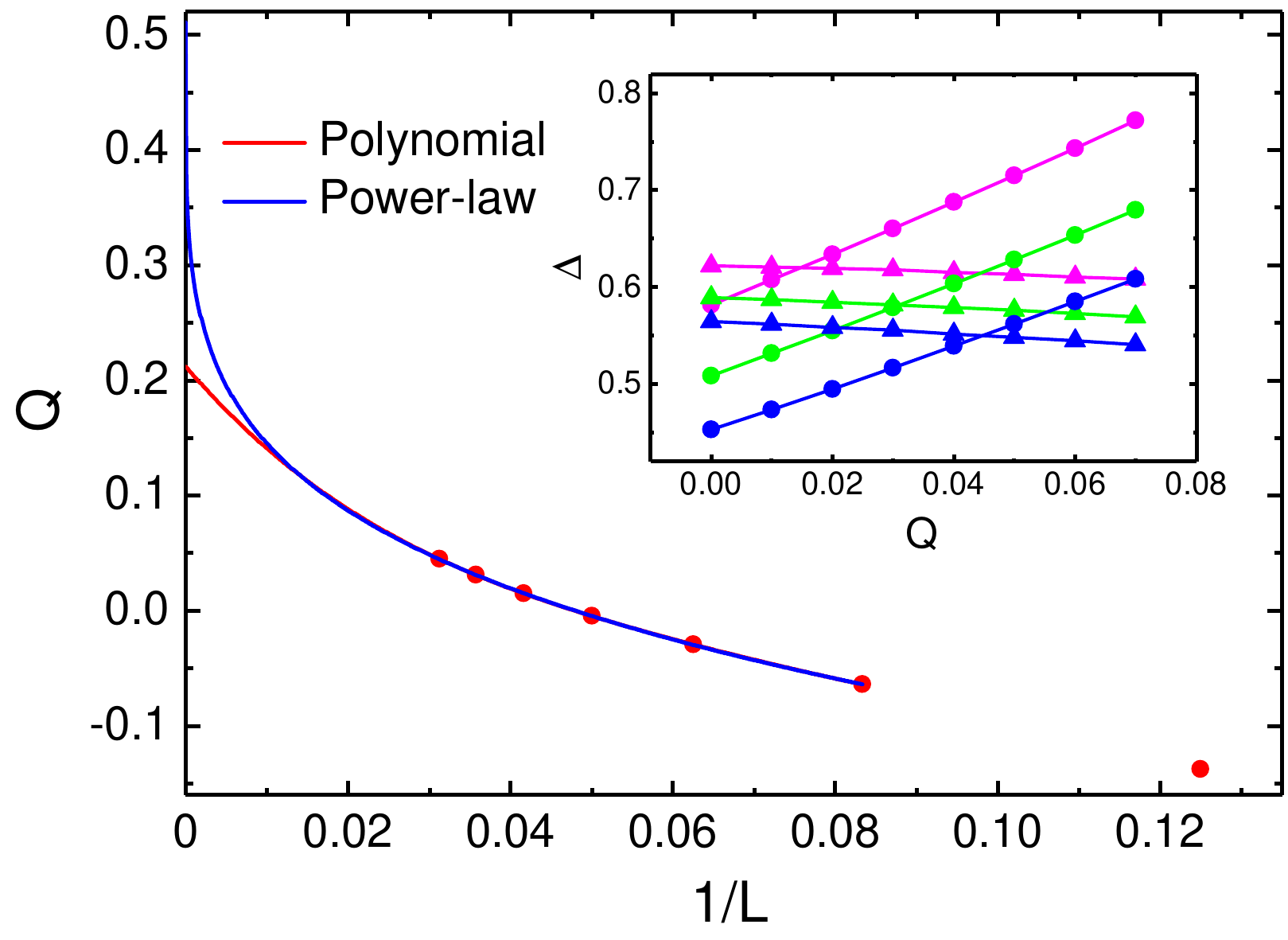}
\caption{Singlet-quintuplet gap crossing points in systems with $\alpha=1.6$, along with two types of fits to the $L \ge 12$ data. The blue curve
shows a power-law correction, $\propto L^{-b}$ with $b \approx 0.2$. The red curve shows a fourth-order polynomial fit. The inset shows gaps
to the states with $S=2,k=0$ (circles) and $S=0,k=\pi$ (triangles) for $L=24$, $28$, and $32$ (top to bottom).}
\label{a16hlevel}
\end{figure}

In addition to the difficulties in extrapolating the gap crossings when $\alpha^{-1} \agt 0.6$, we also in some cases encounter problems with the cumulant method
discussed in Sec.~\ref{sub:cumulant}. As illustrated in the insets of Fig.~\ref{a4a25}, the extrapolated cumulant crossing points appear to be inconsistent with the
level crossing points. To test the stability of the extrapolated cumulant crossing points, in Fig.~\ref{a4a25}(a) we also show a fit where the largest-$L$ point
(obtained from $L=192$ and $2L=384$ data) is left out. We observe that the extrapolated value is significantly lower when this data point is included, and
it is also visually clear that there is an accelerating downward curvature in the data points as the system size increases.

The simplest explanation of all these results is that, for some range of $\alpha$, roughly $0.6 \alt \alpha \alt 0.7$, neither the QMC nor the ED results
have reached sufficienty large system sizes to be in the asymptotic regime where reliable finite-size analysis is possible. It does appear, however, that the
trends for the two calculations for all $\alpha \in [0,1]$ are such that the results approach each other as the system size increases. Below we will further show
that, not only does the extrapolated critical $Q$ value decrease when the system size is increased in the cumulant method, but also, for the $\alpha$ values where
the size dependence is non-monotonic, the values extracted with level crossing method increase.

The difficulty in reaching the asymptotic limit is why in Fig.~\ref{Fig.QMC} we have shown the two extracted boundaries to the VBS phases (the QLRO--VBS1 as well
as the AFM--VBS2 boundary) and consistently used $L=32$ and $L=256$, respectively, for the ED and QMC calculations. For extrapolating the gap crossing points
we have used a power-law correction $\propto L^{-b}$ for $\alpha \ge 3$, where we do not observe non-monotonic behaviors (and the exponent $b$ always is
close to $2$, as expected). For smaller $\alpha$ we use polynomial fits. We will show further evidence that the curves in Fig.~\ref{Fig.QMC} bound the actual
boundary to the VBS phases in such a way that the blue (ED) curve is closer when $\alpha^{-1} \alt 0.6$ while the red (PQMC cumulant) curve is better
for $\alpha^{-1} \agt 0.7$.

\subsubsection{AFM--VBS2 transition}

Observing that the singlet-triplet crossing points in Fig.~\ref{Fig.QMC} follow the same trend as the QMC computed boundary to the two VBS phases for
the whole range of $\alpha^{-1} \in [0,1]$, we conjecture that this level crossing applies not only to the QLRO--VBS1 transition but also to the
to the AFM--VBS2 transition. In Sec.~\ref{sub:LCQMC} we will confirm this by studying level crossings for larger system in this regime using data from
SSE simulations. We will also show that the level-crossing boundary indeed converges well toward the phase boundary obtained with the cumulant method with
increasing system size in the range of $\alpha$ values corresponding to the AFM--VBS2 transition.

\begin{figure}[t]
\includegraphics[width=7.25cm]{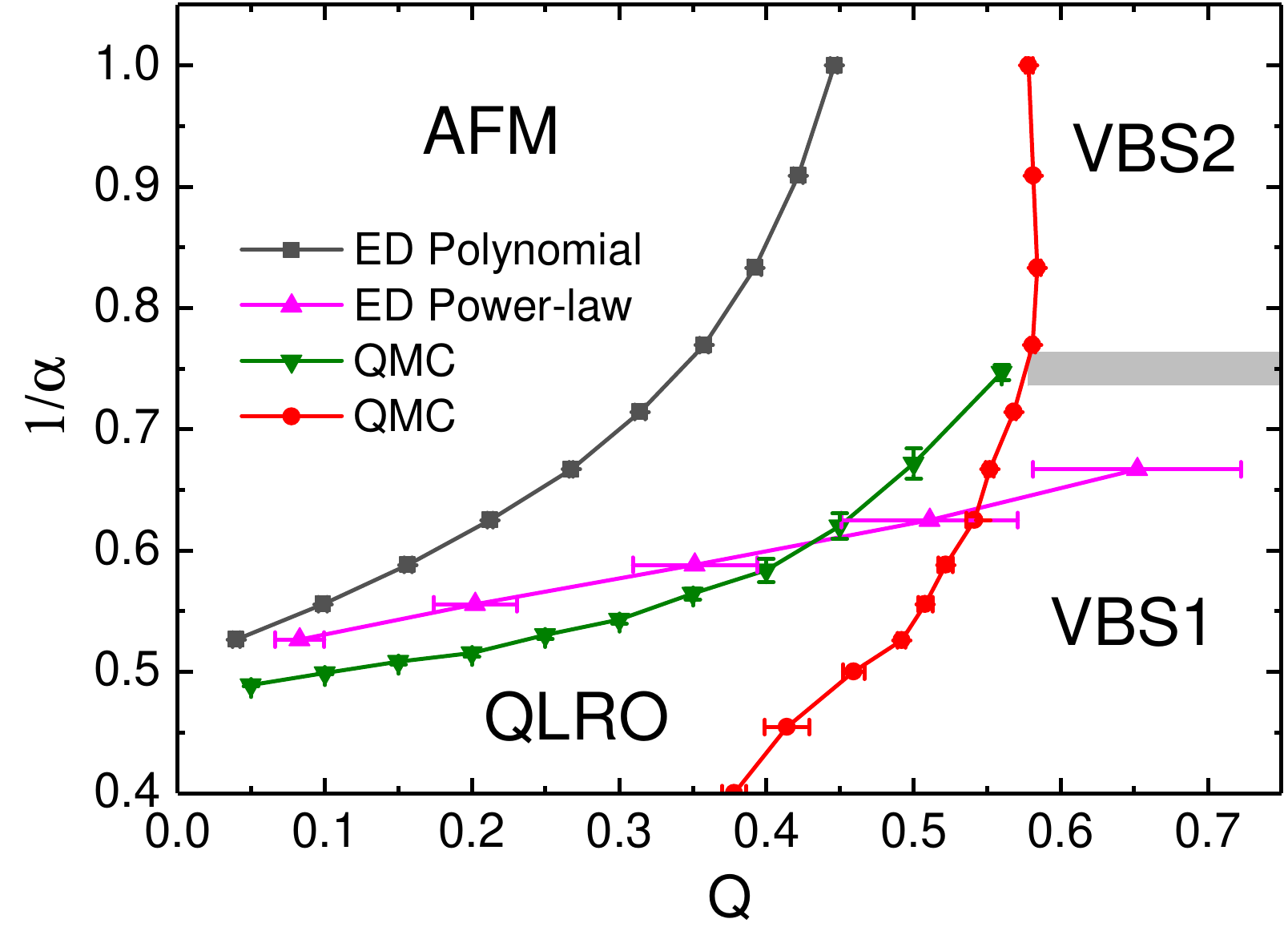} 
\caption{Magnification of the upper-left part of the phase diagram in Fig.~\ref{Fig.QMC}, with added curves showing extrapolated crossing points between the
lowest levels with quantum numbers $(k=\pi,S=0)$ and ($k=0,S=2$). The black squares and red circles are based on polynomial and power-law fits, respectively,
to Lanczos data for $L \le 32$ (as illustrated in Fig.~\ref{a16hlevel}). The disparity in the extrapolated values when $Q$ increase shows that the system
sizes used are not large enough for reliable analysis of this level crossing.}
\label{AFQLROboundary}
\end{figure}

\subsubsection{AFM--QLRO transition}

We next turn to the AFM-QLRO phase transition. At a similar transition in both the unfrustrated and frustrated long-range Heisenberg chains, Eqs.~(\ref{hlong})
and (\ref{hfrustlong}), previous calculations have identified a level crossing between states with $(k=0,S=2)$ and $(k=\pi,S=0)$, the former being lower in
the AFM state and the latter being lower in the QLRO state \cite{Sandvik10b,Sandvik10d,S2comment,Wang18}. Note that, in both the AFM phase and the
QLRO phase, the lowest excitation has $(k=\pi,S=1)$, and the level crossing at the AFM--QLRO transition is, thus, between higher, but still low-lying excitations
that become gapless as $L \to \infty$ (see the supplemental material of Ref.~\cite{Wang18}). We here present evidence for the same kind of level crossing
in the long-range $J$-$Q$ chain, again studying the system for fixed $\alpha$ and scanning the gaps versus $Q$.

Figure \ref{a16hlevel} shows results for $\alpha=1.6$. Here the small number of data points makes it difficult to discern a definite asymptotic
scaling form, and we can either fit to a power-law correction or a polynomial. The extrapolated critical $Q$ values from such fits deviate significantly
from each other. In Fig.~\ref{AFQLROboundary} we show results for the phase boundary based on the two types of extrapolations. The two curves approach each
other when $Q \to 0$, but deviate strongly from each other as $Q$ is increased. The power-law extrapolated curve is consistently closer to the QLRO--AFM
boundary obtained from the QMC cumulant method and eventually crosses into the VBS phase, which may indicate that this level crossing also applies to the
phase boundary separating the two types of VBS phases. However, it should be noted that the exponent $b$ of the power-law correction $\propto L^{-b}$
becomes very small in this region; already in the case shown in Fig.~\ref{a16hlevel} we have $b \approx 0.2$, the smallness of which casts some doubt on
the applicability of this form to describe the data. The completely different shape of the curve extracted from the polynomial fit may instead point to its
eventual morphing toward both the QLRO--AFM and AFM--VBS2 phase boundaries when the system size is further increased.

Although the AFM--QLRO phase boundary is almost horizontal in Fig.~\ref{Fig.QMC}, we still find it better to scan the gaps versus $Q$ for fixed $\alpha$,
instead of keeping $Q$ fixed and scanning in the vertical $\alpha$ direction. With the latter approach the levels also cross each other, but at very small values 
of $\alpha$, outside the relevant range $\alpha \ge 1$ of the phase diagram. The extrapolated critical $\alpha^{-1}_c$ values are still close to (a bit higher
than) those of the horizontal scans in Fig.~\ref{Fig.QMC} for $Q \alt 0.2$, while for higher $Q$ the extrapolations become very unreliable and do
not produce meaningful results.

\begin{figure}[t]
\includegraphics[width=7cm]{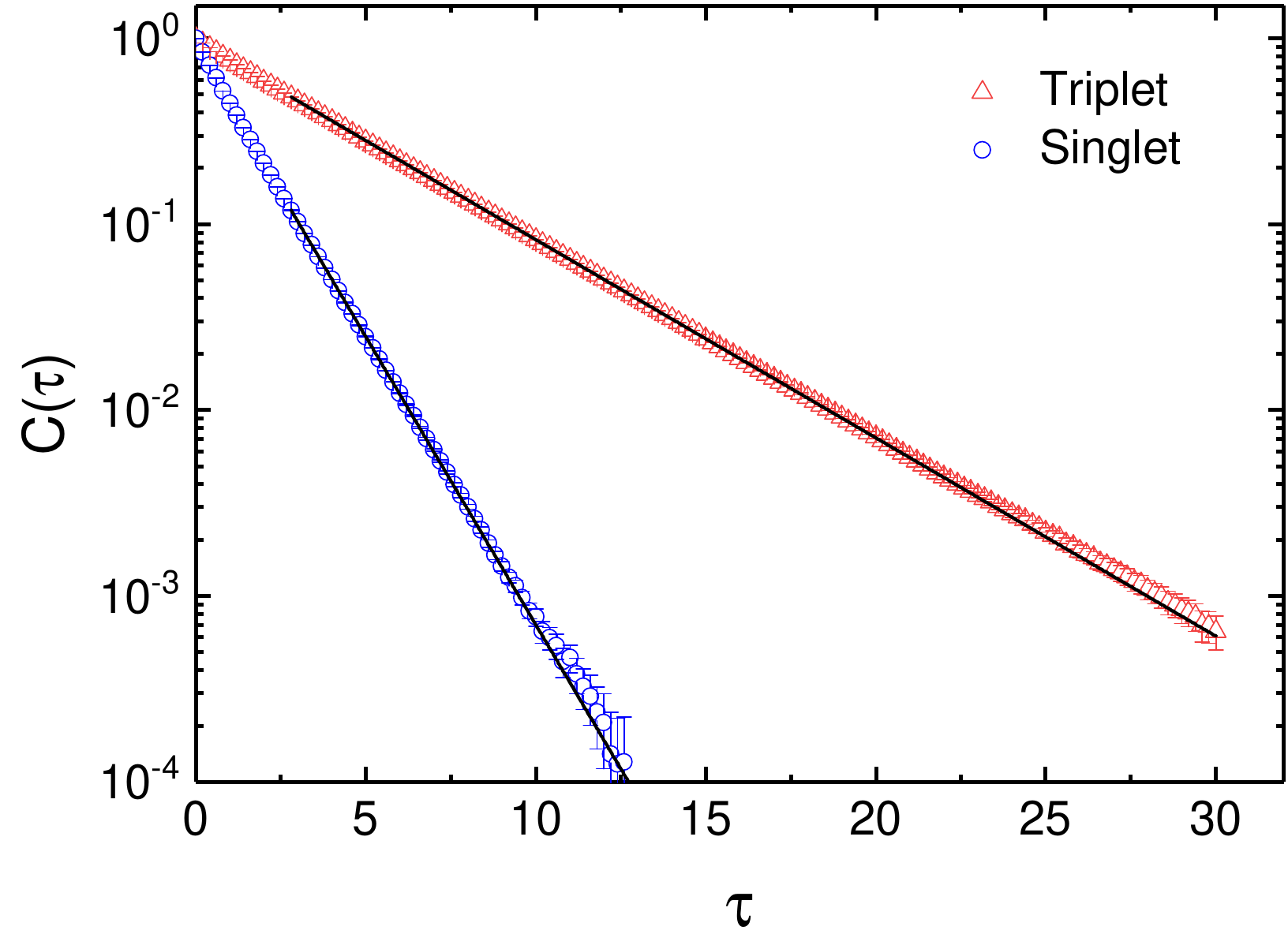}
\caption{The $q=\pi$ time-dependent correlation functions defined in Eqs.~(\ref{ctqdef}) and (\ref{csqdef}) for an $L=32$ chain with $\alpha=1$ and $Q=0.21$.
Both correlation functions have been normalized to $1$ at $\tau=0$. The blue circles and red squares show  SSE results and the lines correspond to fits to
the expected asymptotic exponentially decaying forms, which deliver the gaps $\Delta_T=0.24556(4)$ and $\Delta_S=0.7094(6)$. These results agree
with the exact gaps computed with the Lanczos ED method; $\Delta_T=0.245499$ and $\Delta_S=0.709422$.}
\label{32SSE}
\end{figure}

\subsection{SSE QMC Approach}
\label{sub:LCQMC}

Given the uncertainties in the Lanczos ED results in some regions of the phase diagram, it would clearly be useful to have level-crossing data also for larger
system sizes. Here we use the SSE QMC method to compute imaginary-time ($\tau$) correlation functions $C(q,\tau)$ of operators $O_q$ that excite states with
suitable quantum numbers when acting on the ground
state $|0\rangle$;
\begin{equation}
C(q,\tau)=\langle 0|O_{-q}(\tau)O_{q}(0)|0\rangle.
\label{cOtaudef}
\end{equation}
Here $q$ is the momentum transfer and $O_q(\tau)={\rm e}^{\tau H}O_q {\rm e}^{-\tau H}$ with $\tau \in [0,\beta]$, $\beta$ being the inverse temperature.
The asymptotic exponential decay form $C(q,\tau) \propto {\rm e}^{-\tau \Delta_q(L)}$ gives the corresponding finite-size gap $\Delta_q(L)$. In principle we could
also use the PQMC method for these calculations \cite{Sen15}, but SSE has the advantage of time-periodicity, allowing averaging of the time-dependent correlations
to reduce the statistical errors. We ensure a sufficiently large $\beta$, so that ground state results as in Eq.~(\ref{cOtaudef}) are obtained for the range of
imaginary-time values required to analyse the asymptotic behavior. Since these calculations are computationally very expensive, we only consider a small
number of points on the phase boundaries, to further test the convergence properties observed in Sec.~\ref{Subsec:LCED}.

\begin{figure*}[t]
\includegraphics[width=17cm]{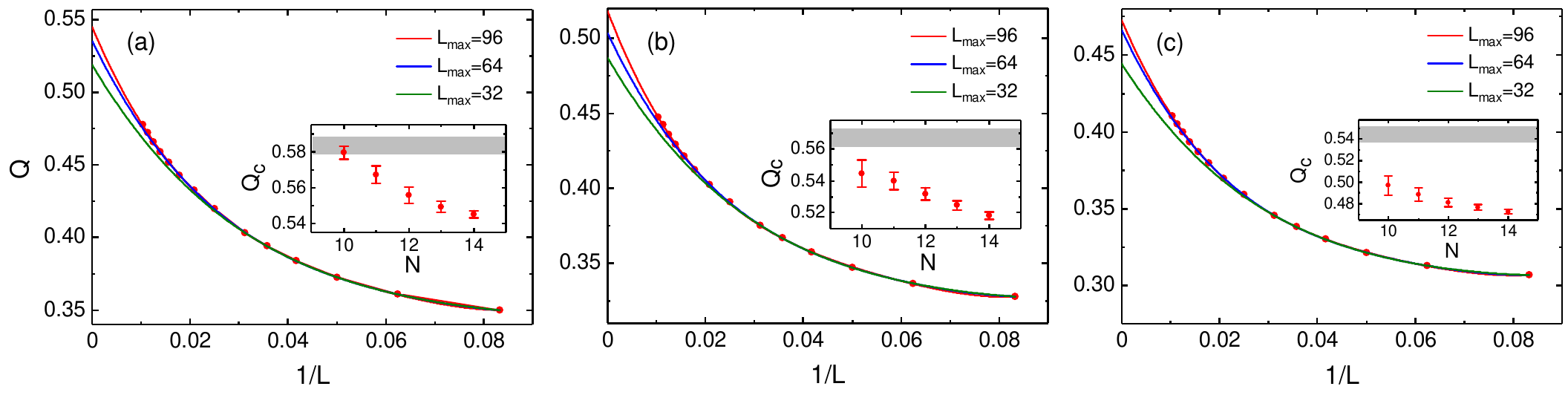}
\caption{Singlet--triplet gap crossing $Q$ values at (a) $\alpha=1.2$, (b) $\alpha=1.4$, and (c) $\alpha=1.6$. Gaps for $L\le 32$ were calculated using
Lanczos ED, and the results for larger systems were obtained from the asymptotic exponential decay of the correlation functions defined in Eqs.~(\ref{cstcorr}).
The three curves in each panel show fourth-order polynomial fits including system sizes up to $L=32$, $64$, and $96$. The insets show extrapolated $Q_c$
values obtained from fits to the data for the $N$ largest system sizes, with the gray horizontal lines indicating the corresponding transition points with
their estimated statistical errors, obtained from the VBS Binder cumulant crossing points for $L$ up to $256$ (Fig.~\ref{Fig.QMC}).}
\label{64and96}
\end{figure*}

To excite levels with the required quantum numbers $k$ and $S$ we use the operators
\begin{subequations}
\begin{eqnarray}
  \mathbb{T}_q&=&\frac{1}{L^{1/2}}\sum_{r}S^z_{r}e^{iqr}, \label{topdef}\\
  \mathbb{S}_q&=&\frac{1}{L^{1/2}}\sum_{r}S^z_{r}S^z_{r+1}e^{iqr}, \label{sopdef}
\end{eqnarray}
\end{subequations}
and compute the imaginary-time correlation functions
\begin{subequations}
\begin{eqnarray}
C_T(q,\tau)&=&\langle \mathbb{T}_{-q}(\tau)\mathbb{T}_{q}(0) \rangle, \label{ctqdef} \\
C_S(q,\tau)&=&\langle \mathbb{S}_{-q}(\tau)\mathbb{S}_{q}(0) \rangle. \label{csqdef}
\end{eqnarray}
\label{cstcorr}
\end{subequations}
The operator $\mathbb{T}_q$ excites triplets with momentum $k=q$ when acting on the $k=0,S=0$ ground state. The operator $\mathbb{S}_q$
excites both singlets and quintuplets, and our method will detect whichever of these levels that is the lower one.

Implementation of the SSE algorithm with loop updates for the $J$-$Q$ model is described in Ref.~\cite{Sandvik10d}. As in the case of the PQMC method discussed
in Sec.~\ref{Sec:Binder}, the computational effort required for sampling the configuration space in the $S^z$ basis scales as $L\ln(L)$ for long-range
interactions when simple tricks for treating the diagonal terms are incorporated \cite{Sandvik03}.

We have tested the method by extracting gaps
for $L=32$ systems, which can be compared with exact Lanczos results. Fig.~\ref{32SSE} shows two examples of correlation functions that deliver the singlet
and triplet gaps at $k=\pi$. In the exponential fits we have systematically excluded data for small $\tau$ until statistically sound fits are obtained.
The results for both gaps agree well with Lanczos ED calculations.

\subsubsection{QLRO--VBS transitions}

At the phase transitions into one of the VBS states, we need the singlet and triplet gaps at $k=\pi$, and since the $S=2$ state with
this momentum is always above the singlet in all cases considered, we can extract both the required gaps from the decay of the correlation functions in
Eq.~(\ref{cstcorr}) with $q=\pi$. Fig.~\ref{64and96} shows gap crossing points at $\alpha=1.2$, $1.4$, and $1.6$, where the results for $L \le 32$ are from
Lanczos calculations and for larger sizes, up to $L=96$, they are extracted from the correlation functions. Here we show three different polynomial fits, carried
out with maximum system size $L=32$, $64$, and $96$, to demonstrate that larger sizes systematically lead to larger critical $Q$ values. In the insets of
all the panels of Fig.~\ref{64and96} we show extrapolated $Q_c$ values obtained with the maximum size $L=96$ as a function of the number of data points included
in the fit (always excluding sizes from the low-$L$ side). Here we observe that $Q_c$ increases when more of the small systems are excluded, and the results
approach the approximate critical values obtained in Sec.~\ref{sub:cumulant} using the VBS cumulant crossing method. We observe such strong size dependence
for all $\alpha$ values in the range $\approx [1,1.6]$, while for still higher values of $\alpha$ the larger systems do not significantly change the extrapolated
$Q_c(\alpha^{-1})$ boundary from the curve in Fig.~\ref{Fig.QMC}. This improving convergence trend for $\alpha \agt 1.6$ is also is reflected in the good agreement
between results from different gap scans in Fig.~\ref{Q0304test}, where the extrapolated $\alpha$ values are above $1.7$.

\begin{figure}[t]
\includegraphics[width=6.5cm]{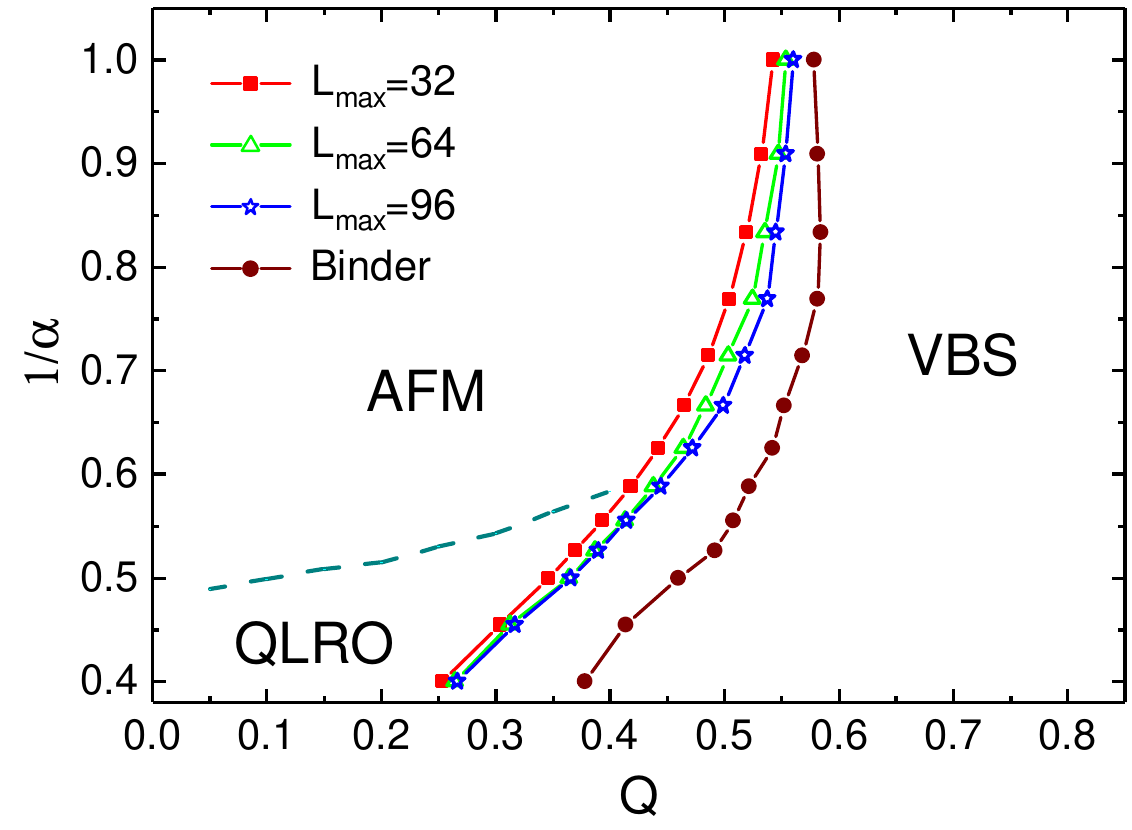}
\caption{AFM--VBS phase boundaries obtained from singlet--triplet gap crossing points extrapolated to infinite size based on data sets with
data for maximum sizes $L=32$, $64$, and $96$, compared with the cumulant-crossing results from Fig.~\ref{Fig.QMC}.} 
\label{EDQMC}
\end{figure}

To further illustrate the systematic shift of the gap-crossing boundary with increasing system size, in Fig.~\ref{EDQMC} we show results based on three different 
maximum system sizes in the polynomial fits and compare with the cumulant-based phase boundary from Fig.~\ref{Fig.QMC}. Recall that this boundary is based on 
system sizes up to $L=256$, and tests with larger $L$ up to $512$ in Sec.~\ref{sub:cumulant} indicated only minor shifts when $\alpha$ is in the range of the 
AFM--VBS2 transition. In Fig.~\ref{EDQMC}, when $\alpha^{-1} \agt 0.5$ the gap-crossing points shift significantly toward the AFM--VBS2 boundary with increasing
maximum $L$ in the fit, and it appears plausible that the two methods will deliver the same phase boundary if sufficiently large systems are used. For smaller
$\alpha^{-1}$, the gap-crossing results are better converged while the cumulant crossings have significant finite-size corrections left, in spite of the 
larger system sizes used [as evidenced, e.g., in Fig.~\ref{a4a25}(a)]. For $0.6 \alt \alpha^{-1} \alt 0.7$ both boundaries determined here are quantitative 
unreliable, but the system-size trends indicate that the true phase boundary falls between the two estimates.

There is a natural physical explanation for the more difficult extrapolations with the cumulant method for smaller $\alpha^{-1}$: The QLRO--VBS transition 
pertaining in this regime is associated with exponentially small VBS order close to the phase boundary \cite{Affleck87}, making the cumulant change very slowly
when moving across the transition. The gap crossings do not have this problem. At higher $\alpha^{-1}$, the AFM--VBS transition has (as we will show in 
Sec.~\ref{Sec:AFMVBS}) more concentional power-law scaling of the order parameters.

\subsubsection{QLRO--AFM transition}

To study the crossing of the gaps in the sectors $(k=\pi,S=0)$ and ($k=0,S=2$) relevant to the QLRO--AFM transition, the procedure for fitting
the imaginary-time correlation has to be modified, because the operator $\mathbb{S}_0$ in Eq.~(\ref{sopdef}) excites both $S=0$ and $S=2$ states. Since the ground state
is in the sector ($k=0,S=0$), there is a constant contribution in addition to the asymptotic exponential decay from which the target $S=2$ gap is obtained.
Fitting to the form $C_S(\tau)={\rm constant} + {\rm e}^{-\Delta_S\tau}$, we find excellent agreement with Lanczos ED calculations for $L=32$. We additionally
studied system sizes $L=48$ and $L=64$, in order to test the stability of the fits based on Lanczos ED results for $L \le 32$ in Sec.~\ref{Subsec:LCED}.

\begin{figure}
\includegraphics[width=6.5cm]{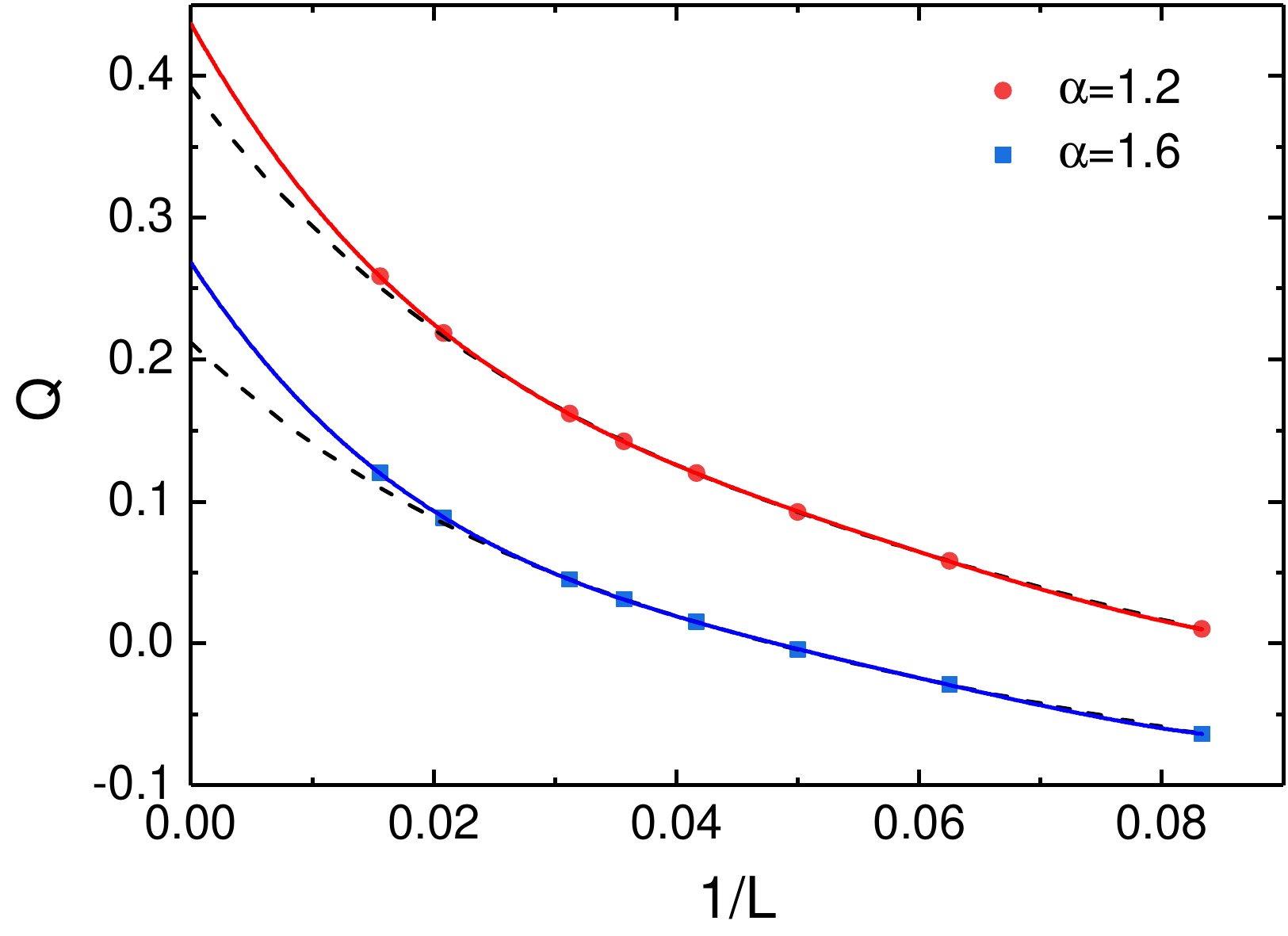}
\caption{Crossing points between the lowest levels with $(k=\pi,S=0)$ and ($k=0,S=2$) at two values of $\alpha$.
The curves are fourth-order polynomials in $1/L$. The points for the largest two system sizes, $L=48$ and $64$, were obtained using
SSE-computed correlation functions, while those for $L=32$ and smaller are from Lanczos ED. The dashed curves are from the fits including
only the $L \le 32$ data.} 
\label{QMCS0S2}
\end{figure}

In Figure \ref{QMCS0S2} we show gap crossing points along with polynomial fits for $\alpha=1.2$ and $1.6$. Compared to the previous Lanczos ED results
for $L$ up to $32$ (also shown in the figure for reference), the extrapolated crossing $Q$ values increase significantly from the results in
Fig.~\ref{AFQLROboundary} (black squares). We have also fitted to a power law, and, as previously in Fig.~\ref{a16hlevel} , the extrapolated values are
then much higher (even higher than with the previous $L\le 32$ fits). The exponent of the power law is now even smaller than in Fig.~\ref{a16hlevel};
$b \approx 0.15$, making this fitting form seem even more implausible than before.

Given that the larger system sizes shift the crossing points significantly toward the QLRO--AFM and AFM--VBS2 phase boundaries in Fig.~\ref{QMCS0S2},
the most likely scenario appears to be that the crossing point between the $S=0$ and $S=2$ levels eventually, as $L \to \infty$, will coincide with both
those boundaries. Results for larger system sizes will be required to definitely confirm this.  

\section{Critical behavior at the AFM-VBS transition}
\label{Sec:AFMVBS}

The most intriguing aspect of the phase diagram in Fig.~\ref{Fig.QMC} is the putative direct AFM--VBS2 transition. Given that the excitations
of the AFM state are magnons carrying spin $S=1$ and that the VBS2 should have deconfined $S=1/2$ spinons, a direct continuous transition between
the two ground states would imply a new type of 1D DQCP. We here provide evidence for the transition indeed being continuous, by studying various critical
properties and extracting critical exponents. We also present evidence for an emergent deformed O(4) symmetry of the AFM and VBS order parameters at the phase
transition. We will mainly focus here on the case $\alpha=1.2$, which in the middle of the range of the direct AFM--VBS2 transition. We have also studied
$\alpha=1.1$ carefully and will report the exponents there, and in addition we carried out limited tests at other points. The critical exponents on the AFM--VBS2
boundary may in principle be varying (as they are at the QLRO--VBS1 transition \cite{Laflorencie05}), but our studies do not indicate any dramatic changes.

\subsection{Order parameters and critical exponents}
\label{Subsec:OP}

\begin{figure}[t]
\includegraphics[width=6.5cm]{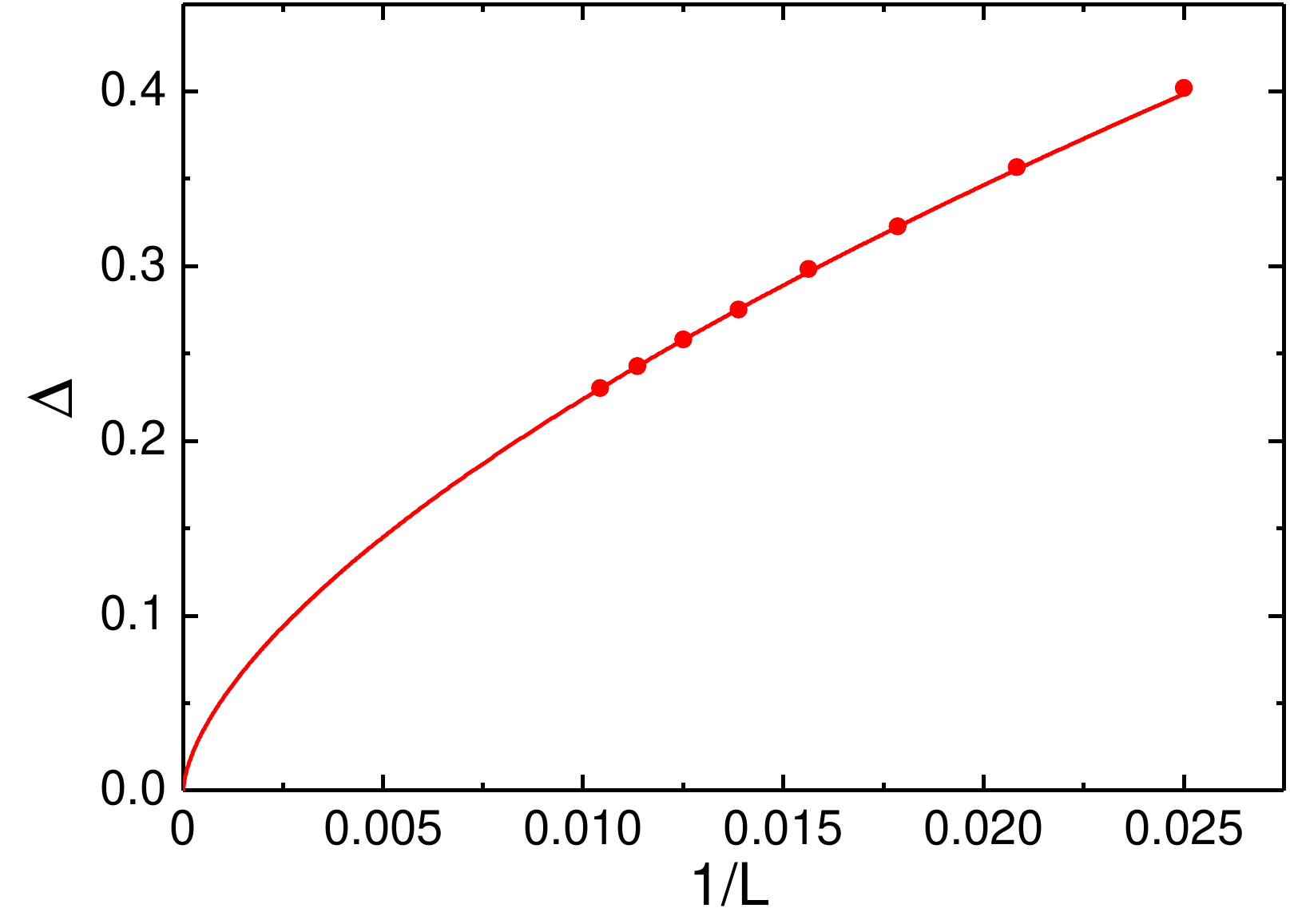}
\caption{Inverse-size dependence of the common singlet--triplet gap at the $Q$ value where the two levels cross each other at fixed $\alpha=1.2$. The gaps
were computed on a grid of $Q$ values using the imaginary-time correlation method discussed in Sec.~\ref{sub:LCQMC}, with chains of length from $L=40$ to
$L=96$ in steps of $8$, and interpolated for the crossing point defining the gap values $\Delta(L)$. Error bars are smaller than the graph symbols. The curve
shows a power-law fit, $\Delta(L) \propto L^{-z}$, to the data for the six largest system sizes, which delivers $z=0.63(1)$.}
\label{a12gap}
\end{figure}

We begin by extracting the dynamic exponent $z$, the value of which affects definitions of other exponents through the quantum to classical correspondence
in scaling forms where the real-space dimensionality $d$ of the quantum system is replaced by $d+z$. Previously Laflorencie et al.~\cite{Laflorencie05} found
a varying dynamic exponent $z \in [3/4,1]$ at the QLRO--AFM transition. Here at the putative direct AFM--VBS2 transition in the long-range
$J$-$Q$ model we extract $z$ from the gap scaling form, $\Delta(L) \propto L^{-z}$, using the gap at the singlet--triplet level crossing point as the finite-size
definition $\Delta(L)$. This method is preferable here, since we only have to analyze a single gap value when, by definition, the singlet and triplet
gaps are the same. To minimize finite-size effects we only use the SSE correlation function approach described in Sec.~\ref{sub:LCQMC}, with
system sizes between $L=40$ and $96$. Results at $\alpha=1.2$, based on interpolations of data close to the gap grossing points, are shown in Fig.~\ref{a12gap}.
A fit to a power law in $1/L$ gives $z=0.63(1)$. It is reasonable that the exponent is less than unity, considering the previous results at the QLRO--AFM
transition and the fact that the spin-wave dispersion relation in the AFM phase is known to be sublinear \cite{Yusuf04}.

\begin{figure}[t]
\includegraphics[width=7.75cm]{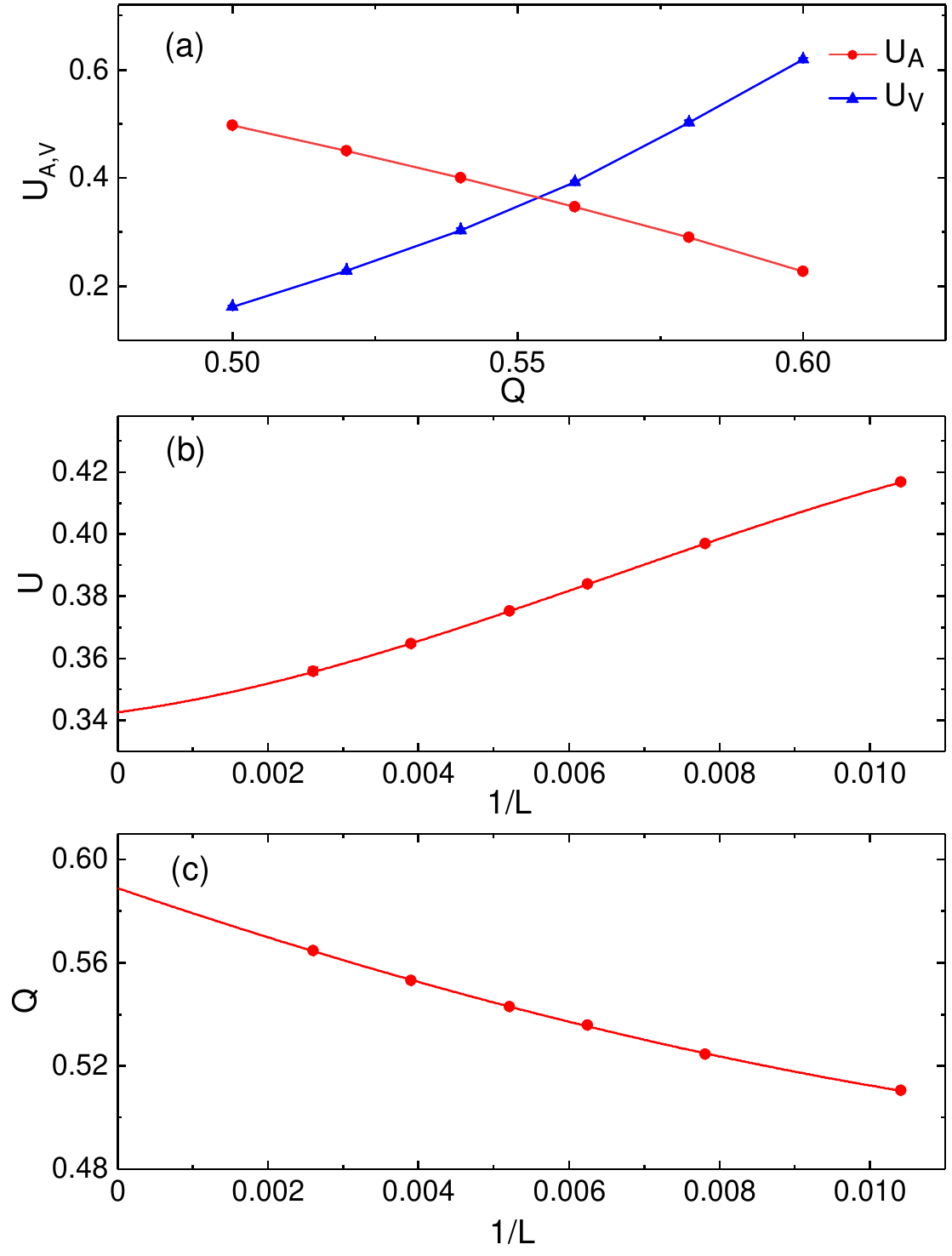}
\caption{(a) $Q$ dependence of the AFM and VBS Binder cumulants for system size $L=256$ at $\alpha=1.2$. Polynomial fits (the curves shown) to the QMC data
  points are used to extract the crossing point between $U_{\rm A}$ and $U_{\rm V}$. The inverse-size dependence of the cumulants at the crossing point and the
  crossing $Q$ value are shown in (b) and (c), respectively. Both quantities are fitted to the form $a+bL^{-c}$. The extrapolated crossing cumulant value is
  $U_c=0.343(2)$ and the critical point is $Q_c=0.589(3)$.}
\label{a12cumucross}.
\end{figure}

Next we consider the squared critical AFM and VBS order parameters, $\langle {m}_s^2 \rangle$ and $\langle D^2 \rangle$.
For a 1D system, they should decay with the system size according to
\begin{equation}\label{m2d2}
  \langle {m}_s^2 \rangle\propto L^{-(z-1+\eta_{\rm A})},~~~~ \langle D^2 \rangle\propto L^{-(z-1+\eta_{\rm V})}.
\end{equation}
Here we consider the scaling at the infinite-size extrapolated transition point, and also at finite-size critical points $Q_c(L)$, as we did above in
the case of the gaps. While the gap calculations are very expensive, and we only went up to $L=96$ above, the static quantities are relatively cheaper
to compute. Here we analyze data for $L$ up to $256$.

\begin{figure}[t]
\includegraphics[width=6.75cm]{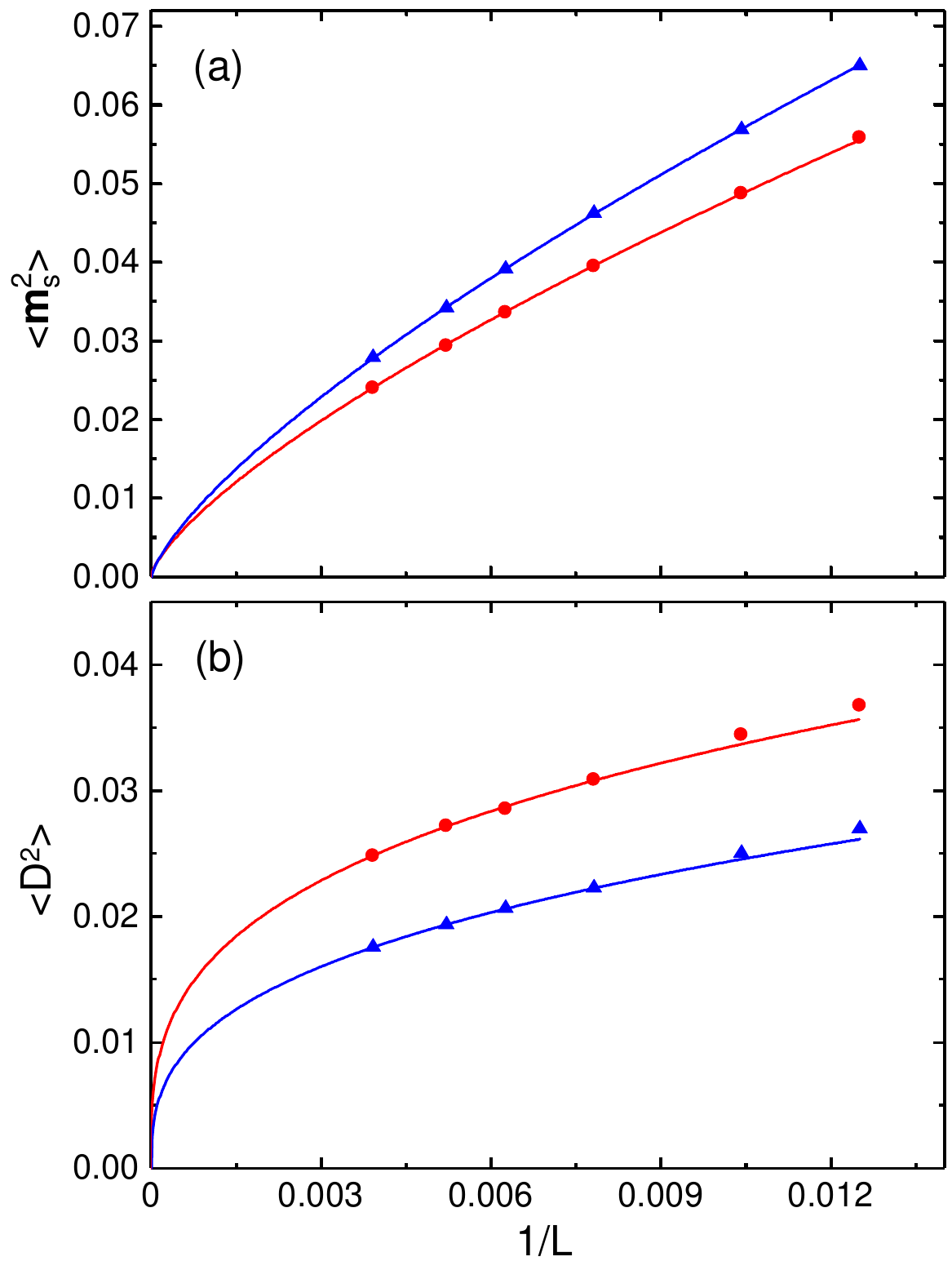}
\caption{Squared AFM (a) and VBS (b) order parameters graphed vs the inverse system size. We show values both at the infinite-size extrapolated critical
point (red circles) extracted in Fig.~\ref{a12cumucross}(c) and at the finite-size $U_{\rm A}=U_{\rm V}$ crossing points in the same figure (blue triangles).
Fits to power laws $\propto L^{-b_{{\rm A,V}}}$ to the data at $Q_c(\infty)$ give exponents $b_{\rm A}=0.72(1)$ in (a) and $b_{\rm V}=0.31(2)$ in (b), and
similar results are obtained with the $Q_c(L)$ points. In all cases only the data for the largest four system sizes were included in the fits. The larger
deviations of the extended fitted curves from the data for the smallest two sizes show that subleading corrections are more prominent in (b) than in (a).}
\label{m2d2crit}
\end{figure}

A useful single-size estimate of the critical point to consider here is the $Q$ value where the AFM and VBS cumulants cross each other. If there
is a direct transition between the two ordered phases, with no intervening QLRO phase, this crossing point should flow with increasing $L$ to the same
unique critical point that we extracted in the preceding sections using the VBS cumulant for two different system sizes used to generate the phase boundary
in Fig.~\ref{Fig.QMC}. In Fig.~\ref{a12cumucross} we illustrate the analysis of the crossing points of the two same-size cumulants. The crossing points flow
to $Q_c=0.589(3)$, which is consistent with our other results reported in previous sections.

Figure \ref{m2d2crit} shows data for the critical order parameters graphed versus $1/L$. We analyze data both at the size-dependent cumulant crossing
points $Q_c(L)$ and at the infinite-size extrapolated crossing point $Q_c(\infty)$ obtained above. Though the overall magnitudes of the order parameters are
clearly different in the two cases, the main difference is an overall factor and the exponents obtained from fits do not differ significantly
(in all cases by less than twice the error bars reported in the caption of Fig.~\ref{m2d2crit}). With the exponents $b_{\rm A,V}$
obtained in the fits to the $Q_c(\infty)$ data (which are less affected by scaling corrections, provided that the estimate of $Q_c$ is reliable) and
using the exponent definitions in Eq.~(\ref{m2d2}), we obtain the values $\eta_{\rm A}=1.09(2)$ and $\eta_{\rm V}=0.68(2)$. Though the error bars here are
purely statistical and do not reflect potential effects of scaling corrections, the values are sufficiently far from each other to conclude that remaining
corrections would not be able to render $\eta_{\rm A}=\eta_{\rm V}$. This inequality of the anomalous dimensions will be of great relevance in the context of
possible emergent symmetries, discussed below in Sec.~\ref{Sec:Symmetries}.

\begin{figure}[t]
\includegraphics[width=6.5cm]{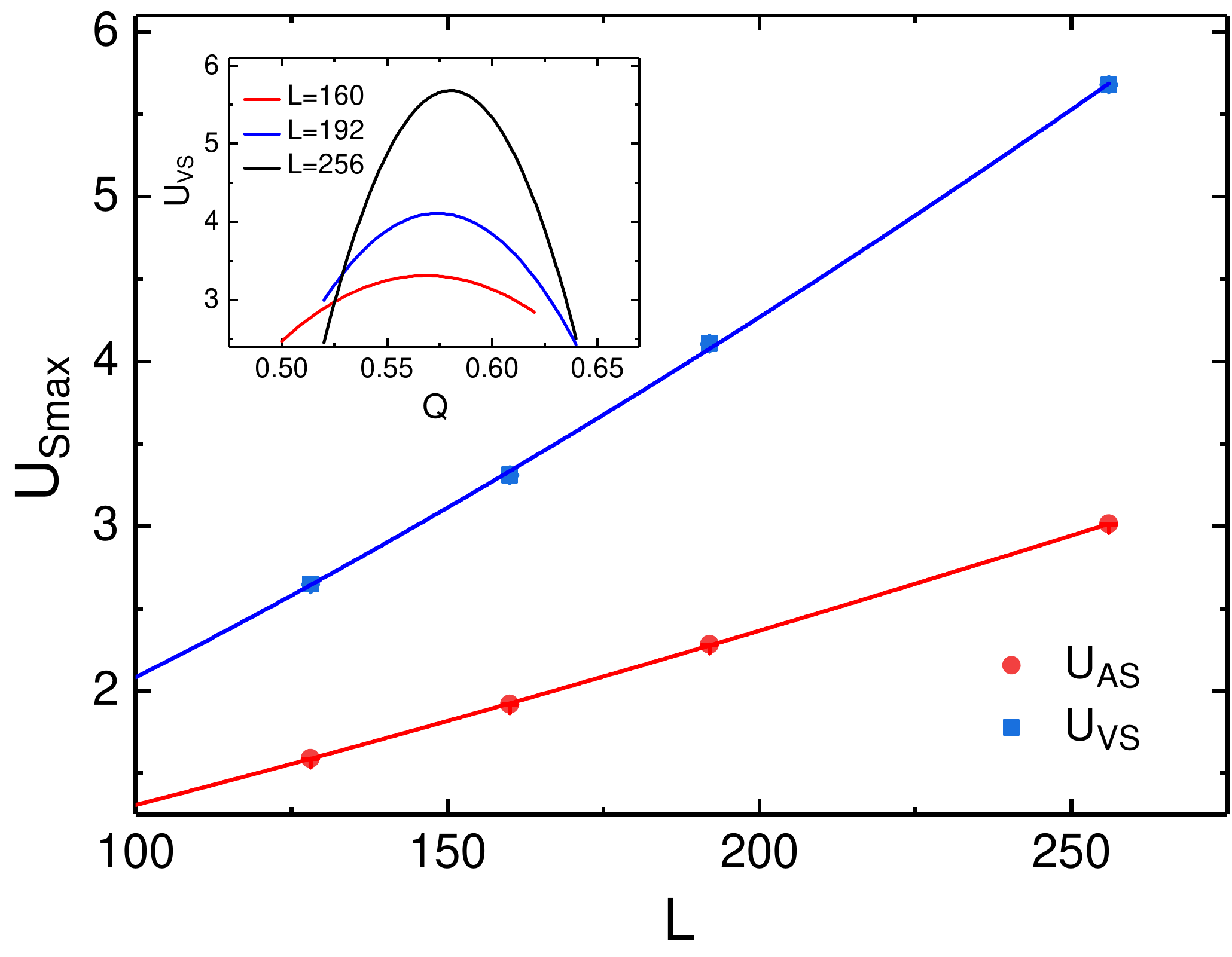}
\caption{Size dependence of the maximum slopes of the AFM and VBS cumulants, obtained by polynomial fitting to data sets such as those shown in
Fig.~\ref{Fig.Bindercross}. The curves show fitted power laws, $U_{\rm smax} \propto L^{1/\nu}$, which give the exponents $1/\nu_{\rm A} \approx 1.21(8)$
for the AFM order and $1/\nu_{\rm V} = 1.34(16)$ for te VBS order. The inset shows examples of the first $Q$ derivative of the fitted polynomial,
from which the maximum values are extracted.} 
\label{Nuexponent}
\end{figure}

Finally we consider the correlation length exponents $\nu_{\rm A}$ and $\nu_{\rm V}$, which also dictate the size of the critical region in finite-size
scaling. We use the dimensionless Binder cumulants, with expected finite-size scaling forms $U_{\rm A,V}(Q,L)=U_{\rm A,V}[(Q-Q_c)L^{1/\nu_{\rm A,V}}]$. One convenient
way to use this scaling form is to take the derivative with respect to $Q$,
\begin{equation}
U_{\rm SA,SV} = \frac{dU_{\rm A,V}}{dQ} = L^{1/\nu_{\rm A,V}}U'_{\rm A,V}[(Q-Q_c)L^{1/\nu_{\rm A,V}}],
\end{equation}
where $U'_{\rm A}(x)$ and $U_{\rm V}(x)$ are the derivatives of the above scaling functions $U_{\rm A,V}(x)$ with respect to $x=(Q-Q_c)L^{1/\nu_{\rm A,V}}$.
The point of maximal slope of a cumulant can be considered as a finite-size definition of the critical point, and therefore we take the first derivative
of polynomials fitted data such as those in Fig.~\ref{Fig.Bindercross} for different system sizes and locate the maximums. We refer to the maximum slopes of
the AFM and VBS cumulants as $U_{\rm SmaxA}$ and $U_{\rm SmaxV}$. These quantities should scale with the system size as $L^{1/\nu_{\rm A,V}}$ according to
the above forms.

\begin{figure*}[t]
\includegraphics[width=\textwidth]{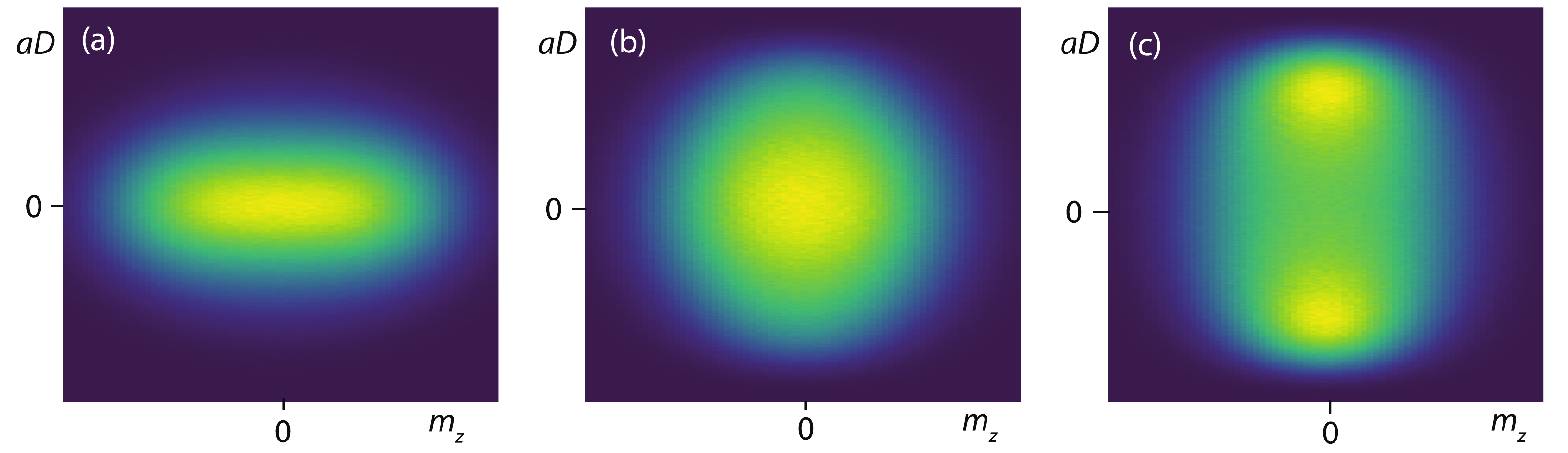}
\caption{Joint order parameter distributions $P(m_z,aD)$ accumulated in PQMC simulations of $L=256$ systems with $\alpha=1.2$ at three different $Q$ points;
(a) Inside the AFM phase at $Q=0.49$. (b) At the critical point, defined as the point $Q_c(256)=0.553$  where the AFM and VBS cumulants
cross each other for this system size (see Fig.~\ref{a12cumucross}). (c) At $Q=0.58$, inside the VBS phase. In all panels the scale factor $a$ is fixed at
its critical value, i.e., based on the data in (b). We here focuse on the shapes of the distributions and therefore include neither axis
markings nor scales for the colors used to represent the probability densities (linearly) from lowest (dark purple) to highest (bright yellow).}
\label{PmzD}
\end{figure*}

Figure \ref{Nuexponent} shows results for $U_{\rm SmaxA}$ and $U_{\rm SmaxV}$ graphed versus the system size. In the inset we show examples of the derivatives of
polynomials fitted to data for VBS cumulants (such as those shown in Fig.~\ref{Fig.Bindercross}). To eliminate finite-size effects as much as possible, we only
use data for system sizes from $L=128$ to $256$, for which we find statistically good fits to power law divergencies. The exponents extracted for these fits are
$\nu_{\rm A}=0.83(5)$ and $\nu_{\rm V}=0.75(7)$, which are equal within the error bars.

We have analyzed data for $\alpha=1.1$ in the same ways as described above for $\alpha=1.2$. The raw data sets have very similar appearances and we only
summarize the results for the exponents: $z=0.62(1)$, $\eta_{\rm A}=1.04(2)$, $\eta_{\rm V}=0.69(2)$, $\nu_{\rm A}=0.74(7)$, and $\nu_{\rm V}=0.70(7)$.
These exponents can not be statistically distinguished from those at $\alpha=1.2$, suggesting that the exponents are constant on the AFM--VBS2 boundary or,
at the very least, exhibit very small
variations.

\subsection{Emergent Symmetries}
\label{Sec:Symmetries} 

As discussed in Sec.~\ref{sub:spinons}, the 2D DQCP may be associated with an emergent SO(5) symmetry, which corresponds to the O(3) symmetric AFM order
parameter and the two VBS components combining into a five-dimensional vector transforming with the said symmetry. Evidence of the higher symmetry has been
detected in different ways in simulations of the $J$-$Q$ model \cite{Suwa16} and in 3D classical loop models \cite{Nahum15b}, though it is not yet clear
\cite{Wang17} whether the symmetry is truly asymptotically exact or eventually breaks down to $O(3) \times U(1)$ (provided that the transition is a DQCP
and not a weak first-order transition), where $U(1)$ is the lower emergent symmetry of the microscopically $Z_4$ symmetric VBS order parameter.

In 1D systems, an emergent O(4) symmetry of the AFM and VBS order parameters is presumably present in the entire QLRO phase, though log corrections
marginally violate the symmetry away from the QLRO--VBS phase transition. The predicted asymptotically exact O(4) symmetry within the Wess-Zumino-Witten
conformal field theory has been confirmed by studying non-trivial relationships between correlation functions in the $J$-$Q$ chain with only
short-range interactions \cite{Patil18}. In this case $\eta_{\rm A}=\eta_{\rm V}$, a condition which normally is regarded as a prerequisite for emergent symmetry
of the two critical order parameters. In the long-range $J$-$Q$ chain, we have demonstrated that $\eta_{\rm A}\not=\eta_{\rm V}$, and then it might appear that
no emergent symmetry should exist. The broken symmetry in the AFM state is O(3), in the VBS it is $Z_2$, and at the critical point, if the
two order parameters fluctuate independently of each other, the symmetry would be O(3)$\times Z_2$. We will demonstrate that the order parameters do not,
in fact, fluctuate independently but are correlated in a non-trivial way corresponding to a deformed O(4) symmetry.

We consider the joint probability distribution of the $m_z$ component of the AFM order parameter in Eq.~(\ref{msdef}) and the VBS order parameter $D$
defined in Eq.~(\ref{ddef}). To construct the distribution $P(m_z,D)$, we save a large number of point pairs $(m_z,D)$ generated in
PQMC simulations. For visualization, we construct histograms, and we also use the original sets of point pairs to construct quantitative measures of the
structure of the correlations between $m_z$ and $D$. Similar ways of analyzing distributions were previously used to detect DQCP SO(5) symmetry in 3D loop
models \cite{Nahum15b}and O(4) symmetry at unusual first-order transitions in $J$-$Q$ \cite{Zhao19} and loop models \cite{Serna19}. Here we follow mainly
the approach developed in Ref.~\cite{Zhao19}.

To detect an emergent symmetry of the two order parameters, their overall arbitrary magnitudes related to their definitions have to be taken into account.
Thus we define the ratio of the squared order parameters as
\begin{equation}
a^2 = \frac{\langle m_z^2\rangle}{\langle D^2\rangle}.
\label{ascale}
\end{equation}
Assuming the O(4) spherical symmetry of the three AFM components and single VBS component, when projecting down to the 2D distribution $P(m_z,D)$
we expect a circular-symmetric distribution once the difference in scales, quantified by the above ratio $a$, has been taken into account. Thus,
we define properly rescaled point pairs $(m_z,aD)$.
Here one can either compute $a$ for each value of the control parameter or fix it at its value at the critical point. In the former case, even
if there is no symmetry, per definition the rescaling will draw out or compress the distribution in the $D$ direction so that the second moments
in both directions will be the same. This effect will weaken the deviations from circular symmetry. We therefore use the second approach (as in
Ref.~\cite{Zhao19}) of fixing $a$ to its critical value. In this case we still have a choice of how to define the finite-size critical point. In
the present case it is convenient to use the same definition as we did above in Sec.~\ref{Subsec:OP};
the $Q(L)$ point at which the two Binder cumulants cross each other (as exemplified un Fig.~\ref{a12cumucross}).

\begin{figure*}[t]
  \includegraphics[width=17cm]{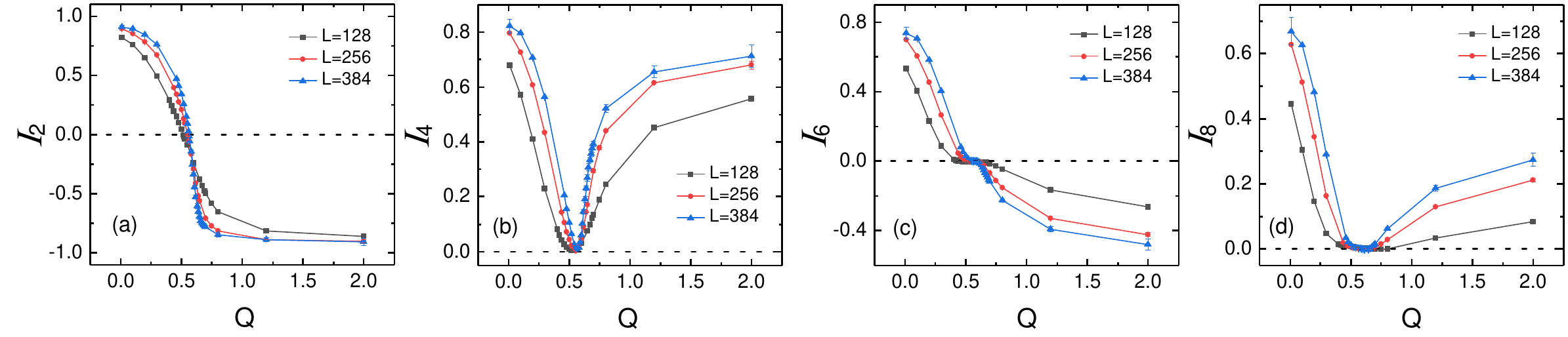}
  \caption{Angular integrals $I_q$, defined in Eq.~(\ref{iqint}), of the joint critical order parameter distributions $P(m_z,aD)$ for three different
  system sizes [the one for $L=256$ is shown in Fig.~\ref{PmzD}(b)], with (a)-(d) corresponding to $q=2,4,6,8$. The dashed lines are drawn at $I_q=0$
  to emphasize the value corresponding to the emergent symmetry.}
  \label{Indist}
\end{figure*}

\begin{figure*}
  \centering
  \includegraphics[width=17cm]{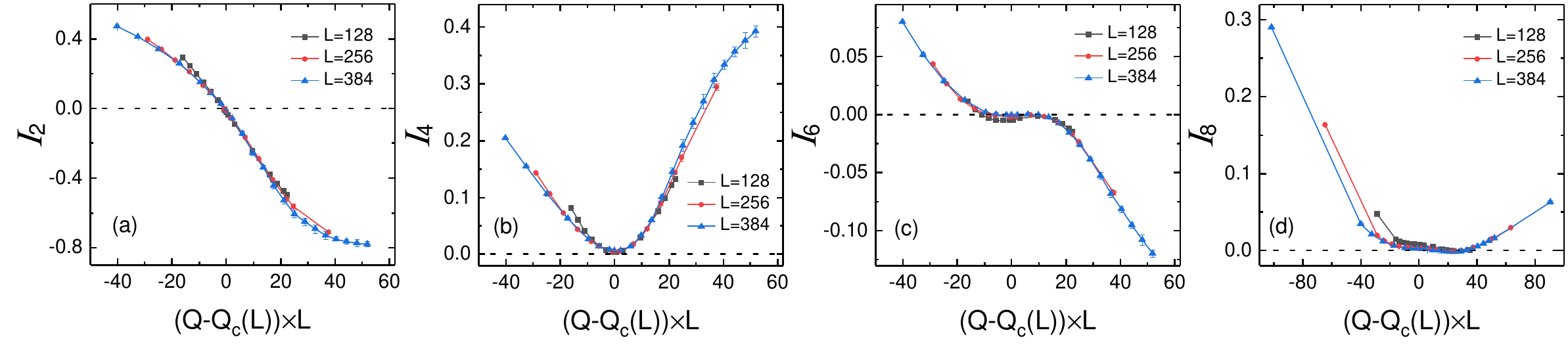}
  \caption{The same angular integrals as in Fig.~\ref{Indist}, here with the $Q$ values rescaled on the horizontal axis according to the expected critical form.
    The critical values $Q_c(L)$ are the cumulant crossing points shown in Fig.~\ref{a12cumucross}(c). The scaling of $Q-Q_c(L)$ by $L$ corresponds
    to the exponent $\nu_4=1$, which is in the range of values $\approx [0.9,1.1]$ for which good collapse of the $L=256$ and $L=384$ data is observed
    when $I_q \approx 0$.}
  \label{Indistcollapse}
\end{figure*}

Examples of distributions are shown in Fig.~\ref{PmzD}, where panel (a) is inside the AFM phase, (b) is at the critical point, and (c) inside
the VBS2 phase. Visually, the critical distribution in panel (b) exhibits perfect rotational symmetry, while those in (a) and (c) have developed the features
expected in the two phases. In the AFM phase, O(3) order projected down to one dimension gives a line segment (with the end-points reflecting
the magnitude of the order parameter), which here is broadened in both directions because of finite-size fluctuations. In the VBS2 phase
the two-fold degeneracy produces two maximums on the vertical axis, again with finite-size fluctuations producing surrounding weight.

To quantify the symmetry at the critical point, we use the angular integrals \cite{Zhao19}
\begin{eqnarray}
I_q &=& \int \text{d}{m}_z \text{d}(aD) P({m}_z,{aD}) \cos \bigl (q \phi(m_z,{aD}) \bigr )\nonumber \\
& = &  \frac{1}{M}  \sum_{i=1}^{M}\cos(q \phi(|{m}_z|, |{aD}|)_i),
\label{iqint}
\end{eqnarray}
where on the second line the integral has been converted to a sum over the point pairs generated in the simulation, with the scale
factor $a$ again fixed to its critical value for given system size. The angles $\phi(|{m}_z|, |{aD}|)_i \in [0,\pi/2]$ are computed
for each data point $({m}_z,{aD})$ and are for convenience restricted to the first quadrant (as allowed by symmetry).

We consider $I_2,I_4,I_6$, and $I_8$, and present results in Fig.~\ref{Indist}. We observe that all four integrals are very close to zero at points corresponding
closely to the size-dependent critical values $Q_c(L)$, thus supporting a circular symmetry of the distributions. This circular symmetry of the 2D distribution
directly translates into an O(4) symmetry of the vector $(m_x,m_y,m_z,aD)$ [or possibly SO(4), because our method can not address the existence of a physical
reflection operator with negative determinant] . Moreover, as shown in Fig.~\ref{Indistcollapse}, we can rescale the $Q$ axis according to a finite-size
scaling form, $I_q=I_q[(Q-Q_c)L^{1/\nu_4}]$, where for $Q_c$ we take the cumulant-crossing  points $Q_c(L)$ to account for the still present drift
of the critical point with the system size. We cannot determine the exponent $\nu_4$ very precisely, but we observe
good data collapse for the largest system sizes roughly in the window $\nu_4 \in [0.9,1.1]$. Thus, $\nu_4$ is marginally larger than the values
we found for the correlation lengths $\nu_{\rm A} \approx \nu_{\rm B}$, though the differences are not sufficiently large within the error bars
to definitely conclude that the are different.

\begin{figure}[t]
\includegraphics[width=7cm]{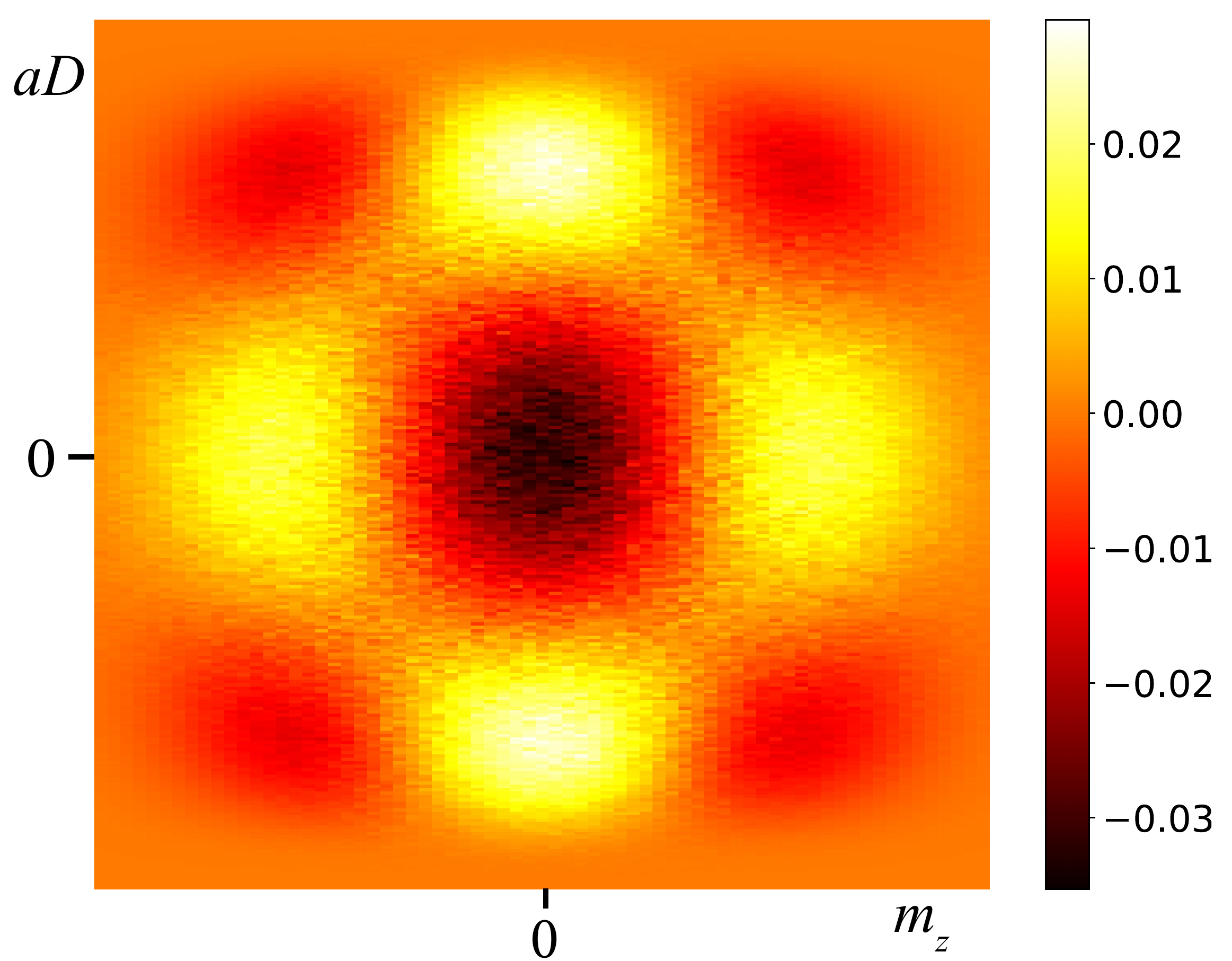}
\caption{Difference $\Delta_P(m_z,aD)$ between the joint probability distribution $P(m_z,aD)$ at the critical point, Fig.~\ref{PmzD}(b), and
the distribution product $P(m_z)P(aD)$ obtained from the same data set.}
\label{DeltaQc}
\end{figure}

\begin{figure}[t!]
  \includegraphics[width=7.5cm]{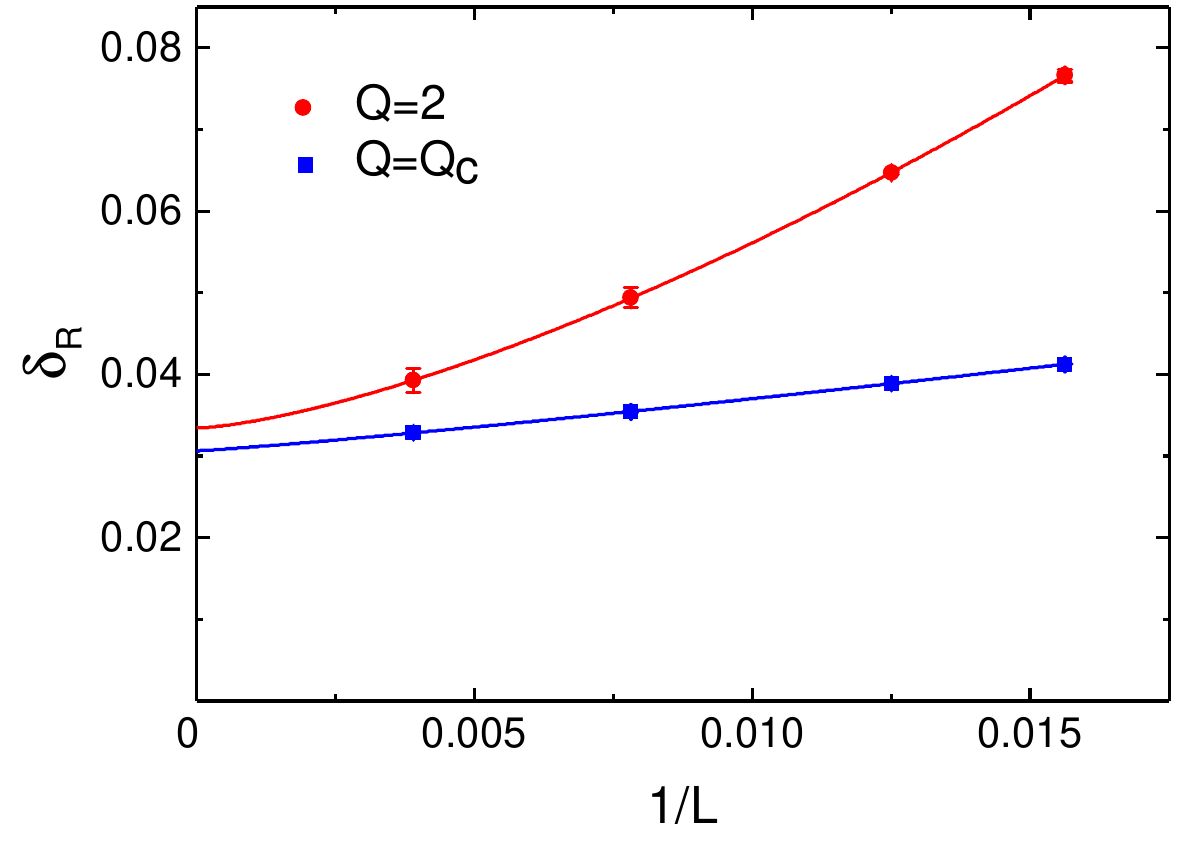}
  \caption{The integrated RMS deviation defined in Eq.~(\ref{deltaRMSdef}) for the system with $\alpha=1.2$ at the finite-size critical points
  $Q_c(L)$ and at the fixed value $Q=2$ inside the VBS2 phase. The dependence on $1/L$ in both cases is fitted to the form $\delta_R=a+L^{-b}$.} 
\label{RMSDelta}
\end{figure}

One way in which a circular symmetric distribution could arise is if both order parameters are normal-distributed and independent. Then, regardless of the
standard deviations of the individual distributions, the distribution $P(m_z,aD)$ with rescaled VBS order parameter would be circular symmetric
(following a 2D normal distribution). The individual distributions $P(m_z)$ and $P(aD)$ are not consistent with Gaussians, however. To further test whether
the distributions are independent or not, we consider the difference between the joint distribution $P(m_z,aD)$ and the product distribution
$P({m}_z)P({aD})$, defining
\begin{equation}
\Delta_P({m}_z,{aD})=P({m}_z,{aD})-P({m}_z)P({aD}).
\label{Deltapdef}
\end{equation}
Figure \ref{DeltaQc} shows a color-coded plot of this quantity at the critical point for $L=256$. Here we observe a four-fold symmetry with an
interesting structure of negative and positive deviations. To test whether these deviations survive in the thermodynamic limit, i.e., whether the
correlations between the two order parameters vanish or not, we define the root-mean-square (RMS) integrated difference
\begin{equation}
\delta_R = \sqrt{\int d{m}_z\int d {aD} \Delta^2_P({m}_z,{aD})}.
\label{deltaRMSdef}
\end{equation}
In Fig.~\ref{RMSDelta} this quantity is graphed versus the inverse system size both at the critical point and at a fixed $Q$ value inside the VBS2
phase. In both cases $\delta_R$ clearly does not decay to zero, demonstrating that the order parameters remain correlated in the thermodynamic limit.

In Appendix \ref{app:dist} we show further results for $\delta_R$ over a larger range of $Q$ values and conclude that the two order parameters are
correlated for all values of $Q$ when $\alpha=1.2$, with the maximum correlation at the critical point. Such correlations would not normally be expected
inside the ordered phases, but apparently the long-range interactions have this effect. It should be noted here that one order parameter being small
in the phase where there is long-range order of the other kind does not immediately imply that $\delta_R\to 0$, because the underlying function
$\Delta_P({m}_z,{aD})$ defined in Eq.~(\ref{Deltapdef}) can clearly be large regardless of the values of the arguments. In Appendix \ref{app:dist} 
we also study the model without long-range interactions and show that its order parameters only remain correlated in the thermodynamic limit in the QLRO
phase (including at the QLRO--VBS transition), as expected, while in the VBS phase $\delta_R \to 0$.

Based on all the above results, we conclude that the direct AFM--VBS2 transition is associated with an emergent symmetry, but of a kind that has, to our
knowledge, not been considered previously in the context of quantum phase transitions. Given that the anomalous dimensions $\eta_{\rm A}$ and $\eta_{\rm V}$ of
the two order parameters are different, the scale factor $a$ in Eq.~(\ref{ascale}) does not approach a constant with increasing system size but takes the scaling
form $a(L) \sim L^{\eta_{\rm V} - \eta_{\rm A}} \approx L^{-0.4}$. Thus, the original distribution $P(m_z,D)$ becomes increasingly elongated in the $D$ direction. The
full distribution would then in some sense only have $O(3) \times Z_2$ symmetry. However, a distribution with this lower symmetry can not in general be
rescaled in the way we have done here to obtain an O(4) symmetric distribution. Moreover, the notion of $O(3) \times Z_2$ symmetry would normally imply
independent fluctuations of the two critical order parameters, which we have shown is not the case here. Thus, we conclude that what we have here is a highly
non-trivial deformed O(4) distribution. In the statistics literature, such a deformed multi-dimensional spherical distribution is said to have ''elliptical
symmetry'' \cite{Bentler83,Paindaveine12}. The AFM--VBS2 transition then has emergent elliptical O(4) symmetry.

Beyond demonstrating the emergent symmetry, the angular integrals $I_q$ in Fig.~\ref{Indist} also provide further evidence for the direct, continuous transition
between the AFM and VBS2 phases, with no intervening QLRO phase. The widths of the features seen close to the transition point---the minimums for $q=2$ and $4$
and plateau for $q=3$---narrow roughly as $L^{-1}$ according to the data collapse in Fig.~\ref{Indistcollapse}, and the location also shifts roughly as it does
in the cumulant crossing. We also point out that it is not necessary to fix the scale factor $a(L)$ to its value at $Q_c(L)$ in order to observe the emergent
symmetry in $I_q$ and the critical scaling of the window over which they vary close to the transition; see results in Appendix \ref{app:dist}.

\section{Summary and Discussion}
\label{Sec:summary}

In summary, the $J$-$Q$ chain with long-range Heisenberg interactions presents and intriguing phase diagram that offered us possibilities both to
study previously known quantum phase transitions (QLRO--AFM and QLRO--VBS) in more detail and, most importantly, exhibits a novel direct,
continuous transition between the AFM ground state and a VBS with coexisting algebraic spin correlations (the VBS2 phase). The latter transition
is a clear-cut case of a deconfining transition, in the sense that spinons do not exist as quasi-particles in the AFM phase (the elementary excitations
of which are magnons with sublinear dispersion \cite{Yusuf04,Laflorencie05}) but are deconfined in the VBS2 phase. The VBS2 phase is presumably gapless, and,
if so, the AFM--VBS2 transition takes place between two gapless phases. We were not able to confirm the gapped versus gappless difference between the VBS1 and
VBS2 phases, because the gap can be very small also in the VBS1 phase, due to its generation by a marginal operator \cite{Affleck87}.

Though we did not discuss any results directly probing the spinons here, we have confirmed their existence in the VBS2 phase in the way introduced in
Ref.~\cite{Tang11}, using the PQMC method for $S=1$ states expressed in an extended valence bond basis (with valence bonds and two unpaired spins).
Unlike a 2D VBS, where the spinons are confined into gapped magnons (some times called triplons), which can be regarded as bound states of spinons
\cite{Sulejman17,Tang13}, in the 1D case the VBS2 background does not, even with the long-range interactions, cause a binding potential between the
spinons. Instead the effective potential is weakly repulsive (as it is in the previously studied VBS states with short-range interactions \cite{Tang11}).
It would clearly be interesting to design a 1D model which has confined spinons in the VBS state.

We have studied the critical behavior of the AFM and VBS order parameters in detail at two points on the AFM--VBS2 boundary and confirmed by tests
at other points that the behaviors are generic for the entire phase boundary. The critical exponents may in principle be varying on the boundary, but
we did not find statistically significant differences here between the two points studied, where the dynamic exponent $z \approx 0.6$, the correlation
length exponents corresponding to AFM and VBS order, $\nu_{\rm A}$ and $\nu_{\rm V}$, are in the range $0.7-0.8$ and possibly $\nu_{\rm A}=\nu_{\rm V}$.
The exponents of the critical correlation functions (the anomalous dimensions) are distinctly different from each other, with $\eta_{\rm A} \approx 1.1 $
and $\eta_{\rm V} \approx 0.7$.

The phase diagram Fig.~\ref{Fig.QMC} suggested by the different ways of extracting the phase boundaries still has some remaining uncertainties, especially
in the region $\alpha^{-1} \approx [0.6,0.7]$ of the long-range interaction parameter, where remaining finite size corrections are not well controlled by any
of the methods used, for the range of available system sizes. For $\alpha \alt 0.6$ our tests indicate that the phase boundary from Lanczos ED gap crossings
(blue symbols) is rather accurate in Fig.~\ref{Fig.QMC} while the cumulant crossing method (red symbols) is significantly affected by finite-size effects. In
contrast, for $\alpha \agt 0.7$ the cumulant based results are stable while the extrapolated gap crossings show large size drifts.

\begin{figure}[t!]
  \includegraphics[width=6.5cm]{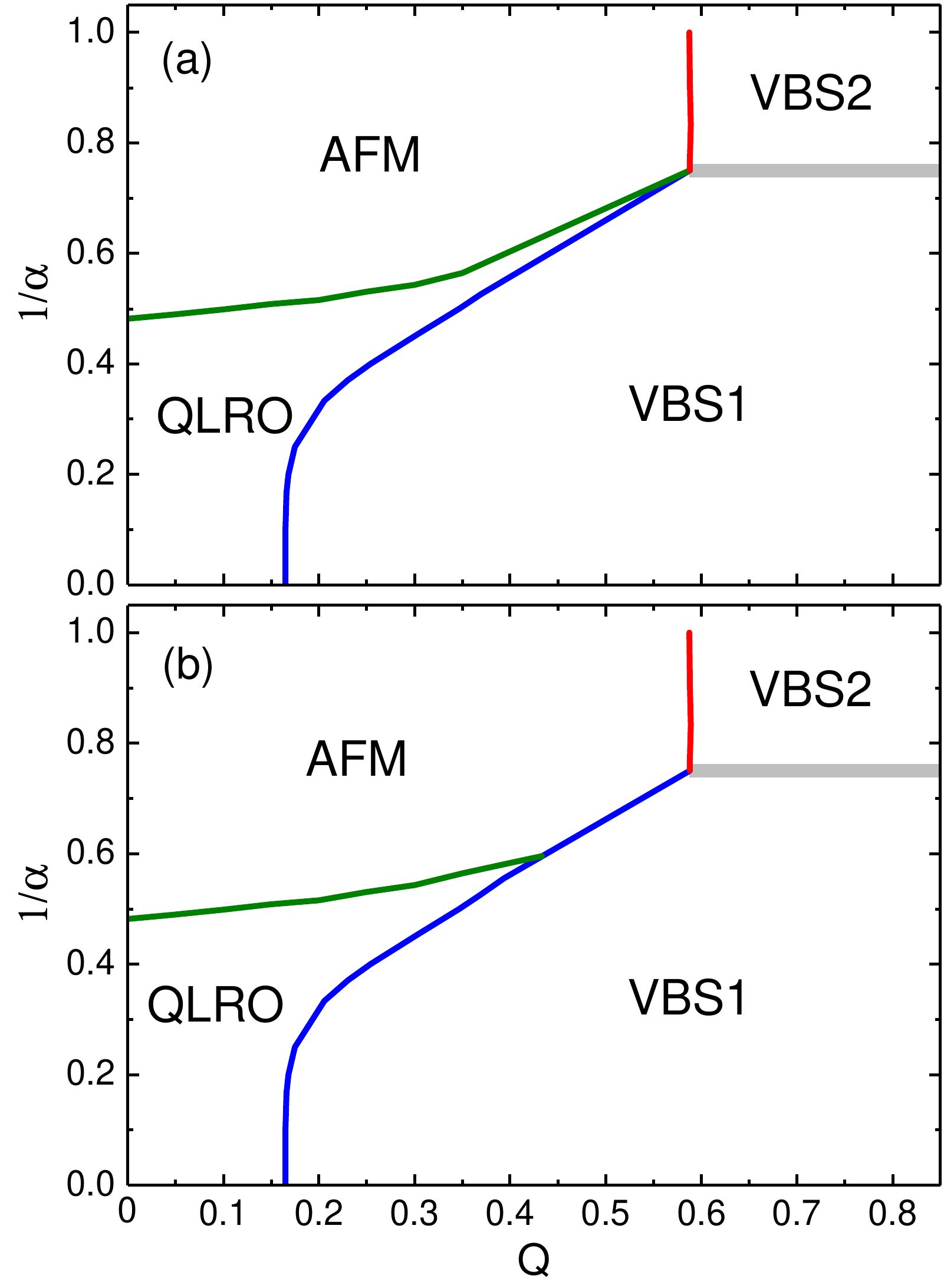}
  \caption{Phase diagrams based on available relaible rresults and two scenarios for how the phases connect in the region $\alpha^{-1} \approx [0.6,0.7]$
    where our results can not provide conclusive results. In (a), the QLRO phase extends toward the VBS2 phase in such a way that there is no extended
    direct AFM--VBS1 phase boundary, while in (b) there is a multi-critical point where only the AFM,QLRO, and VBS phases meet, with a segment of
  direct AFM-VBS1 transitions.} 
\label{twodiagrams}
\end{figure}

The remaining uncertainties leave two possibilities for the most intricate details
of the phase diagram in the region $0.6 \alt \alpha \alt 0.7$ where all the phase boundaries come close to each other.
In Fig.~\ref{twodiagrams} we outline two possible complete phase diagrams based on the reliably determined parts of the phase boundaries and different
ways in which they can connect to each other. In Fig.~\ref{twodiagrams}(a) there is no direct AFM--VBS1 transition, while such a phase boundary exists
in Fig.~\ref{twodiagrams}(b). To distinguish between these scenarios, and to establish the precise location of the VBS1--VBS2 boundary,
additional calculations for larger system sizes will be required. 

One of the most intriguing aspects of the AFM--VBS2 transition is its association with a kind of emergent symmetry not previously discussed in the
context of quantum phase transitions---an elliptical O(4) symmetry, by which the AFM and VBS order parameters fluctuate within an O(4) sphere after
a rescaling of, say, the VBS order parameter by $L^{\eta_{\rm V}-\eta_{\rm A}}$. Elliptical distributions have been studied
in statistics \cite{Bentler83,Paindaveine12}.

Though the emergent O(4) symmetry was here studied in the form of a finite-size property,
in the thermodynamic limit we expect analogous behaviors when the order parameter is observed on a length
scale $\Lambda$, with the spherical symmetry manifested upon rescaling of the VBS order parameter by $\Lambda^{\eta_{\rm V}-\eta_{\rm A}}$. Various cross-correlation
functions should also reflect the symmetry (though useful relationships known in the context of CFTs \cite{Patil18} will not be valid,
because $z\not=1$), though we have not studied those yet. It would be interesting to investigate prospects of elliptical symmetries also within the context of
quantum field theories. It has previously been presumed that $\eta_{\rm V}=\eta_{\rm A}$ is a prerequisit to emergent symmetries of the two order parameters,
such as SO(5) at the 2D DQCP \cite{Nahum15b}.

It is here interesting to note that, because of the different logarithmic corrections to the $r^{-1}$ decaying AFM and VBS correlation in the QLRO
phase the scale factor $a$ also decays to zero with increasing $L$ (or other length scale $\Lambda$) in this case. Calculations in the context of
the conventional Heisenberg chain \cite{Giamarchi89}, which belongs to the QLRO phase, have shown that the multiplicative corrections are
$\ln^{1/2}(r)$ and $\ln^{-3/2}(r)$ for AFM and VBS correlations, respectively. Thus, while the Heisenberg chain is commonly said to host an O(4) symmetry,
it is actually also an elliptical O(4) symmetry, though the deformation of the distribution as a function of the system size $L$ (or scale $\Lambda$)
there is only logarithmic, instead of the stronger algebraic deformation in the model studied here.

In the case of the 2D DQCP, the exponent $\nu'$ governing emergent $U(1)$ symmetry is larger than the correlation-length exponent $\nu$ (i.e., the length scale on
which emergent U(1) symmetry is manifested inside the VBS phase is larger than the correlation length) \cite{Lou09,Levin04}. In the field theory, this reflects
a ``dangerously irrelevant'' operator that was assumed to be responsible for the eventual reduction in symmetry to $Z_4$ inside the VBS phase
\cite{Senthil04a,Levin04}. In the case
at hand here, where the VBS phase breaks $Z_2$ symmetry, it is not clear if some similar mechanism should be at play, or whether the symmetry break-down is
also controlled by $\nu_{\rm A}$ and  $\nu_{\rm B}$, then presumably with $\nu_{\rm A}= \nu_{\rm B}$. The exponent $\nu_4$ that we here extracted from the way
the elliptical O(4) symmetry is violated in the ordered phases is somewhat larger than $\nu_{\rm A}$ and $\nu_{\rm B}$, but not enough so beyond statistical errors
(and possibly some systematical errors due to remaining finite-size corrections) to definitely conclude that they are different.

We envision that the model and results presented here will stimulate further field-theory work on a broader range of 1D DQCP-like phenomena, especially
as regards the nature of emergent symmetry. A nonlinear sigma-model was already constructed for the AFM phase and the QLRO--AFM transition of the long-range
Heisenberg model \cite{Laflorencie05} but how to incorporate the VBS formation within this scheme (or whether there is a better starting point)
remains to be understood.

Beyond working out the remaining unknown aspects of the phase diagram illustrated in Fig.~\ref{twodiagrams}, there is also clearly much room for further computational
work, e.g., more detailed studies of the AFM--VBS2 boundary (and a possible AFM--VBS1 transition), the nature of the coexisting algebraic correlations inside
the VBS2 phase, the nature of the VBS1--VBS2 transition, and the properties of the multi-critical point(s). It would also be interesting to consider anisotropic
interactions, both XY-like and Ising-like.

Experimentally, since it is possible to engineer 1D long-range spin interactions within a variety of the 
platforms currently explored for quantum simulators \cite{Bohnet16,Zeiher17,Nguyen18}, it is also plausible that some short-range interaction could be
realized that competes with AFM ordering and leads to a quantum phase transition into a state with VBS order. Metallic chains have been predicted to host
long-range Heisenberg interactions, and there are variations in the strengths of the short-range and long-range interactions depending on the constituent
elements, including different signs of the short-range couplings \cite{Tung11}. It may also be possible to vary the interactions depending on the nature
of the substrate. Even though the spins live in an itinerant electronic environment, some of the phenomena discussed here may still survive.

Here it is worth recalling the previous work on the frustrated Heisenberg chain with long-range interactions \cite{Sandvik10b}, where a strongly first-order
AFM--VBS transition was found. Thus, the continuous nature of this transition is not guaranteed. The reason for the different transitions induced by the
correlated singlet projectors of the $Q$ term and frustrating second-neighbor Heisenberg interactions $J_2$ may lay in the nature of the VBS2 state. Here,
in the long-range $J$-$Q$ chain we found coexisting algebraic spin correlations at wave-number $\pi$, while in Ref.~\cite{Sandvik10b} dominant correlations
at $\pi/2$ were detected in the long-range Heisenberg model. It should be possible to tune the short-range interactions also in other, experimentally more
accessible ways, e.g., with $J_2$ and $J_3$, in such a way as to change the nature of the algebraic correlations in the VBS2 phase and thereby change the
nature of the quantum phase transition. This aspect of the system could be explored experimentally, as well as computationally in a broader range of models
than the $J$-$Q$ chain studied here.

\acknowledgments

We would like to thank Pranay Patil, Yu-Rong Shu, Jun Takahashi, Han-Qing Wu, Ling Wang, and Cenke Xu for useful discussions. The work at Sun Yat-Sen
University was supported by Grants No.~NKRDPC-2017YFA0206203, No.~NKRDPC-2018YFA0306001, No.~NSFC-11974432, No.~NSFG-2019A1515011337, National
Supercomputer Center in Guangzhou, and Leading Talent Program of Guangdong Special Projects. The work at Boston University was supported by the NSF
under Grant No. DMR-1710170 and by a Simons Investigator Award. Some of the calculations were carried out on the Shared Computing Cluster managed
by Boston University's Research Computing Services.

\begin{figure}[t]
\includegraphics[width=6.5cm]{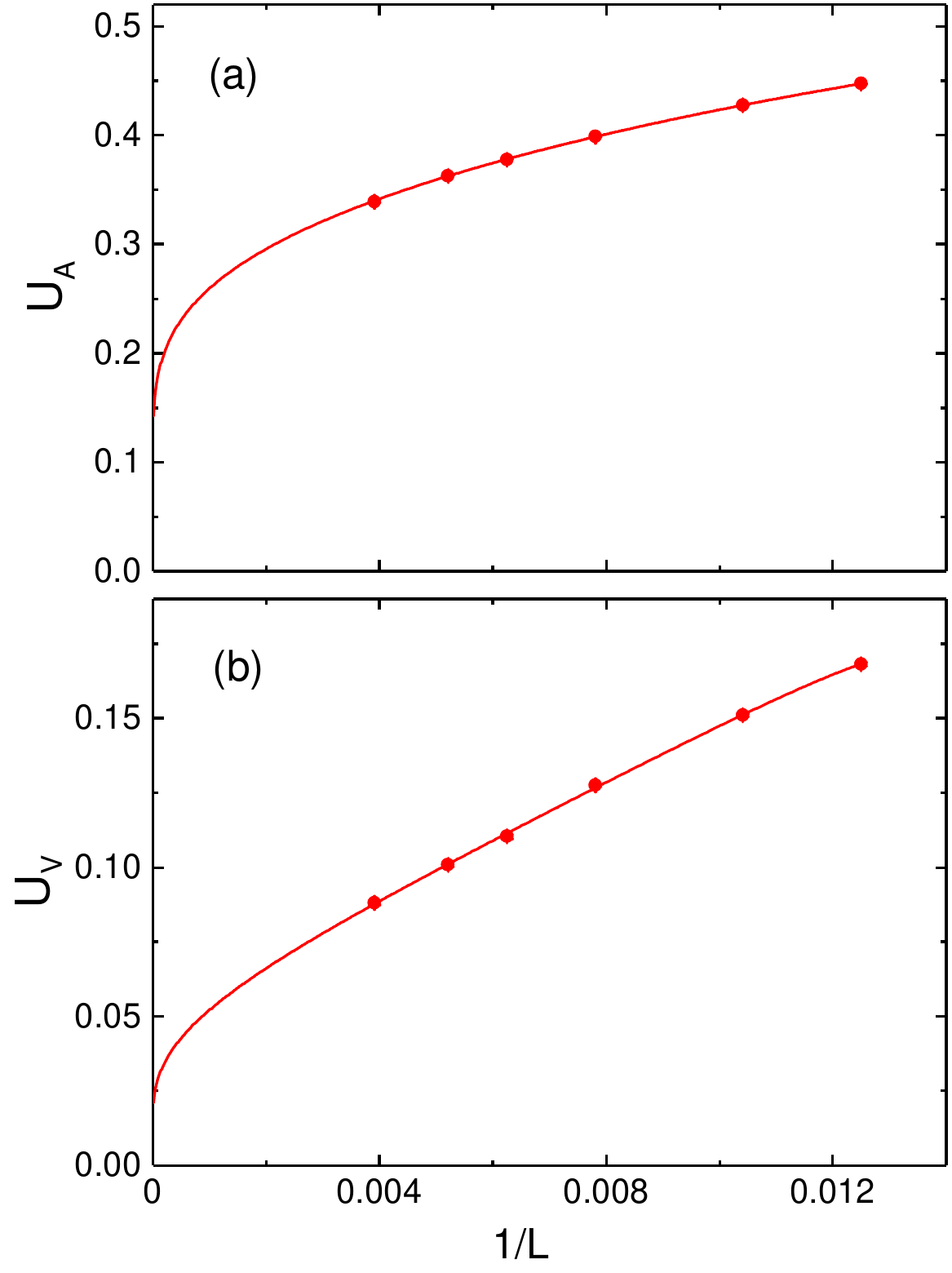}
\caption{AFM (a) and VBS (b) Binder cumulants, defined in Eqs.~(\ref{uadef}) and (\ref{uvdef}), respectively, versus the inverse system size of the
standard Heisenberg chain, i.e., $\alpha=\infty,Q=0$ in the Hamiltonian Eq.~(\ref{hjrjq}). The curves show $U_{\rm A,V} \to 0$ fits to the form
$a\ln^{-1}(L/L_0)$ with adjustable parameters $a$ and $L_0$.}
\label{Heisenberguauv}
\end{figure}

\begin{figure*}[t]
\includegraphics[width=17cm]{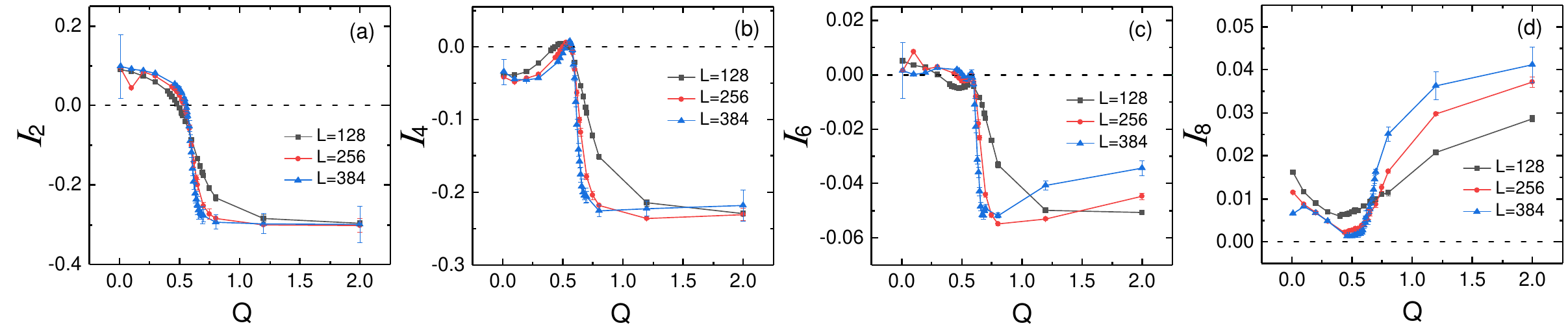}
\caption{Angular integrals as defined in Eq.~(\ref{iqint}) with the re-scaling parameter $a$ in the distribution $P(m_z,aD)$ computed separately for each
$Q$ value, instead of fixing it's value at $Q=Q_c(L)$ as done in Fig.~\ref{Indist}.}
\label{IqaL}
\end{figure*}

\appendix

\section{Binder cumulants in the QLRO state}
\label{app:heisenberguauv}

Both the spin and dimer correlations decay with distance $r$ as $r^{-1}$ in the standard $S=1/2$ Heisenberg chain. These dominant algebraic forms are affected by
multiplicative log corrections, of the form $\ln^{1/2}(r)$ and $\ln^{-3/2}(r)$ for the spin and dimer correlations, respectively \cite{Giamarchi89}.
The log corrections imply related corrections also in the second and fourth powers of the order parameters needed for the Binder cumulants,
Eqs.~(\ref{uadef}) and (\ref{uvdef}), but we are not aware of any predictions for the logs in the fourth powers. It is therefore not clear how the Binder
cumulants will scale with the system size.

Fig.~\ref{Heisenberguauv} shows PQMC results for both cumulants graphed versus $1/L$. We observe that the results can be fitted to simple
inverse-log decays with increasing system size. These forms should apply asymptotically in the entire QLRO phase and lend support to
the applicability of the cumulant-crossing method to extract the phase boundaries of the QLRO phase.

\begin{figure}[t]
\includegraphics[width=6.5cm]{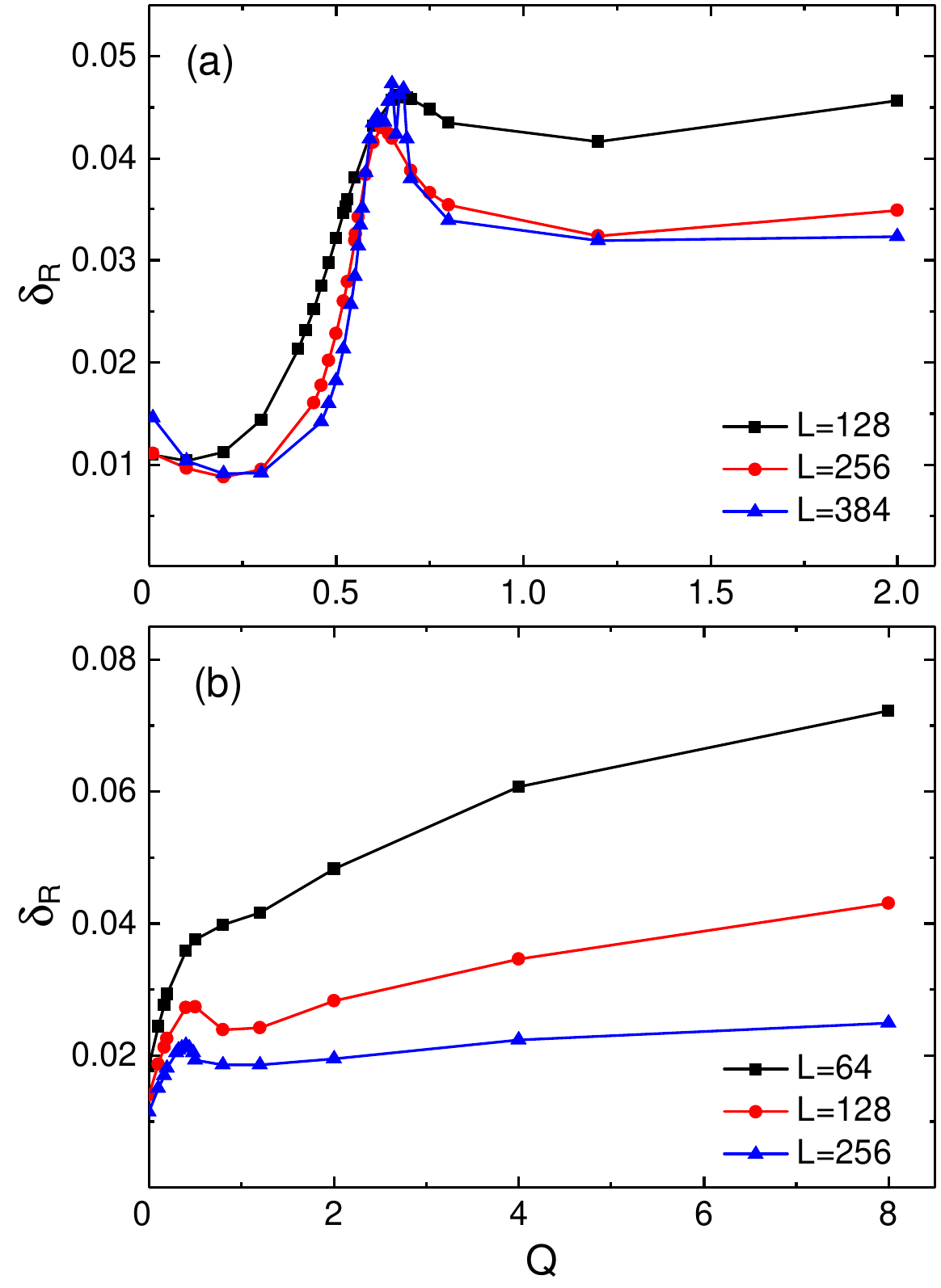}
\caption{The quantity $\delta_R$ defined in Eq.~\ref{deltaRMSdef} to characterize the deviations of the joint (AFM,VBS) order parameter distribution from a
product distribution, graphed versus $Q$ for different system sizes. Panel (a) is for the long-range $J$-$Q$ chain with $\alpha=1.2$ and (b) is for the model
with only nearest-neighbor Heisenberg interactions.}
\label{deltaRvsQ}
\end{figure}

\begin{figure}[t]
\includegraphics[width=6.5cm]{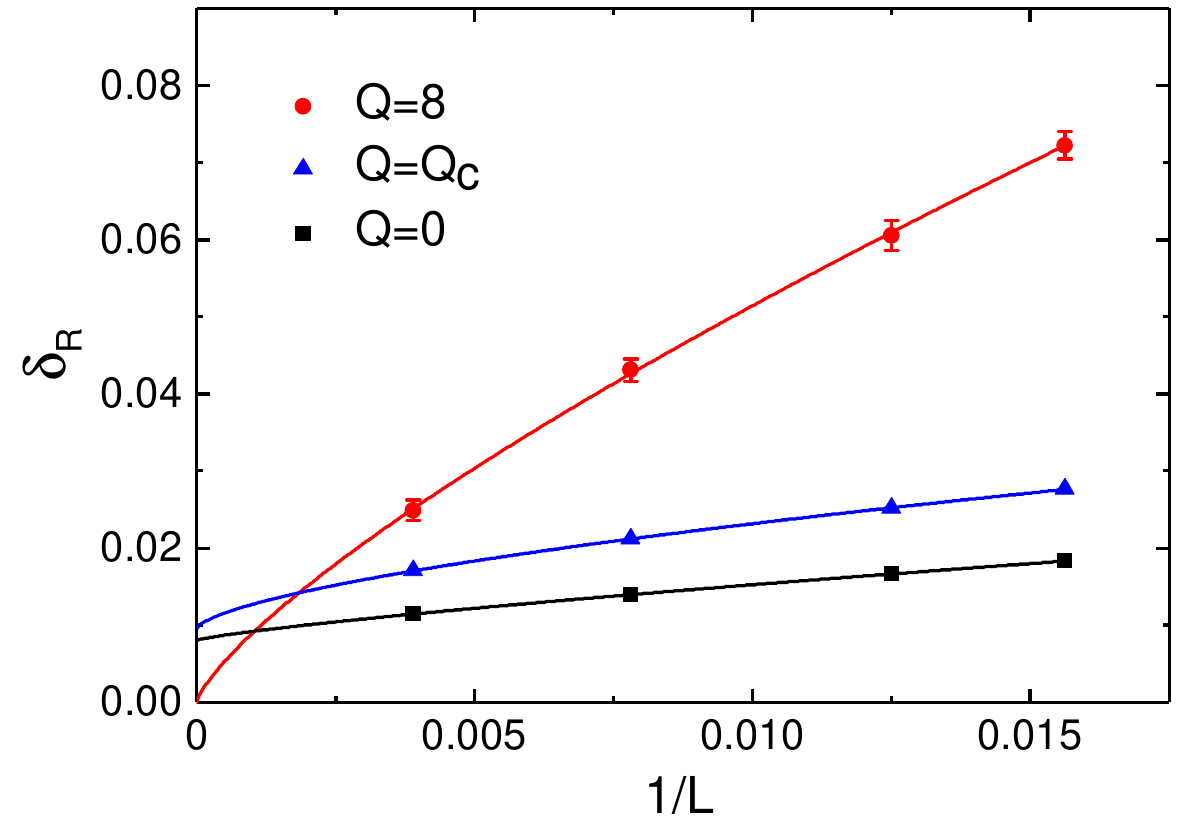}
\caption{Finite-size scaling of $\delta_R$ of the short-range $J$-$Q$ chain ($\alpha=\infty$)
for points inside the VBS1 phase ($Q=8$), at the transition point ($Q_c = 0.16478$), and in
the QLRO phase ($Q=0$) the curves show fits to power laws + constants (with the constant vanishing at $Q=8$).}
\label{nolrdeltaRvsL}
\end{figure}

\section{Additional emergent symmetry analysis}
\label{app:dist}

Here we provide some further results on the emergent elliptical O(4) symmetry in the long-range $J$-$Q$ model. We also show results
supporting conventional emergent O(4) symmetry in the model with only short-range interactions ($\alpha=\infty$).

\subsection{Angular Integrals with floating scale factor}

In Sec.~\ref{Sec:Symmetries} we investigated emergent symmetry with the choice of fixing the value of the scale factor $a=a(L)$ defined in Eq.~(\ref{ascale})
at its value at $Q_c(L)$. When instead using a floating value $a(Q,L)$, the $Q$ dependent angular integrals defined in Eq.~(\ref{iqint}) differ
significantly from those shown in
Fig.~\ref{Indist}. However, as shown in Fig.~\ref{IqaL}, we still observe that these symmetry-detecting integrals vanish at points close to the previously
determined $Q_c(L)$ values and deviate from zero away from these points. An exception is $I_6$, which is small in the whole range $0 \le Q < Q_c(L)$. $I_8$
is also clearly non-zero at its minimum, but appears to decrease to zero with increasing $L$. As before in the fixed-$a$ calculations, we also here
see that the distinct features of the curves in the neighborhood of the transition point become sharper with increasing $L$.

\subsection{Order-parameter correlations}

In Sec.~\ref{Sec:Symmetries}  we analyzed the quantity $\delta_R$ defined in Eq.~(\ref{deltaRMSdef}), which characterizes the overall deviation of the joint
probability distribution $P(m_z,aD)$ from the product distribution $P(m_z)P(aD)$. Here $a$ is the factor defined in Eq.~(\ref{ascale}) that is used to
set the two order parameters on equal scales. We showed results for the long-range $J$-$Q$ chain in Fig.~\ref{RMSDelta} at a critical point and inside the
VBS2 phase, demonstrating that the two order parameters remain correlated in both cases
in the thermodynamic limit. Here in Fig.~\ref{deltaRvsQ}(a) we show results for a wide range of $Q$ values for three different system sizes. We can clearly see
that the order parameters are the most strongly correlated ($\delta_R$ is peaked) at the critical point but do not decay to zero when $L \to \infty$ for any $Q$.

We next consider the $J$-$Q$ chain without the long-range interactions, which undergoes a QLRO--VBS1 transition and which is known to have an emergent
O(4) symmetry at the transition point \cite{Patil18}.
Inside the VBS1 phase the two order parameters should be decoupled (and the AFM correlations decay exponentially).
In the QLRO phase the O(4) symmetry is weakly violated, which in the field-theory description is due to the presence of a marginally irrelevant
operator \cite{Giamarchi89,Affleck87}. Fig.~\ref{deltaRvsQ}(b) shows results for $\delta_R$ versus $Q$. Here we see a less distinct peak in the neighborhood
of the transition point than in Fig.~\ref{deltaRvsQ}(a). Inside the VBS1 phase $\delta_R$ is large but decays rapidly with the system size. The peak and the
values inside the AFM phase also decay clearly with increasing $L$.

In Fig.~\ref{nolrdeltaRvsL} we analyze the finite-size trends in the case of the model without long-range interactions at representative
points inside the two phases and
at the transition point $Q_c=0.16478$ determined in Sec.~\ref{Subsec:LCED} (we used this $Q$ value for all sizes). Here power-law fits show that $\delta_R$
decays to zero inside the VBS1 phase but not inside the QLRO phase and at the transition point. These behaviors are expected, since the entire QLRO phase has
critical AFM and VBS order parameters that are highly correlated according to the CFT description, with only weak deformation of the O(4)
symmetry away from the QLRO--VBS1 boundary \cite{Giamarchi89,Affleck87}.


\newpage

\begin{thebibliography}{99}

\bibitem{Sandvik02}
A. W. Sandvik, S. Daul, R. R. P. Singh, and D. J. Scalapino, 
{\it Striped Phase in a Quantum XY Model with Ring Exchange}, Phys. Rev. Lett. {\bf 89}, 247201 (2002). 

\bibitem{Motrunich04}
O. I. Motrunich and A. Vishwanath, 
{\it Emergent photons and transitions in the O(3) sigma model with hedgehog suppression}, Phys. Rev. B {\bf 70}, 075104 (2004).

\bibitem{Haldane83}
F. D. M. Haldane, 
{\it Nonlinear Field Theory of Large-Spin Heisenberg Antiferromagnets: Semiclassically Quantized Solitons of the One-Dimensional Easy-Axis N\'eel State}, Phys. Rev. Lett. {\bf 50}, 1153 (1983).

\bibitem{Chakravarty89}
S. Chakravarty, B. I. Halperin, and D. R. Nelson, 
{\it Two-dimensional quantum Heisenberg antiferromagnet at low temperatures}, Phys. Rev. B {\bf 39} 2344 (1989).

\bibitem{Read90}
N. Read  and S. Sachdev, 
{\it Spin-Peierls, valence-bond solid, and N\'eel ground states of low-dimensional quantum antiferromagnets}, Phys. Rev. B {\bf 42}, 4568 (1990).

\bibitem{Murthy90}
G. Murthy and S. Sachdev, 
{\it Action of hedgehog-instantons in the disordered phase of the 2+1 dimensional CP$^{N-1}$ model}, Nucl. Phys. B {\bf 344}, 557 (1990). 

\bibitem{Senthil04a}
T. Senthil, A. Vishwanath, L. Balents, S. Sachdev, and M. P. A. Fisher, 
{\it Deconfined Quantum Critical Points}, Science {\bf 303}, 1490-1494 (2004).

\bibitem{Senthil04b}
T. Senthil, L. Balents, S. Sachdev, A. Vishwanath, and M. P. A. Fisher, 
{\it Quantum criticality beyond the Landau-Ginzburg-Wilson paradigm}, Phys. Rev. B {\bf 70}, 144407 (2004).

\bibitem{Sachdev08}
S. Sachdev, 
{\it Quantum magnetism and criticality}, Nature Phys. {\bf 4}, 173-185 (2008).

\bibitem{Sandvik07}
A. W. Sandvik, 
{\it Evidence for Deconfined Quantum Criticality in a Two-Dimensional Heisenberg Model with Four-Spin Interactions}, Phys. Rev. Lett. {\bf 98}, 227202 (2007).

\bibitem{Melko08}
R. G. Melko and R. K. Kaul, 
{\it Scaling in the Fan of an Unconventional Quantum Critical Point}, Phys. Rev. Lett. {\bf 100}, 017203 (2008).

\bibitem{Jiang08}
F.-J. Jiang, M. Nyfeler, S. Chandrasekharan, and U.-J. Wiese, 
{\it From an antiferromagnet to a valence bond solid: evidence for a first-order phase transition}, J. Stat. Mech.: Theory Exp. {\bf 2008}, P02009 (2008).

\bibitem{Lou09}
J. Lou, A. W. Sandvik, and N. Kawashima, 
{\it Antiferromagnetic to valence-bond-solid transitions in two-dimensional SU(N) Heisenberg models with multispin interactions}, Phys. Rev. B {\bf 80}, 180414 (2009).

\bibitem{Sandvik10a}
A. W. Sandvik, 
{\it Continuous Quantum Phase Transition between an Antiferromagnet and a Valence-Bond Solid in Two Dimensions: Evidence for Logarithmic Corrections to Scaling}, Phys. Rev. Lett. {\bf 104}, 177201 (2010).

\bibitem{Kaul11}
R. K. Kaul, 
{\it Quantum criticality in SU(3) and SU(4) antiferromagnets}, Phys. Rev. B {\bf 84}, 054407 (2011).

\bibitem{Harada13}
K. Harada, T. Suzuki, T. Okubo, H. Matsuo, J. Lou, H. Watanabe, S. Todo, and N. Kawashima, 
{\it Possibility of deconfined criticality in SU(N) Heisenberg models at small N}, Phys. Rev. B {\bf 88}, 220408 (2013).

\bibitem{Chen13}
K. Chen, Y. Huang, Y. Deng, A. B. Kuklov, N. V. Prokof’ev, and B. V. Svistunov, 
{\it Deconfined Criticality Flow in the Heisenberg Model with Ring-Exchange Interactions}, Phys. Rev. Lett. {\bf 110}, 185701 (2013).

\bibitem{Block13}
M. S. Block, R. G. Melko, and R. K. Kaul, 
{\it Fate of CP$^{N-1}$ Fixed Points with q Monopoles}, Phys. Rev. Lett. {\bf 111}, 137202 (2013).

\bibitem{Pujari15}
S. Pujari, F. Alet, and K. Damle, 
{\it Transitions to valence-bond solid order in a honeycomb lattice antiferromagnet}, Phys. Rev. B {\bf 91}, 104411 (2015).

\bibitem{Shao16}
H. Shao, W. Guo, and A. W. Sandvik, 
{\it Quantum criticality with two length scales}, Science, {\bf 352}, 213-216 (2016).

\bibitem{Qin17}
Y. Q. Qin, Y.-Y. He, Y.-Z. You, Z.-Y. Lu, A. Sen, A. W. Sandvik, C. Xu, and Z. Y. Meng, 
{\it Duality between the Deconfined Quantum-Critical Point and the Bosonic Topological Transition}, Phys. Rev. X {\bf 7}, 031052 (2017).

\bibitem{Ma18}
N. Ma, G.-Yu Sun, Y.-Z. You, C. Xu, A. Vishwanath, A. W. Sandvik, and Z. Y. Meng, 
{\it Dynamical signature of fractionalization at a deconfined quantum critical point}, Phys. Rev. B {\bf 98}, 174421 (2018). 

\bibitem{Kragset06}
S. Kragset, E. Sm{\o}rgrav, J. Hove, F. S. Nogueira, and A. Sudb{\o}, 
{\it First-Order Phase Transition in Easy-Plane Quantum Antiferromagnets}, Phys. Rev. Lett. {\bf 97}, 247201 (2006).

\bibitem{Sreejith14}
G. J. Sreejith and S. Powell, 
{\it Critical behavior in the cubic dimer model at nonzero monomer density}, Phys. Rev. B {\bf 89} 014404 (2014).

\bibitem{Nahum15a}
A. Nahum, J. T. Chalker, P. Serna, M. Ortu\~no, and A. M. Somoza, 
{\it Deconfined Quantum Criticality, Scaling Violations, and Classical Loop Models}, Phys. Rev. X {\bf 5} 041048 (2015)

\bibitem{Nahum15b}
A. Nahum, P. Serna, J. T. Chalker, M. Ortu\~no, and A. M. Somoza, 
{\it Emergent SO(5) Symmetry at the N\'eel to Valence-Bond-Solid Transition}, Phys. Rev. Lett. {\bf 115} 267203 (2015).

\bibitem{Sreejith19}
G. J. Sreejith, S. Powell, and A. Nahum, 
{\it Emergent SO(5) Symmetry at the Columnar Ordering Transition in the Classical Cubic Dimer Model}, Phys. Rev. Lett. {\bf 122} 080601 (2019).

\bibitem{Kaul12}
R. K. Kaul and A. W. Sandvik, 
{\it Lattice Model for the SU(N) N\'eel to Valence-Bond Solid Quantum Phase Transition at Large N}, Phys. Rev. Lett. {\bf 108}, 137201 (2012)

\bibitem{Dyer15}
E. Dyer, M. Mezei, S. S. Pufu, and S. Sachdev, 
{\t Scaling dimensions of monopole operators in the CP$^{N_b-1}$ theory in 2 + 1 dimensions}, J. High Energ. Phys. {\bf 2015} 37 (2015); Erratum {\it ibid.} {\bf 2016} 111 (2016).

\bibitem{Wang17}
C. Wang, A. Nahum, M. A. Metlitski, C. Xu, and T. Senthil, 
{\it Deconfined Quantum Critical Points: Symmetries and Dualities}, Phys. Rev. X {\bf 7}, 031051 (2017).

\bibitem{Ma19}
R. Ma and C. Wang,
{\it A theory of deconfined pseudo-criticality},
arXiv:1912.12315.

\bibitem{Nahum19}
A. Nahum,
{\it Note on Wess-Zumino-Witten models and quasiuniversality in 2+1 dimensions},
arXiv:1912.13468.

\bibitem{Voit95}
J. Voit, 
{\it One-dimensional Fermi liquids}, Rep. Prog. Phys. {\bf 58}, 977 (1995).
  
\bibitem{Nakamura99}
M. Nakamura, 
{\it Mechanism of CDW-SDW Transition in One Dimension}, J. Phys. Soc. Jpn. 68, 3123 (1999).

\bibitem{Nakamura00}
M. Nakamura, 
{\it Tricritical behavior in the extended Hubbard chains}, Phys. Rev. B {\bf 61}, 16 377 (2000).

\bibitem{Sengupta02}
P. Sengupta, A. W. Sandvik, and D. K. Campbell, 
{\it Bond-order-wave phase and quantum phase transitions in the one-dimensional extended Hubbard model}, Phys. Rev. B {\bf  65}, 155113 (2002)

\bibitem{Sandvik04}
A. W. Sandvik, L. Balents, and D. K. Campbell, 
{\it Ground State Phases of the Half-Filled One-Dimensional Extended Hubbard Model},
Phys. Rev. Lett. {\bf 92}, 236401 (2004).

\bibitem{Tsuchiizu04}
M. Tsuchiizu and A. Furusaki, 
{\it Ground-state phase diagram of the one-dimensional half-filled extended Hubbard model}, Phys. Rev. B {\bf 69}, 035103 (2004).

\bibitem{Tang11}
Y. Tang and A. W. Sandvik, 
{\it Method to Characterize Spinons as Emergent Elementary Particles},
Phys. Rev. Lett. {\bf 107}, 157201 (2011).

\bibitem{Jiang19}
S. Jiang and O. Motrunich, 
{\it Ising ferromagnet to valence bond solid transition in a one-dimensional spin chain: Analogies to deconfined quantum critical points}, Phys. Rev. B {\bf 99}, 075103 (2019).

\bibitem{Roberts19}
B. Roberts, S. Jiang, O. I. Motrunich, 
{\it Deconfined quantum critical point in one dimension}, Phys. Rev. B {\bf 99}, 165143 (2019).

\bibitem{Huang19}
R.-Z. Huang, D.-C. Lu, Y.-Z. You, Z. Y. Meng, and T. Xiang, 
{\it Emergent Symmetry and Conserved Current at a One Dimensional Incarnation of Deconfined Quantum Critical Point},
Phys. Rev. B {\bf 100}, 125137 (2019).

\bibitem{Mermin66}
N. D. Mermin and H. Wagner, 
{\it Absence of Ferromagnetism or Antiferromagnetism in One- or Two-Dimensional Isotropic Heisenberg Models}, Phys. Rev. Lett. {\bf 17}, 1133 (1966).

\bibitem{Tung11}
J. C. Tung, G. Y. Guo, 
{\it Ab initio studies of spin-spiral waves and exchange interactions in 3d transition metal atomic chains}, 
Phys. Rev. B {\bf 83}, 144403 (2011).

\bibitem{Bohnet16}
J. G. Bohnet, B. C. Sawyer, J. W. Britton, M. L. Wall, A. M. Rey, M. Foss-Feig, and J. J. Bollinger, 
{\it Quantum spin dynamics and entanglement generation with hundreds of trapped ions}, 
Science {\bf 352}, 1297 (2016).

\bibitem{Zeiher17}
J. Zeiher, J.-Y. Choi, A. Rubio-Abadal, T. Pohl, R. van Bijnen, I. Bloch, and C. Gross,
{\it Coherent Many-Body Spin Dynamics in a Long-Range Interacting Ising Chain}, 
Phys. Rev. X {\bf 7}, 041063 (2017).

\bibitem{Nguyen18}
T. L. Nguyen, J. M. Raimond, C. Sayrin, R. Corti\~ana, T. Cantat-Moltrecht, F. Assemat, I. Dotsenko, S. Gleyzes, 
S. Haroche, G. Roux, Th. Jolicoeur, and M. Brune, 
{\it Towards Quantum Simulation with Circular Rydberg Atoms}, 
Phys. Rev. X {\bf 8}, 011032 (2018).

\bibitem{Shastry81}
B. S. Shastry and B. Sutherland, 
{\it Excitation Spectrum of a Dimerized Next-Neighbor Antiferromagnetic Chain}, Phys. Rev. Lett. {\bf 47}, 964 (1981).

\bibitem{Faddeev81}
L. D. Faddeev and L. A. Takhtajan, 
{\it What is the spin of a spin wave?}, Phys. Lett. A {\bf 85}, 375 (1981).

\bibitem{Giamarchi89}
T. Giamarchi and H. J. Schulz, 
{\it Correlation functions of one-dimensional quantum systems}, 
Phys. Rev. B {\bf 39}, 4620 (1989).

\bibitem{Majumdar69a}
C. K. Majumdar and D. K. Ghosh, 
{\it On Next-Nearest-Neighbor Interaction in Linear Chain. I}, J. Math. Phys. {\bf 10}, 1388 (1969).

\bibitem{Majumdar69b}
C. K. Majumdar and D. K. Ghosh, 
{\it On Next-Nearest-Neighbor Interaction in Linear Chain. II}, J. Math. Phys. {\bf 10}, 1399 (1969).

\bibitem{Uhrig97}
G. S. Uhrig, 
{\it Symmetry and Dimension of the Magnon Dispersion of Inorganic Spin-Peierls Systems}, 
Phys. Rev. Lett. 79, 163 (1997).

\bibitem{Sandvik99}
A. W. Sandvik and D. K. Campbell, 
{\it Spin-Peierls Transition in the Heisenberg Chain with Finite-Frequency Phonons}, 
Phys. Rev. Lett. {\bf 83}, 195 (1999).

\bibitem{Suwa15}
H. Suwa and S. Todo, 
{\it Generalized Moment Method for Gap Estimation and Quantum Monte Carlo Level Spectroscopy}, 
Phys. Rev. Lett. {\bf 115}, 080601 (2015). 

\bibitem{Sanyal11}
S. Sanyal, A. Banerjee, and K. Damle, 
{\it Vacancy-induced spin texture in a one-dimensional S=$\frac{1}{2}$ Heisenberg antiferromagnet}, Phys. Rev. B {\bf 84}, 235129 (2011)

\bibitem{Patil18}
P. Patil, E. Katz, and A. W. Sandvik, 
{\it Numerical investigations of SO(4) emergent extended symmetry in spin-$\frac{1}{2}$ Heisenberg antiferromagnetic chains}, Phys. Rev. B {\bf 98}, 014414 (2018).

\bibitem{Affleck85}
I. Affleck, 
{\it Critical Behavior of Two-Dimensional Systems with Continuous Symmetries}, Phys. Rev. Lett. {\bf 55}, 1355 (1985).

\bibitem{Affleck87}
I. Affleck and F. D. M. Haldane, 
{\it Critical theory of quantum spin chains}, Phys. Rev. B {\bf 36}, 5291 (1987).

\bibitem{Sulejman17}
T. Sulejmanpasic, H. Shao, A. W. Sandvik, and M. \"Unsal, 
{\it Confinement in the Bulk, Deconfinement on the Wall: Infrared Equivalence between Compactified QCD and Quantum Magnets}, Phys. Rev. Lett. {\bf 119}, 091601 (2017).

\bibitem{Laflorencie05}
N. Laflorencie, I. Affleck, and M. Berciu, 
{\it Critical phenomena and quantum phase transition in long range Heisenberg antiferromagnetic chains}, J. Stat. Mech. {\bf 2005} P12001 (2005).

\bibitem{Sandvik10b}
A. W. Sandvik, 
{\it Ground States of a Frustrated Quantum Spin Chain with Long-Range Interactions}, Phys. Rev. Lett. {\bf 104}, 137204 (2010).

\bibitem{Yusuf04}
E. Yusuf, A. Joshi, and K. Yang, 
{\it Spin waves in antiferromagnetic spin chains with long-range interactions}, Phys. Rev. B {\bf 69} 144412 (2004).

\bibitem{Levin04}
M. Levin and T. Senthil, 
{\it Deconfined quantum criticality and N\'eel order via dimer disorder}, 
Phys. Rev. B {\bf 70}, 220403(R) (2004).

\bibitem{Senthil06}
T. Senthil and M. P. A. Fisher, 
{\it Competing orders, nonlinear sigma models, and topological terms in quantum magnets}, 
Phys. Rev. B {\bf 74}, 064405 (2006).

\bibitem{Suwa16}
H. Suwa, A. Sen, and A. W. Sandvik, 
{\it Level spectroscopy in a two-dimensional quantum magnet: Linearly dispersing spinons at the deconfined quantum critical point}, 
Phys. Rev. B {\bf 94}, 144416 (2016).

\bibitem{Koga00}
A. Koga and N. Kawakami, 
{\it Quantum Phase Transitions in the Shastry-Sutherland Model for SrCu$_2$(BO$_3$)$_2$}, Phys. Rev. Lett. {\bf 84}, 4461 (2000).

\bibitem{Corboz13}
P. Corboz and F. Mila, 
{\it Tensor network study of the Shastry-Sutherland model in zero magnetic field}, Phys. Rev. B {\bf 87}, 115144 (2013).

\bibitem{Zhao19}
B. Zhao, P. Weinberg, and A. W. Sandvik, 
{\it Symmetry-enhanced discontinuous phase transition in a two-dimensional quantum magnet}, Nature Phys. {\bf 15}, 678 (2019).

\bibitem{Lee19}
J. Y. Lee, Y.-Z. You, S. Sachdev, A. Vishwanath, 
{\it Signatures of a Deconfined Phase Transition on the Shastry-Sutherland Lattice: Applications to Quantum Critical SrCu$_2$(BO$_3$)$_2$}, Phys. Rev. X {\bf 9}, 041037 (2019).

\bibitem{Serna19}
P. Serna and A. Nahum, 
{\it Emergence and spontaneous breaking of approximate O(4) symmetry at a weakly first-order deconfined phase transition}, Phys. Rev. B {\bf 99}, 195110 (2019).

\bibitem{Yu19}
J. Yu, R. Roiban, S.-K. Jian, and C.-X. Liu, 
{\it Finite-scale emergence of 2+1D supersymmetry at first-order quantum phase transition}, Phys. Rev. B {\bf 100}, 075153 (2019).

\bibitem{Zayed17}
M. Zayed, Ch. R\"ueegg, J. Larrea, A. M. L\"auchli, C. Panagoplos, S. S. Saxena, M. Ellerby, D. McMorr, Th. Str\"assle,
S. S. Klotz, G. Hamel, R. A. Sadykov, V. Pomjakushin, M. Boehm, M. Jimin\'ez-Ruiz, A. Schneidewin, E. Pomjakushin, M. Stingaciu, 
K. Conder, and H. M. R{\o}nnow, 
{\it 4-spin plaquette singlet state in the Shastry-Sutherland compound SrCu$_2$(BO$_3$)$_2$}, Nature Phys. {\bf 13}, 962 (2017).

\bibitem{Guo19}
J. Guo, G. Sun, B. Zhao, L. Wang, W. Hong, V. A. Sidorov, N. Ma, Q. Wu, S. Li, Z. Y. Meng, A. W. Sandvik, and L. Sun, 
{\it Quantum phases of SrCu$_2$(BO$_3$)$_2$, from high-pressure thermodynamics}, arXiv:1904.09927.

\bibitem{Sandvik10d}
A. W. Sandvik, 
{\it Computational Studies of Quantum Spin Systems}, AIP Conf. Proc. {\bf 1297}, 135 (2010).

\bibitem{S2comment}  
In Refs.~\cite{Sandvik10b} and \cite{Sandvik10d} it was stated that the QLRO--AFM transition is characterized by a level crossing
of two $S=0$ excited states. However, one of the levels had been misidentified and actually has $S=2$ \cite{Wang18}. The conclusions
regarding this level crossing being associated with the QLRO--AFM transition in Refs.~\cite{Sandvik10b} and \cite{Sandvik10d} were
in other respects confirmed  by the DMRG calculations on larger chains in Ref.~\cite{Wang18}.

\bibitem{Wang18}
L. Wang and A. W. Sandvik, 
{\it Critical Level Crossings and Gapless Spin Liquid in the Square-Lattice Spin-1/2 J$_1$-J$_2$ Heisenberg Antiferromagnet}, Phys. Rev. Lett. {\bf 121}, 107202 (2018). 

\bibitem{Kumar13}
M. Kumar and Z. G. Soos, 
{\it Decoupled phase of frustrated spin-1/2 antiferromagnetic chains with and without long-range order in the ground state}, 
Phys. Rev. B {\bf 88}, 134412 (2013).

\bibitem{Sandvik10c}
A. W. Sandvik and H. G. Evertz, 
{\it Loop updates for variational and projector quantum Monte Carlo simulations in the valence-bond basis}, Phys. Rev. B {\bf 82}, 024407 (2010).

\bibitem{Beach06}
K. S. D. Beach and A. W. Sandvik, 
{\it Some formal results for the valence bond basis},
 Nucl. Phys. B {\bf 750}, 242 (2006).

 \bibitem{Sandvik03}
A. W. Sandvik, 
{\it Stochastic series expansion method for quantum Ising models with arbitrary interactions},
Phys. Rev. B {\bf 68}, 056701 (2003).

\bibitem{Liang88}
S. Liang, B. Doucot, and P. W. Anderson, 
{\it Some New Variational Resonating-Valence-Bond-Type Wave Functions for the Spin-1/2 Antiferromagnetic Heisenberg Model on a Square Lattice}, Phys. Rev. Lett. {\bf 61}, 365 (1988).

\bibitem{Binder81}
K. Binder, 
{\it Finite size scaling analysis of ising model block distribution functions}, Z. Phys. B {\bf 43}, 119 (1981).

\bibitem{Liu18}
L. Liu, H. Shao, Y.-C. Lin, W. Guo, and A. W. Sandvik, 
{\it Random-Singlet Phase in Disordered Two-Dimensional Quantum Magnets}, Phys. Rev. X {\bf 8}, 041040 (2018).

\bibitem{dnote}
In the PQMC method $D$ for a given sampled configuration has a simple estimator based on individual transition-graph loops \cite{Liang88,Sandvik10c}, 
and $D^2$ can be computed based on an estimator involving also two loops \cite{Beach06}. However, $D^4$ has a very complicated expression \cite{Beach06}
and it is not feasible to compute it in practice. To compute $U_V$, we therefore simply use the number $D$ obtained from the transition graph and take 
its second and fourth power in each configuration for alternative definitions of $\langle D^2\rangle$ and $\langle D^4\rangle$. These mean values
deviate very little from the correct expectation values and are expected to have the same scaling forms. We use $\langle D^2\rangle$ computed 
with the full two-loop estimator for studying the critical scaling of the VBS order parameter,

\bibitem{Ma18b}
N. Ma, P. Weinberg, H. Shao, W. Guo, D.-X. Yao, and A. W. Sandvik, 
{\it Anomalous Quantum-Critical Scaling Corrections in Two-Dimensional Antiferromagnets}, 
Phys. Rev. Lett. {\bf 121}, 117202 (2018).

\bibitem{Nomura92}
K. Nomura and K. Okamoto, 
{\it Fluid-dimer critical point in S=1/2 antiferromagnetic Heisenberg chain with next nearest neighbor interactions}, 
Phys. Lett. A {\bf 169}, 433 (1992).

\bibitem{Eggert96}
S. Eggert, 
{\it Numerical evidence for multiplicative logarithmic corrections from marginal operators}, 
Phys. Rev. B {\bf 54}, R9612 (1996).

\bibitem{Sen15}
A. Sen, H. Suwa, and A. W. Sandvik,
{\it Velocity of excitations in ordered, disordered, and critical antiferromagnets},
Phys. Rev. B {\bf 92}, 195145 (2015).

\bibitem{Bentler83}
P. M . Bentler, 
{\it Some contributions to efficient statistics in structural models: specification and estimation of moment structures}, 
Psychometrika {\bf 48}, 493 (1983).

\bibitem{Paindaveine12}
D. Paindaveine, 
{\it Elliptical symmetry} in Encyclopedia of Environmetrics, 2nd edition, A. H. El-Shaarawi and W. Piegorsch (eds), pp.~802-807
 John Wiley \& Sons Ltd (Chichester, 2012).

\bibitem{Tang13}
Y. Tang and A. W. Sandvik, 
{\it Confinement and Deconfinement of Spinons in Two Dimensions}, 
Phys. Rev. Lett. {\bf 110}, 217213 (2013).
  
\end{thebibliography}
\end{document}